\documentclass[12pt,reqno]{amsart}

\usepackage{amsmath,amssymb,amsthm}
\usepackage{enumitem}
\usepackage{fullpage}
\usepackage{graphicx}
\usepackage{subcaption}
\usepackage{mathrsfs}

\numberwithin{equation}{section}

\newtheorem{thm}{Theorem}

\newtheorem{prop}{Proposition}

\def\Ref#1{Ref.~\cite{#1}}

\DeclareMathOperator\arctanh{arctanh} 
\DeclareMathOperator\sgn{sgn} 

\def\smallsqrt#1{{\textstyle\sqrt{\vphantom{()} #1}}}

\def\rhs{\text{ r.h.s. }}
\def\const{\text{const.}}

\def\Rnum{{\mathbb R}}

\def\u{\mathbf{u}}
\def\A{\mathbf{A}}

\def\L{\mathcal{L}}
\def\T{\mathcal{T}}

\def\newt{\text{Newt}}
\def\eff{\text{eff}}
\def\circ{\text{circ}}

\def\F{\mathrm{F}}
\def\E{\mathrm{E}}

\def\i0{{i^0}}
\def\I{\mathscr{I}}

\allowdisplaybreaks[3]

\begin{document}

\title{Analogue of a Laplace-Runge-Lenz vector\\ for particle orbits (timelike geodesics)\\ in Schwarzschild spacetime}

\author{
Stephen C. Anco$^1$ 
\lowercase{\scshape{and}}
Jordan A. Fazio$^2$
\\\\
$^1$D\lowercase{\scshape{epartment}} \lowercase{\scshape{of}} M\lowercase{\scshape{athematics and}} S\lowercase{\scshape{tatistics}}\\
B\lowercase{\scshape{rock}} U\lowercase{\scshape{niversity}}\\
S\lowercase{\scshape{t.}} C\lowercase{\scshape{atharines}}, ON L2S3A1, C\lowercase{\scshape{anada}}
\\\\
$^2$D\lowercase{\scshape{epartment}} \lowercase{\scshape{of}} P\lowercase{\scshape{hysics}}\\
U\lowercase{\scshape{niversity}} \lowercase{\scshape{of}} T\lowercase{\scshape{oronto}}\\
T\lowercase{\scshape{oronto}}, ON M5S1A7, C\lowercase{\scshape{anada}}
}

\begin{abstract}
In Schwarzschild spacetime, 
the timelike geodesic equations, which define particle orbits, 
have a well-known formulation as a dynamical system 
in coordinates adapted to the timelike hypersurface containing the geodesic. 
For equatorial geodesics, 
the resulting dynamical system is shown to possess 
a conserved angular quantity and two conserved temporal quantities,
whose properties and physical meaning are analogues of 
the conserved Laplace-Runge-Lenz vector, 
and its variant known as Hamilton's vector, in Newtonian gravity. 
When a particle orbit is projected into the spatial equatorial plane, 
the angular quantity yields the coordinate angle at which the orbit has 
either a turning point (where the radial velocity is zero)
or a centripetal point (where the radial acceleration is zero). 
This is the same property as the angle of the respective 
Laplace-Runge-Lenz and Hamilton vectors in the plane of motion in Newtonian gravity. 
The temporal quantities yield the coordinate time and the proper time 
at which those points are reached on the orbit. 
In general, for orbits that have a single turning point,
the three quantities are globally constant; 
for orbits that possess more than one turning point,
the temporal quantities are just locally constant 
as they jump at every successive turning point,
while the angular quantity similarly jumps only if an orbit is precessing. 
This is analogous to the properties of a generalized Laplace-Runge-Lenz vector 
and generalized Hamilton vector 
which are known to exist for precessing orbits in post-Newtonian gravity. 
The angular conserved quantity can be used to define a direct analogue of these vectors at spatial infinity. 
\end{abstract}

\maketitle

\begin{center}
emails:\
sanco@brocku.ca,\ 
j.fazio@mail.utoronto.ca
\end{center}

\section{Introduction}\label{sec:intro}

The orbits of a massive particle around a spherical black hole 
are described by the timelike geodesic equations in Schwarzschild spacetime, 
where the spacetime motion of the particle lies in a timelike hypersurface. 
By rotational symmetry, this hypersurface can be identified with  
the equatorial hypersurface in adapted polar coordinates $(t,r,\phi)$, 
with the metric $ds^2=-\frac{r-2M}{r}dt^2+\frac{r}{r-2M}dr^2+r^2d\phi^2$. 
Consequently, the spatial motion of the particle can be described as 
belonging to an equatorial plane which is orthogonal to the timelike Killing vector $\partial/\partial_t$. 
It is well known \cite{MisThoWhe,Wal} 
that the coordinate form of the geodesic equations 
then can be integrated with use of the Killing vectors 
$\partial/\partial_t$ and $\partial/\partial_\phi$
to obtain the geodesics $(t(\tau),r(\tau),\phi(\tau))$ parameterized by proper time $\tau$
in an explicit analytical form \cite{Hag,Dar}. 
This yields the particle orbit $(r(\tau),\phi(\tau))$ in the equatorial plane. 
A complete exposition, including a classification of all types of orbits, 
can be found in the comprehensive reference \cite{Cha}. 

In the limit of Newtonian gravity, 
all non-circular particle orbits have an interesting property that 
they possess a conserved quantity known as the Laplace-Runge-Lenz (LRL) vector
\cite{GolPooSaf,Cor}. 
This quantity is a constant of motion for non-circular orbits. 
It lies in the plane of the orbit 
and points from the focus to the periapsis point of the orbit, 
while its magnitude is normalized to depend only on the energy constant of motion 
for the orbit. 
When polar coordinates $(r,\phi)$ are used on the orbital plane, 
the focus of the orbit is located at  $r=0$ and thus the angle of the LRL vector 
in this plane coincides with the angle at which the periapsis point occurs in the orbit. 

More generally, a counterpart of the LRL vector exists for non-circular particle orbits 
in any central force dynamics \cite{BacRueSou,Fra}. 
This conserved vector has the same dynamical features as in the Newtonian case: 
it lies in the orbital plane along the radial line(s) from the origin to the periapsis point(s),
and it is a constant of motion for orbits that are non-precessing. 
But for precessing orbits, it is only a local constant of motion,
because after a particle passes through one periapsis point 
then when the particle reaches the apoapsis point this vector jumps 
such that it lies along a radial line from the origin to the next (up coming) 
periapsis point on the orbit \cite{Cor,AncMeaPas}. 
This piecewise conservation property of the LRL vector 
has been well-studied in classical mechanics \cite{SerSha,BucDen,Per,LeaFle}. 
A variant of the LRL vector can be defined similarly by using the apoapsis point(s). 
Thus, the most general form of the LRL vector is geometrically tied to 
the turning points of an orbit, at which the radial speed is zero. 

There is a related conserved vector, called Hamilton's vector \cite{Cor}, 
which differs by lying on the radial lines(s) from the origin to the point(s) of the orbit
at which the radial acceleration is zero. 
The latter points will be called centripetal points hereafter. 

A case of central force dynamics that is relevant to 
the timelike equatorial geodesics in Schwarzschild spacetime is given by 
the ``revolving'' orbits in Newtonian gravity with a cubic correction \cite{Cha1995,Lyn-Bel}, 
since this describes the post-Newtonian limit of the geodesics. 
A detailed derivation and description of the generalized LRL vector (and its variants) 
in that case has been explored in recent work \cite{AncMeaPas}. 

This motivates the natural question of whether some analogue of a LRL vector 
exists for the Schwarzschild timelike geodesic equations themselves. 
The primary purpose of the present work is to obtain such an analogue. 

Specifically, an angular conserved quantity will be found that 
shares the main properties of the angle of the LRL-vector in post-Newtonian gravity, 
and reduces to this angle in the post-Newtonian limit of timelike geodesics. 
In particular, when a non-circular equatorial timelike geodesic is projected 
into the equatorial plane, giving a spatial orbit $(r(\tau),\phi(\tau))$, 
then the angular conserved quantity will yield 
the coordinate angle of radial lines that intersect the orbit at an apsis point. 
Additionally, two related temporal conserved quantities will be derived, which yield
the coordinate time and proper time at which the apsis point is reached. 
All three quantities are globally constant for orbits that have a single apsis point, 
and otherwise are just locally constant in general. 

The derivation of these LRL-type quantities starts from the well-known integration of
the timelike equatorial geodesic equations in terms of the coordinates $(t,r,\phi)$. 
In standard treatments (cf \cite{MisThoWhe,Cha}), 
the integration is given by first integrals of the equations of motion 
for $(t(\tau),r(\tau),\phi(\tau))$ 
in which the constants of integration are either omitted or arbitrary 
or taken to be initial values for the orbital motion. 
Instead, in the present work, these constants are fixed 
in a way that strictly involves the intrinsic properties of the dynamics of the spatial orbit 
$(r(\tau),\phi(\tau))$. 
This leads to first integrals that do not involve either an arbitrary integration constant 
or an integration constant related to initial values, 
and thus these first integrals differ from the ones appearing in standard treatments. 
They will be referred to as an angular $\phi$-quantity, a temporal $t$-quantity, 
and a proper time $\tau$-quantity. 

In particular, 
the derivation will show how to use turning points or centripetal points of 
the spatial orbit $(r(\tau),\phi(\tau))$ 
to fix the integration constants in the first integrals of the geodesic motion 
for $(t(\tau),r(\tau),\phi(\tau))$. 
These points can be found directly from the Schwarzschild effective potential, 
in terms of the energy and angular momentum constants of motion, 
without the need to have integrated the geodesic equations of motion. 
The resulting $\phi$-quantity will be shown to yield 
the coordinate angle of radial lines that intersect the spatial orbit 
at either a turning point or a centripetal point. 
An important feature of this LRL quantity is that it is only locally conserved 
because it undergoes a jump at every successive apsis point 
in the case of an elliptic-like orbit that precesses. 
For all other non-circular orbits, there is no jump in the $\phi$-quantity. 
The locally conserved $t$-quantity and $\tau$-quantity describe 
the coordinate time(s) and proper time(s) at which 
successive turning/centripetal points are reached in a non-circular orbit. 
In contrast to the $\phi$-quantity, 
they jump at every apsis point for all orbits that possess more than one turning point 
(not just for precessing elliptic-like orbits). 

For orbits that cross the horizon, the integration constants can be fixed alternatively 
by use of a horizon-crossing point instead of either a turning point or centripetal point. 
Indeed, this becomes the only possibility in the case of asymptotically-circular orbits 
that cross the horizon. 

In the case of an orbit with a turning point, which is an apoapsis or a periapsis, 
the resulting three conserved quantities have been derived previously \cite{Kos,GomHorKos} (see additional references therein) 
in work motivated by accurate numerical calculations of 
orbits near a black hole and earth satellite orbits for a global positioning network. 
However, the connection of these quantities to an analogue of the LRL vector 
was not given, nor were their global properties discussed. 
Two new aspects of the present work are that 
the three quantities are generalized using centripetal points as well as horizon-crossing points,
and that the non-trivial global properties of these quantities 
are comprehensively explained for all types of orbits. 

The three conserved $\phi$, $t$, $\tau$-quantities obtained here 
are physically and mathematically important:\\
$\bullet$  
they do not involve initial values, similarly to the energy and angular momentum constants of motion;\\
$\bullet$  
they yield an intrinsic description of an orbit in terms of a complete set of integrals of motion; \\
$\bullet$  
they allow predicting future and past turning/centripetal points on an orbit,
as well as any horizon-crossing points, 
strictly in terms of values of dynamical variables at any point in the orbit; \\
$\bullet$
they indicate a ``hidden'' dynamical symmetry structure in the geodesic equations of motion. 

Moreover, the conserved angular quantity determines, for each orbit, 
a generalized LRL vector at spatial infinity in Schwarzschild spacetime. 
This vector inherits the local and global properties of the angular quantity
and it constitutes a natural analogue of the Newtonian LRL vector
in the case of a turning point, 
and the Newtonian Hamilton vector in the case of a centripetal point. 
There is no Newtonian analogue in the case of a horizon-crossing point. 
Construction of the generalized LRL vector is a main new result of this paper. 

The rest of the paper is organized as follows. 

In section~\ref{sec:eom},
the timelike equatorial geodesic equations are briefly reviewed 
as a first-order dynamical system, 
which will provide a useful framework for deriving intrinsic conserved quantities
without the use of initial conditions. 
The notion of an integral of motion and local versus global conservation 
is also reviewed. 

In section~\ref{sec:firstintegrals},
the conserved $\phi$, $t$, $\tau$-quantities are derived 
starting from the well-known separability \cite{MisThoWhe,Cha} 
of the geodesic equations of motion, 
by adapting recent work \cite{AncMeaPas} 
on LRL-type integrals of motion in general central force dynamics,
which uses turning points and centripetal points of the spatial dynamics. 
Altogether, the resulting five conserved quantities provide a complete set of 
integrals of motion for $(t(\tau),r(\tau),\phi(\tau))$,
and discussion is given of how they differ from the first integrals 
seen in standard treatments.

In section~\ref{sec:properties}, 
the global properties of the $\phi$, $t$, $\tau$-quantities as analogues of
the (post-) Newtonian LRL vector are discussed. 
The $\phi$-quantity is shown to be multi-valued on orbits that precess,
whereas the $t$, $\tau$-quantities are shown to be multi-valued on orbits that possess more than one turning point. 

In section~\ref{sec:LRLvector}, 
the generalized LRL vector at spatial infinity is constructed
and its properties are discussed. 

In section~\ref{sec:remarks}, 
some concluding remarks are given,
including how the $\phi$, $t$, $\tau$-quantities correspond to 
dynamical symmetries of the geodesic Lagrangian. 

Two appendices contain additional results. 
In appendix~\ref{sec:evaluate},
the $\phi$, $t$, $\tau$-quantities are expressed in an explicit analytical form 
in terms of elementary and elliptic functions 
for each different type of timelike equatorial non-circular geodesics 
in Schwarzschild spacetime.
Appendix~\ref{sec:newtonian}
discusses the Newtonian limit of the $\phi,t,\tau$-quantities
and explains their relationship with the LRL vector and Hamilton's vector
in Newtonian gravity. 

Throughout, geometrized units with $c=1$ and $G=1$ are used.

\section{Equations of motion and integrals of motion}\label{sec:eom}

In the Schwarzschild black hole spacetime, 
the metric is given by the line element \cite{MisThoWhe,Wal,Cha}
\begin{equation}\label{metric}
ds^2=-\frac{r-2M}{r}dt^2+\frac{r}{r-2M}dr^2+r^2\sin^2\theta d\phi^2+r^2d\theta^2
\end{equation}
with the use of standard spherical coordinates $(t,r,\phi,\theta)$,
where $M$ denotes the Komar mass of the spacetime,
and $r=2M$ is the spherical null surface constituting the horizon. 

Consider a free massive particle moving on a timelike geodesic outside of the horizon. 
As is well known \cite{MisThoWhe,Wal,Cha}, 
any timelike geodesic lies in a hypersurface, $\Sigma$, 
that is isometric to the equatorial hypersurface $\theta = \tfrac{1}{2}\pi$ 
under a global spatial rotation. 
(In \Ref{Cha}, this hypersurface is called the ``invariant plane''.)
Thus, $\theta$ can be fixed to be $\tfrac{1}{2}\pi$ by spherical symmetry,
and then the metric on the hypersurface containing the timelike geodesic is given by
\begin{equation}
ds^2\big|_\Sigma =-\frac{r-2M}{r}dt^2+\frac{r}{r-2M}dr^2+r^2d\phi^2 .
\end{equation}
This metric has two Killing vectors $\partial_t$ and $\partial_\phi$. 

The particle's 4-velocity is expressed as 
\begin{equation}\label{4vel}
\u = \sigma \partial_t + v \partial_r + \omega \partial_{\phi}, 
\end{equation}
with the notation 
\begin{align}\label{4vel-components}
\sigma = \frac{dt}{d\tau}, 
\quad
v = \frac{dr}{d\tau}, 
\quad
\omega = \frac{d\phi}{d\tau} ,
\end{align}
where $\tau$ denotes proper time of the particle, as defined by 
\begin{equation}\label{propertimeeqn}
g(\u,\u) =-1 .
\end{equation}
The equation of motion of the particle is given by the timelike geodesic equation
\begin{equation}\label{timelikegeodesiceqn}
\frac{d\u}{d \tau} = 0 . 
\end{equation}

In component form, 
the timelike geodesic equation \eqref{timelikegeodesiceqn} 
and the proper time equation \eqref{propertimeeqn} 
are given by 
\begin{subequations}\label{timelikegeodesiceqn-componentform}
\begin{gather}
\frac{d\sigma}{d\tau}+\frac{2M}{r(r-2M)}\sigma v = 0,
\\
\frac{dv}{d\tau} -\frac{M}{r(r-2M)}v^2+\frac{M(r-2M)}{r^3}\sigma^2 -(r-2M)\omega^2 = 0,
\\
\frac{d\omega}{d\tau} +\frac{2}{r}\omega v = 0,
\end{gather}
\end{subequations}
and 
\begin{equation}\label{propertimeeqn-componentform}
\frac{r-2M}{r}\sigma^2 -\frac{r}{r-2M}v^2-r^2\omega^2 =1 . 
\end{equation}
These equations \eqref{timelikegeodesiceqn-componentform}--\eqref{propertimeeqn-componentform} 
for the 4-velocity components \eqref{4vel-components}
comprise a coupled system of nonlinear second-order ODEs 
when expressed in terms of the dynamical variables $(t(\tau),r(\tau),\phi(\tau))$:
\begin{subequations} \label{timelikegeodesiceqn-2ndord}
\begin{align}
\ddot{t} & = -\frac{2M\dot{t}\dot{r}}{r(r-2M)}, 
\label{tddot} 
\\
\ddot{r} & = \frac{M\dot{r}^2}{r(r-2M)} -\frac{M(r-2M)\dot{t}^2}{r^3}  + (r-2M)\dot{\phi}^2 , 
\label{rddot} 
\\
\ddot{\phi} & = -\frac{2\dot{\phi}\dot{r}}{r} , 
\label{phiddot}
\end{align}
\end{subequations}
and 
\begin{equation}\label{propertimeeqn-2ndord}
\frac{r-2M}{r}\dot{t}^2-\frac{r}{r-2M}\dot{r}^2 -r^2\dot{\phi}^2 =1 , 
\end{equation}
where a dot denotes differentiation with respect to $\tau$. 
Note that the proper time equation \eqref{propertimeeqn-2ndord}
can be used to simplify the radial acceleration equation:
\begin{equation}\label{radial-accel}
\ddot{r} = (r-3M) \dot{\phi}^2 - \frac{M}{r^2} .
\end{equation}
The second-order ODEs \eqref{timelikegeodesiceqn-2ndord} 
can be derived from the geodesic Lagrangian \cite{MisThoWhe,Wal,Cha}
\begin{equation}\label{Lagr}
\L = \tfrac{1}{2}g(\u,\u) = \tfrac{1}{2}\Big( -\frac{r-2M}{r}\dot{t}^2 +\frac{r}{r-2M}\dot{r}^2 +r^2\dot{\phi}^2 \Big) ,
\end{equation}
whose Euler-Lagrange equations are given by 
\begin{subequations}\label{ELeqns}
\begin{align}
\delta \L/\delta t = & (1 -2M/r) \big( \ddot t -\rhs\eqref{tddot} \big) ,
\\
\delta \L/\delta r = & {-(r/(r-2M))} \big( \ddot r -\rhs\eqref{rddot} \big) ,
\\
\delta \L/\delta \phi = & {-r^2} \big( \ddot \phi -\rhs\eqref{phiddot} \big) .
\end{align}
\end{subequations}
Each solution $( t(\tau), r(\tau), \phi(\tau) )$ of 
the ODE system \eqref{timelikegeodesiceqn-2ndord}--\eqref{propertimeeqn-2ndord}
describes a timelike equatorial geodesic, parameterized by proper time, 
in Schwarzschild spacetime.

An \emph{integral of motion} of the timelike equatorial geodesic equations 
is a scalar function
\begin{equation}\label{FI}
I(\tau,t,r,\phi,\sigma,v,\omega)
\end{equation}
that is (locally) constant with respect to proper time, $\tau$, 
on (segments of) every geodesic $(t(\tau),r(\tau),\phi(\tau) )$. 
In particular, 
$I$ is not necessarily a global constant on the entirety of a geodesic 
but may be merely piecewise constant and undergo finite jumps at certain points. 
This generality is essential for understanding 
the $t$, $\tau$-integrals of motion for orbits with more than one turning point 
as well as the $\phi$-integral of motion for precessing orbits. 
If an integral of motion does not depend explicitly on $\tau$, namely $I_\tau=0$, 
then it defines a \emph{(local) constant of motion}. 

Integrals of motion in general describe locally conserved quantities. 
In the literature on ODEs and classical mechanics,
such quantities are sometimes called first integrals. 
However, in the literature on super-integrability (see e.g. \Ref{MilPosWin}), 
a ``first integral'' is typically meant to be a globally conserved (single-valued) quantity, 
such as energy and angular momentum. 
This distinction will be important for understanding the status of 
the $\phi,t,\tau$-quantities. 
In particular, their existence does not imply super-integrability, 
but they still are associated with dynamical symmetries, 
as will be outlined in section~\ref{sec:remarks}. 
See also \Ref{AncBalGan} for a general discussion.

\section{LRL quantities for timelike equatorial geodesics}\label{sec:firstintegrals}

It is well known that the timelike equatorial geodesic equations \eqref{timelikegeodesiceqn-2ndord}
are separable in the sense that they can be integrated to obtain the geodesic motion 
$( t(\tau), r(\tau), \phi(\tau) )$ in terms of initial conditions. 
See \Ref{Cha} for a comprehensive treatment. 
A brief summary will be given here as a preliminary step 
in the derivation of the conserved $\phi$, $t$, $\tau$-quantities
discussed in Section~\ref{sec:intro}. 

First, since the proper time equation \eqref{propertimeeqn-componentform} provides 
a relation among the variables $r,\sigma,v,\omega$, 
one of them can be eliminated in terms of the others, 
without loss of generality in an integral of motion \eqref{FI}. 
It is convenient to eliminate
\begin{equation}\label{vsq}
v^2=(1-2M/r)^2\sigma^2- (1-2M/r)(1+r^2\omega^2) . 
\end{equation}
Then every integral of motion is given by a function 
\begin{equation}\label{I}
I(\tau,t,r,\phi,\sigma,\omega)
\end{equation}
that satisfies 
\begin{equation}\label{deteqn}
0 = \frac{dI}{d\tau} =
I_{\tau} + I_t \sigma + I_r v + I_{\phi} \omega 
-I_\sigma \frac{2M\sigma v}{r(r-2M)} - I_{\omega} \frac{2\omega v}{r} 
\end{equation}
locally in $\tau$, for all timelike equatorial geodesics. 
This is a linear first-order PDE which can be turned into a system of ODEs 
via the method of characteristics \cite{Joh}:
\begin{equation}\label{characteristics}
\frac{d \tau}{1} 
= \frac{dt}{\sigma} 
= \frac{dr}{v} 
= \frac{d\phi}{\omega} 
= -\frac{r(r-2M)d\sigma}{2M\sigma v} 
= - \frac{r d\omega}{2\omega v}.
\end{equation}
In particular, each solution of this ODE system defines a characteristic curve 
in the space of variables $(\tau,t,r,\phi,\sigma,\omega)$, 
where the curve corresponds to a geodesic with an arbitrary parameterization. 

Second, the ODE system \eqref{characteristics} can be arranged into a triangular form 
in which the ODEs successively become separable and hence can be integrated.  
To begin, 
$\dfrac{dr}{v} = -\dfrac{r d\omega}{2 \omega v}$ is separable,
which yields the integral of motion 
\begin{equation}\label{I1}
I_1=\omega r^2 .
\end{equation}
This quantity coincides with the constant of motion given by the rotation Killing vector,
which is the angular momentum, $L=I_1$. 
Likewise, 
$\dfrac{dr}{v} = -\dfrac{r (r-2M) d \sigma}{2 M \sigma v}$ is separable, 
and hence it yields the integral of motion 
\begin{equation}\label{I2}
I_2 =\frac{r-2M}{r}\sigma ,
\end{equation}
which coincides with the energy constant of motion 
given by the time-translation Killing vector, $E=I_2$. 
Then the relation \eqref{vsq} can be rewritten in terms of the integrals of motion $I_1$ and $I_2$,
giving 
\begin{equation}\label{vsquared}
v^2 = I_2{}^2  - \frac{r - 2M}{r^3}I_1{}^2 - \frac{r - 2M}{r} 
\end{equation}
which expresses $v$ as a function only of $r$. 
Next, through expressions \eqref{vsquared} and \eqref{I1}, 
$\dfrac{dr}{v} =\dfrac{d\phi}{\omega} = \dfrac{r^2 d\phi}{I_1}$ is separable. 
This yields the integral of motion 
\begin{equation}\label{I3}
I_3 = \phi - I_1 \int\frac{dr}{r^2 v} ,
\end{equation}
provided that $v$ is not identically zero. 
Similarly, with the use of $I_2$, 
$\dfrac{dr}{v} =\dfrac{dt}{\sigma} = \dfrac{(r - 2M) d t}{r I_2}$ is separable, 
thus yielding one more integral of motion
\begin{equation}\label{I4}
I_4 = t - I_2 \int\frac{rdr}{(r-2M)v} , 
\end{equation}
again provided that $v$ is not identically zero. 
Note that $v\equiv 0$ corresponds to a circular timelike equatorial geodesic. 

The four integrals of motion \eqref{I1}, \eqref{I2}, \eqref{I3}, \eqref{I4} 
do not contain $\tau$ explicitly,
and therefore they are local constants of motion
for non-circular geodesics. 
A fifth integral of motion, which involves $\tau$ explicitly, 
comes from integrating 
$\dfrac{d\tau}{1}=\dfrac{dr}{v}$. 
This yields 
\begin{equation}\label{I5}
I_5 = \tau - \int \frac{dr}{v} . 
\end{equation}
The proper time equation  \eqref{propertimeeqn-componentform} 
can be viewed as being a sixth integral of motion which serves as a constraint. 

For non-circular geodesics, 
the six integrals of motion 
\eqref{I1}, \eqref{I2}, \eqref{I3}, \eqref{I4}, \eqref{I5}, and \eqref{propertimeeqn-componentform} 
constitute a complete, functionally independent set of first integrals. 
In particular, the timelike equatorial geodesic equations in first-order form 
\eqref{4vel-components} and \eqref{timelikegeodesiceqn-componentform}
for $( t(\tau), r(\tau), \phi(\tau), \sigma(\tau), v(\tau), \omega(\tau) )$,
or equivalently in second-order form \eqref{timelikegeodesiceqn-2ndord} 
for $(t(\tau),r(\tau),\phi(\tau) )$,
have a total of six dynamical degrees of freedom 
related by one constraint. 

A worthwhile remark is that this set of first integrals 
has a natural alternative derivation 
in the Hamiltonian-Jacobi framework \cite{MisThoWhe} for the geodesic equations. 
More specifically, as shown in \Ref{GomHorKos}, 
an explicit solution $S$ of the Hamilton-Jacobi equation is easy to obtain by separation of variables and use of the constants of motion $\L$, $I_1$, $I_2$, 
through which there is a canonical transformation from the variables $q=(t,r,\phi)$ 
and their canonical momenta $p=\partial \L/\partial \dot q = (-I_2,\frac{rv}{r-2M},I_1)$
to new canonical momenta $\tilde p = (\L,I_2,I_1)$
and corresponding new variables $\tilde q = \partial S/\partial \tilde p = (I_5,I_4,I_3)$, 
where both $\tilde q$ and $\tilde p$ are conserved quantities. 

To proceed, it will be useful to rewrite 
each of the integrals of motion \eqref{I3}, \eqref{I4}, \eqref{I5} 
in a mathematically equivalent form 
in which the implicit integration constants are made explicit, 
namely 
$I_3 = \Phi + \Phi_0$, 
$I_4 = T+T_0$, 
$I_5 = \T+\T_0$, 
where $\Phi_0$, $T_0$, $\T_0$ denote arbitrary constants of integration. 
These constants can be conveniently absorbed into an arbitrary value $r_0$ 
for one endpoint of the integration, whereby 
\begin{equation}\label{Theta}
\Phi \equiv \phi - L\int_{r_0}^{r}\frac{dr}{r^2v} 
\end{equation}
is an angular integral of motion, 
and similarly 
\begin{align}
T & \equiv t - E\int_{r_0}^{r}\frac{rdr}{(r - 2M)v} ,
\label{T}
\\
\T & \equiv \tau - \int_{r_0}^{r}\frac{dr}{v} 
\label{Tau}
\end{align}
are temporal integrals of motion,
with \begin{equation} \label{vsquared-EL}
v = \sgn(v)\sqrt{ E^2  - \Big(1 -\frac{2M}{r}\Big)\Big(1+\frac{L^2}{r^2}\Big) } . 
\end{equation}
Like the first integrals $I_1$, $I_2$, $I_3$, 
these integrals \eqref{Theta}, \eqref{T}, \eqref{Tau} 
are (locally) conserved quantities for timelike equatorial geodesic motion 
$(t(\tau),r(\tau),\phi(\tau))$. 

In standard treatments (cf \cite{MisThoWhe,Cha}) of 
integrating the geodesic equations, 
the focus is on the solution for $(t(\tau),r(\tau),\phi(\tau))$
rather than on identifying a complete set of integrals of motion. 
More specifically, 
$\Phi$, $T$, $\T$, $r_0$ 
either are taken to be arbitrary constants or are fixed in terms of initial values
$(t(\tau_0),r(\tau_0),\phi(\tau_0)$ at $\tau=\tau_0$ by the relations
\begin{equation}\label{standard.I3.I4.I5}
\Phi = \phi(\tau_0), 
\quad
T=t(\tau_0), 
\quad
\T = \tau_0,
\quad 
r_0 = r(\tau_0) . 
\end{equation}

By comparison, the main goal here is to fix the constant $r_0$ 
in the integrals of motion \eqref{Theta}, \eqref{T}, \eqref{Tau}  
by finding a dynamically distinguished value of $r$ 
directly from the radial velocity equation \eqref{vsquared-EL}, 
without using any features that would be specific to a particular geodesic,
such as initial conditions. 
Two general possibilities consist of
a \emph{turning point} (TP), 
defined as a radial value $r$ at which the radial velocity $v$ vanishes, 
and a \emph{centripetal point} (CP), 
defined as a radial value $r$ at which the radial acceleration $\dot v$ vanishes. 
These radial values are dynamically distinguished and can be determined entirely
in terms of the angular momentum and energy constants of motion. 
The physical meaning of these points, 
with respect to stationary observers at spatial infinity, is that 
turning points are the local extrema of the radial location of a particle in the orbit,
namely periapsis and apoapsis points, 
and that centripetal points are the local extrema of the particle's radial Doppler shift in the orbit. 

This will be the key step in the derivation of the conserved $\phi$, $t$, $\tau$-quantities,
which is explained in detail next.

\subsection{Conserved $\phi$, $t$, $\tau$ -quantities} 

In terms of the constants of motion \eqref{I1} and \eqref{I2},
which are respectively equal to angular momentum and energy 
\begin{align}
L & = g(\u,\partial/\partial_\phi) = r^2 \omega , 
\label{L}
\\
E & = g(\u,\partial/\partial_t) = \dfrac{r-2M}{r}\sigma , 
\label{E}
\end{align}
the radial velocity equation \eqref{vsquared} 
and the radial acceleration equation \eqref{radial-accel} are respectively given by 
\begin{equation} \label{radial-vel-EL}
v^2 = E^2  - \Big(1 -\frac{2M}{r}\Big)\Big(1+\frac{L^2}{r^2}\Big) 
\end{equation}
and 
\begin{equation}\label{radial-accel-EL}
\dot{v} = \frac{(r-3M)L^2}{r^4} - \frac{M}{r^2} .
\end{equation}

All turning points are given by the positive real roots $r=r_*$ of 
equation \eqref{radial-vel-EL} with $v=0$; 
all centripetal points are given by the positive real roots $r=r^*$ of 
equation \eqref{radial-accel-EL} with $\dot{v}=0$. 
In particular, the turning point equation can be expressed as a cubic 
\begin{equation}\label{tp-eqn}
(E^2-1){r_*}^3 +2M {r_*}^2 - L^2(r_* - 2M) = 0,
\end{equation}
and similarly the centripetal point equation is a quadratic 
\begin{equation}\label{ip-eqn}
M{r^*}^2 - L^2 (r^*-3M) =0 . 
\end{equation}
A classification of the positive real roots of these two equations 
can be found as a by-product of the well-known classification \cite{Cha} of 
the different types of timelike equatorial geodesics, 
summarized in Appendix~\ref{sec:evaluate}. 
The latter classification shows that at least one turning point exists 
whenever 
\begin{equation}\label{tp-case1}
E\leq 1 \text{ and } L^2 <16M^2
\end{equation}
or 
\begin{equation}\label{tp-case2}
E\leq E_\circ \text{ and } L^2\geq 16M^2 
\end{equation}
where $E_\circ$ is the energy of the unstable or marginally stable circular orbit;
and that at least one centripetal point exists whenever 
\begin{equation}\label{ip-case}
E\geq E_\circ \text{ and } L^2\geq 12M^2 
\end{equation}
where $E_\circ$ is the energy of the stable or marginally stable circular orbit. 

The classification of timelike equatorial geodesics also shows that 
neither a turning point nor a centripetal point exists when 
\begin{equation}\label{no-tp-ip-case}
E>1 \text{ and } L^2 <12M^2 . 
\end{equation}
Geodesics in this case either plunge into the horizon or escape to infinity. 
An alternative for a distinguished radial value is then the location of the horizon, 
given by $r = 2M$. 
Hereafter, this will be called a {\em horizon crossing point} (HP). 

The preceding discussion establishes a preliminary result. 

\begin{thm}\label{thm:firstintegrals}
For all timelike equatorial non-circular geodesics parameterized by proper time $\tau$, 
the integrals of motion \eqref{L}, \eqref{E}, \eqref{Theta}, \eqref{T}, \eqref{Tau} 
expressed as functions of $(t,r,\phi,\dot{t},\dot{r},\dot{\phi})$ 
yield five locally conserved quantities 
\begin{align}
L & = r^2 \dot{\phi} , 
\label{angularmomentum}
\\
E & = \dfrac{r-2M}{r}\dot{t} , 
\label{energy}
\\
\Phi & = \phi - L\int_{r_0}^{r}\frac{\sgn(v)\, dr}{r^2\sqrt{ E^2  - (1-2M/r)(1+L^2/r^2) }} 
\mod 2\pi,
\label{LRLangle}
\\
T & = t - E\int_{r_0}^{r}\frac{\sgn(v)r\, dr}{(r - 2M) \sqrt{ E^2  - (1- 2M/r)(1+L^2/r^2) }} ,
\label{LRLtime}
\\ 
\T & = \tau - \int_{r_0}^{r}\frac{\sgn(v)\, dr}{\sqrt{ E^2  - (1- 2M/r)(1+L^2/r^2) }} , 
\label{LRLtau}
\end{align}
where $r_0$ is taken to be 
either a turning point $r_*$ (when either condition \eqref{tp-case1} or condition \eqref{tp-case2} holds), 
a centripetal point $r^*$ (when condition \eqref{ip-case} holds),
or the horizon-crossing point $2M$ (when condition \eqref{no-tp-ip-case} holds). 
Each of these five quantities \eqref{angularmomentum}--\eqref{LRLtau}
can be evaluated locally in terms of the values of
$(t(\tau),r(\tau),\phi(\tau),\dot{t}(\tau),\dot{\phi}(\tau))$ 
at proper time $\tau$ on a given geodesic. 
These values also determine the value of $\dot{r}(\tau)$ 
through the proper time equation \eqref{propertimeeqn-2ndord}. 
\end{thm} 

Note that these integrals of motion do not involve initial values of an orbit 
and therefore are different than the first integrals found in standard treatments 
for integrating the timelike equatorial geodesic equations. 
Most importantly, the three integrals of motion $\Phi$, $T$, $\T$ 
are intrinsic conserved quantities which have the same local dynamical status 
as $E$ and $L$. 

In the classical mechanics and astronomy literature, 
a point at which $r$ is a local extremum on a given orbit is commonly called an \emph{apsis};
a local minimum point is called a \emph{periapsis}, 
and a local maximum point is called an \emph{apoapsis}. 
For timelike equatorial geodesics, 
the same terminology can be used for the spatial motion $(r(\tau),\phi(\tau))$,
where the radial extrema are measured by stationary observers at spatial infinity. 
Note that every apsis on an orbit yields a turning point, 
but the set of all turning points $r=r_*$ 
may include values of $r$ that do not occur on a given orbit,
since turning points are determined just by the radial velocity equation \eqref{radial-vel-EL}. 
The same consideration applies to centripetal points $r=r^*$. 
Note that centripetal points always come in pairs corresponding to $v$ being positive or negative on different parts of an orbit. 
Their physical meaning for timelike equatorial geodesics is that 
the radial Doppler shift for a particle in a spatial orbit, 
as measured by stationary observers at spatial infinity, is a local extremum. 

Thus, the following main result holds. 

\begin{thm}\label{thm:LRLanalogs}
For a given non-circular spatial orbit $(r(\tau),\phi(\tau))$, 
the angular integral of motion $\Phi$ is the coordinate angle (in the invariant plane) 
of the point $r=r_0$ on the orbit; 
the two temporal integrals of motion $T$ and $\T$ 
are respectively the coordinate time and the proper time
at which these points are reached. 
When $r_0$ is a turning point, 
$\Phi$ represents an analogue of the angle of the LRL vector;
when $r_0$ is a centripetal point, 
$\Phi$ represents an analogue of the angle of Hamilton's vector. 
$\Phi$ has no Newtonian analogue when $r_0$ is a horizon-crossing point. 
\end{thm}

The angular momentum $L$ and the energy $E$ 
are well known physical quantities which are globally constant 
on any timelike equatorial geodesic; 
namely, they are global constants of motion. 
In contrast, 
the angular quantity $\Phi$ and the temporal quantities $T$ and $\T$ are not widely known. 
As will be shown in section~\ref{sec:properties}, 
they are constants of motion locally on non-circular geodesics,
whereas globally 
$T$ and $\T$ will be multi-valued whenever the geodesic describes an orbit that has more than one apsis, 
and $\Phi$ will be multi-valued whenever the geodesic describes an orbit that precesses.
Their analogues in Newtonian gravity with cubic corrections, 
which have been discussed recently in \Ref{AncMeaPas}, 
are generalizations of the well-known Newtonian LRL vector 
and the less familiar Newtonian Hamilton vector \cite{GolPooSaf,Cor}. 
Explicit formulas for $\Phi$, $T$, $\T$ in the case of orbits that have an apoapsis point 
have been derived in \Ref{Kos,GomHorKos} 
with the motivation of obtaining efficient and numerically accurate expressions for use
in studying particle orbits near a black hole 
and earth satellite orbits for a global positioning network.

\section{Physical properties of the analogue LRL conserved quantities}\label{sec:properties}

The physical meaning of the angular integral of motion \eqref{LRLangle}
and the two temporal integrals of motion \eqref{LRLtime} and \eqref{LRLtau}
will now be discussed for all of the different types of spacetime orbits
$(t(\tau),r(\tau),\phi(\tau))$ given by non-circular equatorial timelike geodesics. 
In each case, the choice of $r_0$ as either 
a turning point (TP), a centripetal point (CP), or a horizon-crossing point (HP)
will be considered,
and the resulting properties of the integrals of motion 
as analogues of the LRL angle and LRL time in (post-) Newtonian gravity
will be described. 

To begin the discussion, 
a short summary of the well-known classification of orbits \cite{Cha} is presented
together with the possibilities for $r_0$ for each type of orbit.

\subsection{Types of orbits and choices of $r_0$}

The radial acceleration equation \eqref{radial-accel} can be expressed in the physical form 
\begin{equation}\label{eff-force}
\ddot{r} = -\dfrac{M}{r^2} + \dfrac{L^2(r-3M)}{r^4} = F_\eff =-\dfrac{d V_\eff}{dr}
\end{equation}
through the use of the angular momentum \eqref{angularmomentum}. 
The effective radial force $F_\eff$ defines a central force 
for which the associated effective potential $V_\eff$ can be obtained directly 
by expressing the proper time equation \eqref{vsquared-EL} in the oscillator form \cite{MisThoWhe}
\begin{equation}\label{VE-eqn}
\tfrac{1}{2} v^2 + V_\eff = \tfrac{1}{2}(E^2 -1),
\end{equation}
where $v=\dot{r}$ is the radial velocity. 

Hereafter, it will be convenient to introduce the reciprocal radial variable 
\begin{equation}\label{u}
u = \frac{2M}{r} . 
\end{equation}
This differs by the factor $2M$ compared to the notation used in \Ref{Cha} 
for classifying orbits 
and has the advantage that $u=1$ corresponds to the horizon $r=2M$. 
An overbar will denote a quantity or a variable divided by $2M$. 
In particular, 
\begin{equation}\label{Lbar}
\bar{L} = \frac{L}{2M} . 
\end{equation}

The effective potential is given by 
\begin{equation}\label{effpotential}
V_\eff= -\frac{M}{r} + \frac{L^2}{2r^2} - \frac{ML^2}{r^3}
=\tfrac{1}{2}\big( (1 - u)( 1+ \bar{L}^2 u^2) -1 \big) 
= \tfrac{1}{2}\big( E^2 -1 -Q(u)),
\end{equation}
where 
\begin{equation}\label{Q}
Q(u) = \bar{L}^2 (u^3 - u^2) +  u+ E^2-1= v^2
\end{equation}
is a cubic polynomial given by the radial velocity equation \eqref{radial-vel-EL}. 

This potential \eqref{effpotential} has the following features \cite{MisThoWhe,Cha}:\\
$\bullet$ 
no extrema when $\bar{L}^2<3$; \\
$\bullet$ 
an inflection when $\bar{L}^2=3$, with $V_\eff=\tfrac{8}{9}$;\\
$\bullet$ 
a maximum and a minimum when $\bar{L}^2>3$,
with $V_\eff = E_+^2$ and $V_\eff = E_-^2$, respectively.\\
Here  
\begin{equation}\label{E+-}
E_\pm^2 = \tfrac{2}{3}+\tfrac{2}{27}\bar{L}^2\big( 1 \pm \smallsqrt{(1-3/\bar{L}^2)^3} \big) ,
\end{equation}
which have the ranges
$\tfrac{8}{9}< E_-^2 \leq 1$ and $\tfrac{8}{9}< E_+^2 <\infty$
for $3<\bar{L}^2<\infty$. 

Tables~\ref{orbits-bounded} and~\ref{orbits-unbounded} 
summarize all of the types of non-circular orbits 
that arise from the shape of the effective potential \eqref{effpotential}
(cf Appendix~\ref{sec:evaluate}). 
For each type of orbit, the possibilities for $u_0=1/r_0$ are listed.
A horizon-crossing point (HP) refers to $u=1$, namely $r=2M$; 
a centripetal point (CP) refers to 
\begin{equation}\label{ip-u}
u^*= u_\pm
\quad\text{ if } \bar{L}^2\geq 3
\end{equation} 
as given by the two positive roots of the quadratic equation $Q'(u^*)=0$ (cf \eqref{ip-eqn}); 
and a turning point (TP) refers to 
\begin{equation}\label{tp-u}
u_*=\begin{cases}
u_1,u_2,u_3 &
\text{ if }\quad
\bar{L}^2\geq 4,\  1>E\geq E_-; 4> \bar{L}^2 \geq 3,\ E_+\geq E\geq E_-
\\
u_2,u_3 &
\text{ if }\quad
\bar{L}^2\geq 4,\  E_+\geq E\geq 1
\\
u_1 &
\text{ if }\quad
\bar{L}^2 < 3,\ E<1; \bar{L}^2>3,\ E_->E; 4>\bar{L}^2>3,\ 1>E>E_+
\end{cases}
\end{equation} 
as given by the positive root(s) of the cubic equation $Q(u_*)=0$ (cf \eqref{tp-eqn}).
Additionally, a superscript indicates the multiplicity of choices.

\begin{table}[h!!]
\centering
\caption{
Types of bounded non-circular orbits.
}
\label{orbits-bounded} 
\begin{tabular}{l||c|c|c|c|c}
\hline
Orbit type & $\bar{L}^2$ & $E^2$ & \parbox{0.5in}{\centering Root\\case\strut} & Range of $u$ & \parbox{1in}{\centering Distinguished\\ radial points}
\\
\hline\hline
horizon-crossing
& 
$<3$
& 
$<1$
& 
(4)
&
$\geq u_1$ 
& 
TP, 
HP
\\
&
$>3$
& 
$<E_-^2$
& 
(4)
&
$\geq u_1$ 
& 
TP,
HP
\\
& 
$>3$
& 
$\geq E_-^2$ and $<E_+^2$
& 
(2a),(3)
&
$\geq u_3$ 
& 
TP,
HP
\\
&
$\geq3$ and $<4$
& 
$>E_+^2$ and $<1$
& 
(4)
&
$\geq u_1$ 
& 
TP,
CP$^2$,
HP
\\
\hline
asymptotic circular
& 
$\geq3$ 
& 
$E_+^2$
& 
(1), (2b)
&
$> u_3$ 
& 
HP
\\
horizon-crossing
&
&
&
&
&
\\
\hline
asymptotic circular
& 
$>3$ and $<4$
& 
$E_+^2$
&
(2b)
& 
$\geq u_1$ and $< u_2$
& 
TP, 
CP
\\
\hline
elliptic-like
&
$>3$ and $<4$
&
$>E_-^2$ and $<E_+^2$
&
(3)
&
$\geq u_1$ and $\leq u_2$
&
TP$^2$, 
CP
\\
&
$\geq 4$
&
$>E_-^2$ and $<1$
&
(3)
&
$\geq u_1$ and $\leq u_2$
&
TP$^2$, 
CP
\\
\hline
\end{tabular}
\end{table}

\begin{table}[h!!]
\centering
\caption{
Types of unbounded orbits.
}
\label{orbits-unbounded} 
\begin{tabular}{l||c|c|c|c|c}
\hline
Orbit type & $\bar{L}^2$ & $E^2$ & \parbox{0.5in}{\centering Root\\case\strut} & Range of $u$ & \parbox{1in}{\centering Distinguished\\ radial points}
\\
\hline\hline
horizon-crossing
& 
$<3$
& 
$\geq 1$
& 
(4)
&
$> 0$ 
& 
HP
\\
& 
$\geq3$ and $<4$
& 
$>1$
&
(4)
& 
$> 0$ 
& 
CP$^2$, 
HP
\\
& 
$\geq4$
& 
$>E_+^2$
&
(4)
& 
$> 0$ 
& 
CP$^2$, 
HP
\\
\hline
asymptotic circular
& 
$4$
& 
$1$
&
(2b)
& 
$> 0$ and $< u_2$ 
& 
CP
\\
parabolic-like 
&&&&&
\\
\hline
asymptotic circular
& 
$>4$
& 
$E_+^2$
& 
(2b)
&
$> 0$ and $< u_2$ 
& 
CP
\\
hyperbolic-like
&&&&&
\\
\hline
parabolic-like 
& 
$>4$
& 
$1$
& 
(3)
&
$> 0$ and $\leq u_2$
& 
TP, 
CP
\\
\hline
hyperbolic-like
& 
$>4$
& 
$>1$ and $<E_+^2$
&
(3)
& 
$> 0$ and $\leq u_2$
& 
TP, 
CP
\\
\hline
\end{tabular}
\end{table}

\subsection{Analogue LRL quantities for non-circular orbits and their local/global properties}

With the change of variable \eqref{u}, 
the angular integral of motion \eqref{LRLangle} is given by 
\begin{equation}\label{LRLangle-u}
\Phi = \phi +\sgn(v \bar{L}) I^\Phi(u;u_0) \mod 2\pi , 
\quad
I^\Phi(u;u_0) = \int_{u_0}^{u}\frac{du}{\sqrt{Q(u)}} , 
\end{equation}
and the temporal integrals of motion \eqref{LRLtime} and \eqref{LRLtau} are given by 
\begin{align}
\bar{T} & = \bar{t}+ \frac{\sgn(v) E}{|\bar{L}|} I^T(u;u_0), 
\quad
I^T(u;u_0) = \int_{u_0}^{u}\frac{du}{(1-u)u^2\sqrt{Q(u)}} , 
\label{LRLtime-u}
\\
\bar{\T} & = \bar{\tau} +\frac{\sgn(v)}{|\bar{L}|} I^\T(u;u_0), 
\quad
I^\T(u;u_0) = \int_{u_0}^{u}\frac{du}{u^2\sqrt{Q(u)}} , 
\label{LRLtau-u}
\end{align}
where $Q(u)$ is the cubic polynomial \eqref{Q}. 

For each type of orbit, 
the explicit evaluation of $\Phi$, $\bar{T}$, $\bar{\T}$ 
can be given in terms of the quadratures provided in Appendix~\ref{sec:evaluate}
(cf 
\eqref{tripleroot-integrals}, \eqref{doublesingleroot-integrals}, \eqref{singledoubleroot-integrals}, \eqref{distinctroot-integrals-range1}, \eqref{distinctroot-integrals-range2}, \eqref{ccroot-integrals})
with the various choice(s) of $u_0$ shown in 
Tables~\ref{orbits-bounded} and~\ref{orbits-unbounded}. 
The global nature of these integrals of motion, 
specifically whether they are global conserved quantities 
or only piecewise conserved quantities, 
will now be stated along with their physical meaning. 

The results are presented in 
Table~\ref{FIs-orbits-bounded} for bounded non-circular orbits
and Table~\ref{FIs-orbits-unbounded} for unbounded orbits. 
Both tables are divided into separate cases for orbits that lie outside of the horizon
and orbits that cross the horizon. 
Derivations and figures for each case will be given afterwards. 

Orbits that do not cross the horizon possess at least one centripetal point $r=r^*=2M/u^*$,
and consequently, this point provides a universal choice of $u_0=u^*=2M/r^*$. 
The resulting conserved quantities are analogues of the angle and the time 
corresponding to Hamilton's vector in (post-) Newtonian gravity,
which is a variant of the LRL vector. 
If the orbit is either bounded or not asymptotically circular,
then at least one turning point $r=r_*=2M/u_*$ exists,
which can be used alternatively as a choice of $u_0=u_*=2M/r_*$. 
The resulting conserved quantities are analogues of the LRL angle and time 
in (post-) Newtonian gravity. 

Orbits that cross the horizon necessarily possess a horizon-crossing point. 
This provides a universal choice of $u_0=2M/r_0=1$. 
The resulting conserved quantities have no analog in the Newtonian case.

\begin{table}[h!!]
\centering
\caption{
Conserved angular quantity \eqref{LRLangle-u}
and conserved temporal quantities \eqref{LRLtime-u}, \eqref{LRLtau-u} 
for \emph{bounded non-circular} orbits.
}
\label{FIs-orbits-bounded} 
\begin{tabular}{l||c||l|c|c|l}
\hline
Orbit 
& 
\parbox{1.5in}{\centering $I^\Phi(u;u_0)$\\$I^T(u;u_0)$, $I^\T(u;u_0)$}
&
\parbox{0.5in}{\centering $u_0$}
&
$\Phi$
& 
$\bar{T}$, $\bar{\T}$
&
\parbox{0.7in}{Physical Meaning}
\\
\hline\hline
elliptic-like
&
\eqref{distinctroot-integrals-range2}
&
TP $u_2$
&
multi-val.
&
multi-val.
&
periapsis
\\
&
&
TP $u_1$
&
multi-val.
&
multi-val.
&
apoapsis
\\
&
&
CP $u_-$
&
multi-val.
&
multi-val.
&
Doppler
\\
&&&&&
max/min
\\
\hline
asymptotic
& 
\eqref{singledoubleroot-integrals}
&
TP $u_1$
&
global
&
global
&
apoapsis
\\
circular
&
&
CP $u_-$
&
double-val.
&
double-val.
&
Doppler
\\
&&&&&
max/min
\\
\hline
\hline
asymptotic
& 
\eqref{singledoubleroot-integrals},
$\bar{L}^2>3$
&
HP $1$
&
global
&
$\infty$, global
&
horizon
\\
circular
&
\eqref{tripleroot-integrals}, 
$\bar{L}^2=3$
&&&&
\\
horizon-crossing 
&&&&&
\\
&&&&&
\\
\hline
horizon-crossing
& 
\eqref{ccroot-integrals}, $E^2<E_-^2$
&
TP $u_3$
&
global
&
global
&
apoapsis
\\
&
\eqref{doublesingleroot-integrals}, $E^2=E_-^2$
& 
HP $1$
&
double-val.
& 
$\infty$, double-val.
& 
horizon
\\
& 
\eqref{distinctroot-integrals-range1}, $E_-^2<E^2<E_+^2$
& 
& 
&
&
\\
\cline{2-6}
&
\eqref{ccroot-integrals}, $E_+^2<E^2<1$
& 
TP $u_3$
&
global
&
global
&
apoapsis
\\
&
&
CP$^2$ $u_\pm$
& 
double-val.
&
double-val.
& 
Doppler
\\
&&&&&
max/min
\\
&
& 
HP $1$
&
double-val.
& 
$\infty$, double-val.
& 
horizon
\\
\hline
\end{tabular}
\end{table}

\begin{table}[h!!]
\centering
\caption{
Conserved angular quantity \eqref{LRLangle-u}
and conserved temporal quantities \eqref{LRLtime-u}, \eqref{LRLtau-u} 
for \emph{unbounded} orbits.
}
\label{FIs-orbits-unbounded} 
\begin{tabular}{l||c||l|c|c|l}
\hline
Orbit 
& 
\parbox{1.5in}{\centering $I^\Phi(u;u_0)$\\$I^T(u;u_0)$, $I^\T(u;u_0)$}
&
\parbox{0.5in}{\centering $u_0$}
&
$\Phi$
& 
$\bar{T}$, $\bar{\T}$
&
\parbox{0.7in}{Physical Meaning}
\\
\hline\hline
hyperbolic-like
&
\eqref{distinctroot-integrals-range2}
&
TP $u_2$
&
global
&
global
&
periapsis
\\
&
&
CP $u_-$
&
double-val.
&
double-val.
&
Doppler
\\
&&&&&
max/min
\\
\hline
parabolic-like
& 
\eqref{distinctroot-integrals-range2}
&
TP $u_2$
&
global
& 
global
&
periapsis
\\
&
&
CP $u_-$
&
double-val.
& 
double-val.
&
Doppler
\\
&&&&&
max/min
\\
\hline
asymptotic circular
& 
\eqref{singledoubleroot-integrals}
&
CP $u_-$
&
global
&
global
&
Doppler
\\
hyperbolic-like 
&
&
&
&
&
max/min
\\
\hline
asymptotic circular
& 
\eqref{singledoubleroot-integrals}
&
CP $u_-$
&
global
&
global
&
Doppler
\\
parabolic-like 
&
& 
& 
&
& 
max/min
\\
\hline
\hline
horizon-crossing 
&
\eqref{ccroot-integrals}, $\bar{L}^2<3$
&
HP $1$
&
global
&
$\infty$, global
&
horizon
\\
\cline{2-6}
&
\eqref{ccroot-integrals}, $\bar{L}^2\geq 3$
&
CP$^2$ $u_\pm$
&
global
&
global
&
Doppler
\\
&&&&&
max/min
\\
&
&
HP $1$
&
global
&
$\infty$, global
&
horizon
\\
\hline
\end{tabular}
\end{table}

\subsection{Global versus local conservation properties}

For any orbit, 
the angular and temporal integrals of motion \eqref{LRLangle-u}, \eqref{LRLtime-u}, \eqref{LRLtau-u} 
are locally constant on each part of the orbit that does not contain a turning point. 
This can be seen directly from their expressions:
the mathematical condition for occurrence of a jump is when 
the factor $\sgn(v)$ which multiplies the quadratures 
$I^\Phi(u;u_0)$, $I^T(u;u_0)$, $I^\T(u;u_0)$ 
changes sign as the orbit goes through a turning point, $u=u_*\neq u_0$. 

Hence, these three integrals of motion will be globally constant 
for orbits with no turning points. 
Their global properties for orbits with a turning point
crucially depend on whether or not the turning point undergoes precession. 
A turning point $u=u_*$ is said to \emph{precess} if 
there are multiple angles $\phi=\phi_1,\phi_2,\ldots$ (mod $2\pi$)
at which the orbit reaches $u=u_*$. 
In particular, the multiplicity may be finite or infinite. 

\begin{prop}\label{prop:global-cases}
(i) The angular and temporal integrals of motion \eqref{LRLangle-u}, \eqref{LRLtau-u}, and \eqref{LRLtime-u} 
in the TP case $u_0=u_*$ 
are global conserved quantities on a orbit 
(namely, they are constant on the entirety of the orbit) 
iff the orbit has no turning points that precess. 
(ii) In the CP and HP cases $u_0=u^*$ 
and $u_0=1$,
the angular and temporal integrals of motion \eqref{LRLangle-u}, \eqref{LRLtau-u}, and \eqref{LRLtime-u} 
are global conserved quantities iff the orbit has no turning points. 
\end{prop}

This result will be now be applied to determine the global nature of 
the angular and temporal integrals of motion \eqref{LRLangle-u}, \eqref{LRLtau-u}, and \eqref{LRLtime-u} 
for each of the different types of orbits listed in Tables~\ref{orbits-bounded} and~\ref{orbits-unbounded}. 
For this discussion, it will be helpful to revert to using $r$ (and $r_0$) 
instead of $u=2M/r$ (and $u_0=2M/r_0)$. 

The number of turning points for an orbit can be 
established through consideration of the equation
\begin{equation}\label{orbit-shape-eqn}
\phi =\Phi -\sgn(v\bar{L}) I^\Phi(2M/r;2M/r_0) \equiv f(r;\Phi,r_0)
\end{equation}
which determines the local shape for all orbits. 
Specifically, the shape will refer to the spatial curve defined by $(r(\tau),\phi(\tau))$
in the equatorial plane, coordinatized by $(r,\phi)$ with $2M\leq r<\infty$, $0\leq\phi<2\pi$. 
The orbit shape equation \eqref{orbit-shape-eqn} has an important 
reflection symmetry property which is related to existence of turning points,
as will now be explained. 

For orbits that possess at least one turning point, 
the apses consist of the set of points $(r_*,\phi_*)$ 
where $\phi_*$ is the angle $\phi$ at which each turning point $r=r_*$ occurs on the orbit. 
A radial apsis line refers to the radial line that connects the horizon to a given apsis on the orbit in the spatial surface $(r,\phi)$, 
namely $\phi=\phi_*$ with $2M\leq r\leq r_*$. 

The local shape of an orbit near an apsis can be found, up to a global rotation, 
by taking $r_0=r_*$ and $\Phi=\phi_*$ in the orbit shape equation \eqref{orbit-shape-eqn}:
$\phi =\phi_* -\sgn(v\bar{L}) I^\Phi(2M/r;2M/r_*)$. 
Since the radial velocity $v$ changes sign at the apsis, 
note that $\phi -\phi_* = - \sgn(\bar{L}) I^\Phi(2M/r;2M/r_*)$ $\longleftrightarrow$ $-(\phi -\phi_*) = \sgn(\bar{L}) I^\Phi(2M/r;2M/r_*)$. 
Hence, the orbit will be locally reflection symmetric with respect to the radial apsis line
\begin{equation}\label{reflect-symm}
(r,\phi) \to (r,2\phi_*-\phi)
\end{equation}
for $r$ in some radial interval that has $r_*$ being an endpoint. 

Global information about the orbit shape can be obtained from 
the local reflection-symmetry \eqref{reflect-symm} combined with knowledge of 
whether the orbit is bounded or unbounded, 
whether it crosses the horizon,
and whether it possesses a circular asymptote. 

Consider, first, unbounded orbits that do not cross the horizon and are not asymptotically circular. 
Clearly, these orbits must possess at least one turning point which is the periapsis. 
Starting at spatial infinity,
$r$ will decrease from $\infty$ to $r_*$ which is a local periapsis. 
The reflection-symmetry \eqref{reflect-symm} will thus hold for $r_*\leq r<\infty$
and this implies that $r$ will increase from $r_*$ to $\infty$. 
Hence, the periapsis $r=r_*$ is the only turning point of the orbit. 
Moreover, this point does not precess, 
namely there is only a single angle $\phi_*$ at which $r=r_*$ is reached on the orbit.
Additionally, $|v|$ will go through a local maximum at some point $r=r^*$
with $r_*<r^*<\infty$, 
and hence the orbit possesses a symmetric pair of centripetal points,
with $v$ having opposite signs. 
Their angular separation is given by the integral 
\begin{equation}\label{CP-angle}
\Delta\phi = \sgn(\bar{L}) 2 I^\Phi(2M/r^*,2M/r_*)
\end{equation}
where $I^\Phi$ is the quadrature \eqref{distinctroot-Phi-integral-range2}.

A similar argument shows that unbounded orbits that are asymptotically circular 
cannot possess any turning points. 
Likewise, unbounded orbits that cross the horizon cannot possess any turning points,
because the horizon can only be entered once. 
Hence, all of these orbits have no precession.
As shown by the local extrema of the effective potential, 
the horizon crossing ones possess two different centripetal points,
while the asymptotically circular ones possess a single centripetal point. 

Consider, next, bounded orbits that are asymptotically circular. 
These orbits must possess at least one turning point which is the apoapsis. 
Starting near the asymptotic circle, $r$ will increase until a local apoapsis $r=r_*$ is reached. 
The reflection-symmetry \eqref{reflect-symm} thereby holds for $r_*\geq r>r_\circ$,
where $r_\circ$ is the location of the limiting circle. 
This implies that $r$ will decrease from $r_*$ to $r_\circ$ asymptotically. 
Therefore, the apoapsis $r=r_*$ is the only turning point of the orbit, 
and this point does not precess. 
For these orbits, 
similarly to the case of unbounded orbits with a single periapsis, 
there exist a symmetric pair of centripetal points given by $r=r^*$, 
with $r_\circ<r^*<r_*$, where $|v|$ goes through a local maximum. 
Their angular separation is given by the previous integral \eqref{CP-angle}
with $I^\Phi$ now being the quadrature \eqref{singledoubleroot-Phi-integral}. 

A similar argument applies to bounded orbits that cross the horizon but are not asymptotically circular. 
Bounded orbits that are asymptotically circular and cross the horizon
cannot possess any turning points, because the horizon can only be entered once. 
Therefore, all of these orbits have no precession. 
The local extrema of the effective potential shows that 
the asymptotically circular ones do not possess any centripetal points,
while the others possess two different centripetal points. 

The remaining type of bounded orbits to consider is elliptic-like orbits. 
These orbits will possess at least two turning points 
which are the apoapsis $r=r_*^{\max}$ and the periapsis $r=r_*^{\min}$. 
The reflection-symmetry \eqref{reflect-symm} thus can be applied to each segment of the orbit with $r_*^{\min}\leq r\leq r_*^{\max}$. 
This implies that the global orbit shape is obtained by successive composition of these segments
such that the respective angles $\phi_*$ 
at which $r=r_*^{\min}$ and $r_*^{\max}$ are reached 
either comprise a finite sequence or an infinite sequence, mod $2\pi$. 
The change in angle between two successive apoapses or periapses on the orbit 
is given by the integral 
\begin{equation}\label{precess-angle}
\Delta\phi = \sgn(\bar{L}) 2 I^\Phi(2M/r_*^{\max},2M/r_*^{\min})
\end{equation}
where $I^\Phi$ is the quadrature \eqref{distinctroot-Phi-integral-range2}. 
The orbit is physically precessing if and only if $\Delta\phi \neq 0$ mod $2\pi$. 
A precessing orbit is periodic (closed) when $\Delta\phi/(2\pi)$ is a rational number, 
and otherwise the orbit is non-periodic when $\Delta\phi/(2\pi)$ is an irrational number. 
In all cases, 
the corresponding changes in proper time and coordinate time
are given by the integrals
\begin{equation}\label{precess-time}
\Delta\bar\tau = (2/\bar L) I^\T(2M/r_*^{\max},2M/r_*^{\min}) ,
\quad
\Delta\bar t = (2E/\bar L) I^T(2M/r_*^{\max},2M/r_*^{\min}) ,
\end{equation}
where $I^\T$ and $I^T$ are the quadratures \eqref{distinctroot-Tau-integral-range2} and \eqref{distinctroot-T-integral-range2}.
Additionally, between any two successive apsis points, 
there will be a centripetal point $r=r^*$. 

The preceding discussion, combined with Proposition~\ref{prop:global-cases}
and Tables~\ref{FIs-orbits-bounded} and~\ref{FIs-orbits-unbounded}, 
establishes the following classification results
for the global properties of the angular and temporal integrals of motion \eqref{LRLangle-u}, \eqref{LRLtime-u}, \eqref{LRLtau-u}. 

\begin{thm}\label{thm:globalproperties.TP}
With $u_0=2M/r_*$ (TP),
the angular and temporal integrals of motion \eqref{LRLangle-u} and \eqref{LRLtime-u}, 
and the proper-time integral of motion \eqref{LRLtau-u},
are:
\\
(i) single-valued global quantities
for unbounded orbits that are either hyperbolic-like or parabolic-like,
and for bounded orbits other than elliptic-like ones;
\\
(ii) multi-valued piecewise quantities
for elliptic-like non-circular orbits,
which undergo a jump \eqref{precess-angle}--\eqref{precess-time}
at every apoapsis if $u_0=2M/r_*^{\min}$, or at every periapsis if $u_0=2M/r_*^{\max}$. 
\end{thm}

\begin{thm}\label{thm:globalproperties.CP}
With $u_0=2M/r^*$ (CP),
the angular and temporal integrals of motion \eqref{LRLangle-u} and \eqref{LRLtime-u},
and the proper-time integral of motion \eqref{LRLtau-u}, are:
\\
(i) single-valued global quantities
for unbounded orbits that are either asymptotically circular or horizon crossing;
\\
(ii) double-valued piecewise quantities
for unbounded orbits that are either hyperbolic-like or parabolic-like,
and for bounded orbits other than elliptic-like ones, 
which undergo a jump at the apsis; 
\\
(iii) multi-valued piecewise quantities
for elliptic-like non-circular orbits,
which undergo a jump at every apsis. 
\end{thm}

\begin{thm}\label{thm:globalproperties.HP}
With $u_0=1$ (HP),
the angular and temporal integrals of motion \eqref{LRLangle-u} and \eqref{LRLtime-u},
and the proper-time integral of motion \eqref{LRLtau-u}, are:
\\
(i) single-valued global quantities
for unbounded horizon-crossing orbits,
and for bounded horizon-crossing orbits that are asymptotically circular; 
\\
(ii) double-valued piecewise quantities
for bounded horizon-crossing orbits other than asymptotically circular ones,
which undergo a jump at the apoapsis. 
\end{thm}

In the HP case, 
note that the angular $\phi$-quantity \eqref{Theta} 
and the temporal $\tau$-quantity \eqref{Tau} are finite at $r=2M$; 
the temporal $t$-quantity \eqref{Theta} is infinite at $r=2M$, 
which is a consequence of the well-known property that $t$ breaks down 
as a coordinate at the horizon. 
(This issue could be avoided by going to coordinates that are non-singular across the horizon \cite{MisThoWhe,Wal,Cha}.)

\subsection{Figures: locally conserved LRL angle $\Phi$}

The notation used in all figures is stated in Table~\ref{fig.notation}. 

\begin{table}[h!!]
\centering
\caption{
Notation for figures. 
}
\label{fig.notation} 
\begin{tabular}{|c|c|}
\hline
Symbol
& Meaning
\\
\hline
\hline
red line
& orbit
\\
dashed line 
& horizon
\\
dotted line 
& circular orbit
\\
\hline
solid brown diamond 
& centripetal point (CP)
\\
solid blue box 
& turning point (TP)
\\
solid red circle 
& horizon point (HP)
\\
\hline
solid black box
& turning point $r=r_0$ where $\phi=\Phi$ 
\\
solid black diamond
& centripetal point $r=r_0$ where $\phi=\Phi$ 
\\
solid black circle 
& horizon point $r=r_0$ where $\phi=\Phi$ 
\\
\hline
multiple black symbols 
& $\Phi$ multi-valued (jumps)
\\
\hline
\end{tabular}
\end{table}

\begin{figure}[ht!]
\centering
\includegraphics[trim=2cm 12cm 4cm 1cm,clip,width=0.44\textwidth]{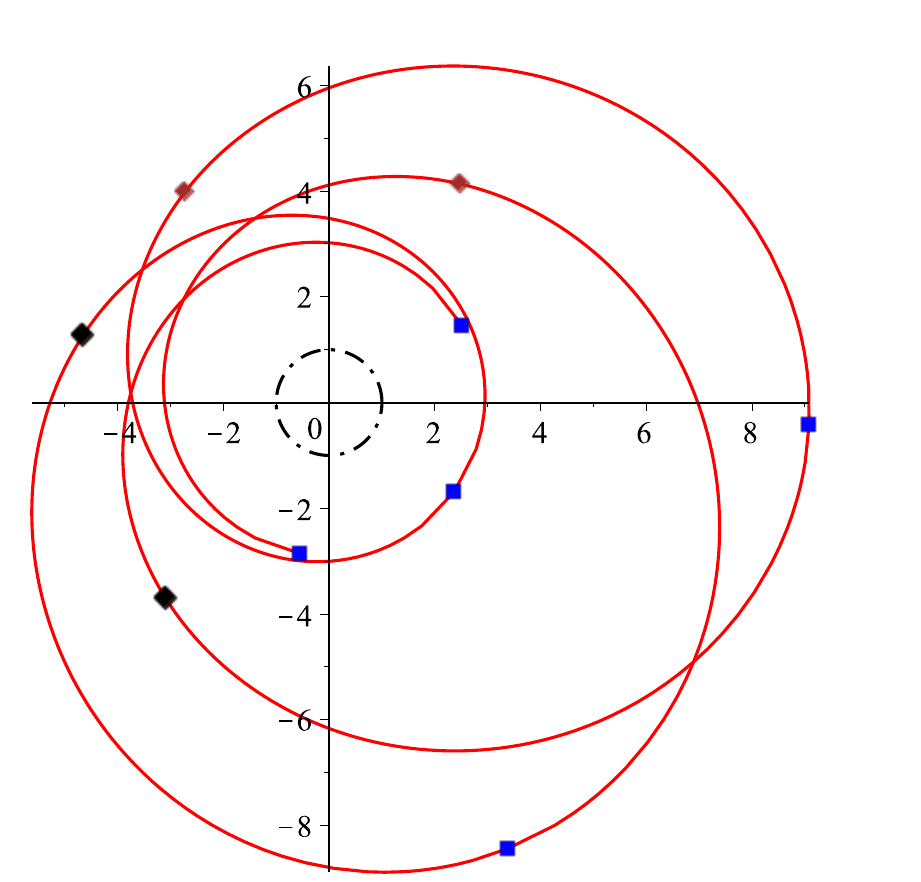}
\includegraphics[trim=2cm 12cm 4cm 1cm,clip,width=0.49\textwidth]{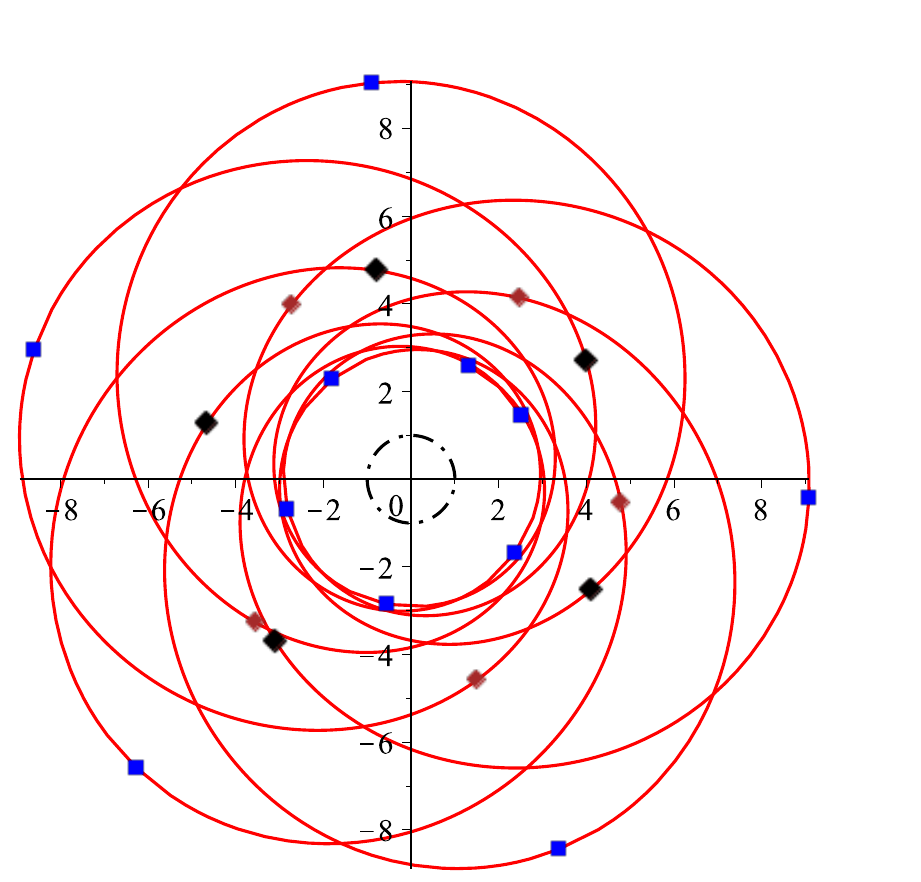}
\caption{Precessing elliptic-like orbit: LRL-CP angle $\Phi$}\label{fig:IP-ellip}
\end{figure}

\begin{figure}[ht!]
\centering
\includegraphics[trim=2cm 12cm 4cm 1cm,clip,width=0.45\textwidth]{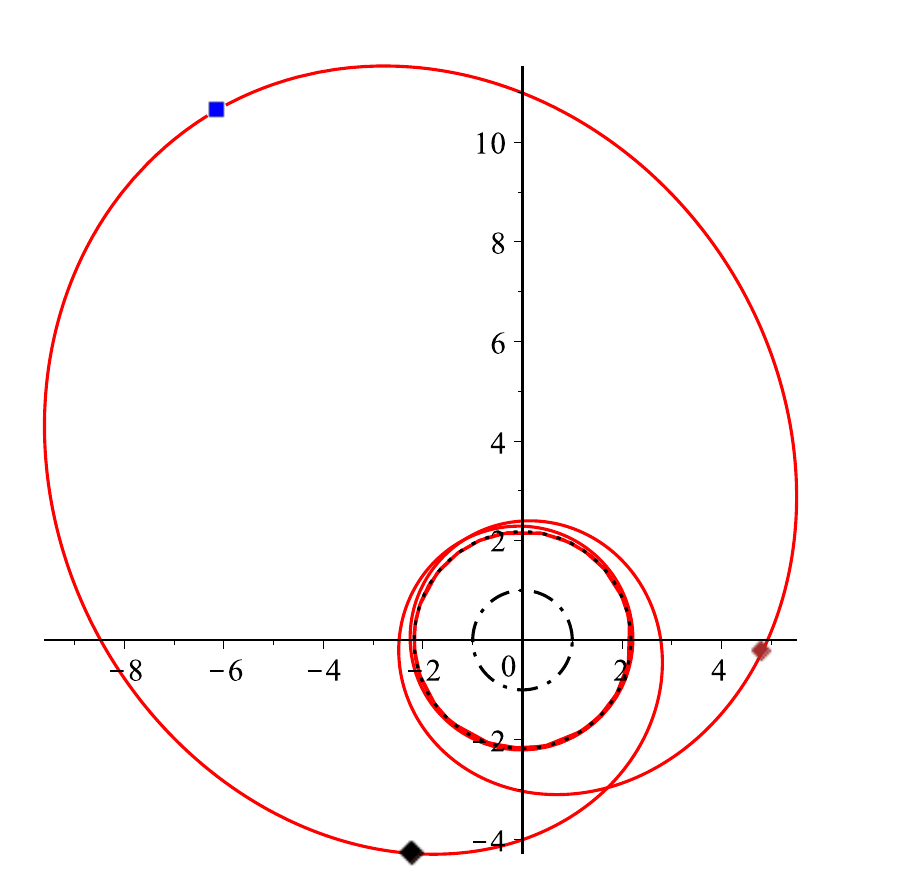}
\includegraphics[trim=2cm 12cm 4cm 1cm,clip,width=0.49\textwidth]{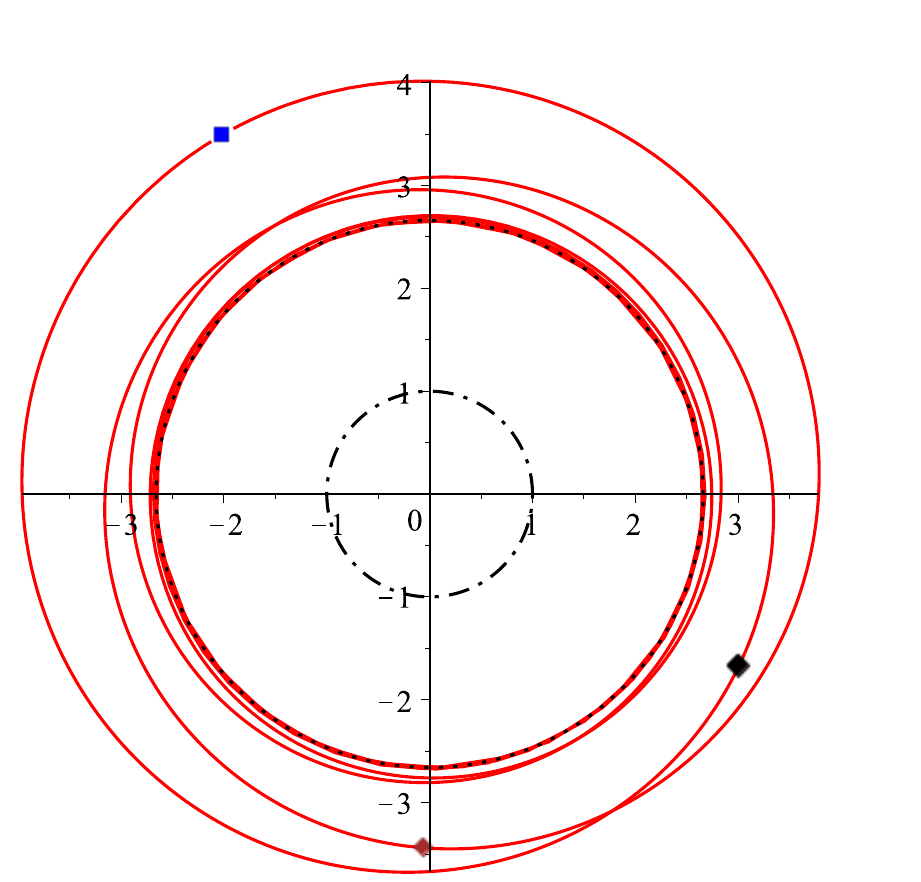}
\caption{Bounded asymptotic circular orbits: LRL-CP angle $\Phi$}\label{fig:IP-acbdd}
\end{figure}

\begin{figure}[ht!]
\centering
\includegraphics[trim=2cm 12cm 4cm 1cm,clip,width=0.45\textwidth]{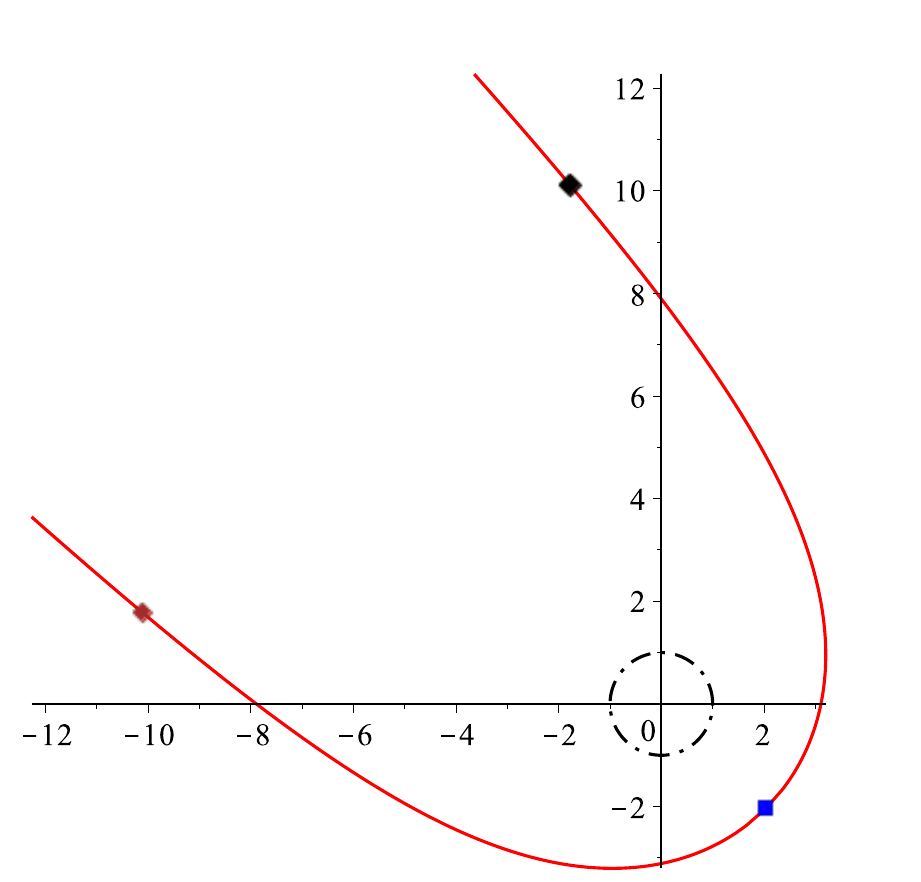}
\includegraphics[trim=2cm 12cm 4cm 1cm,clip,width=0.45\textwidth]{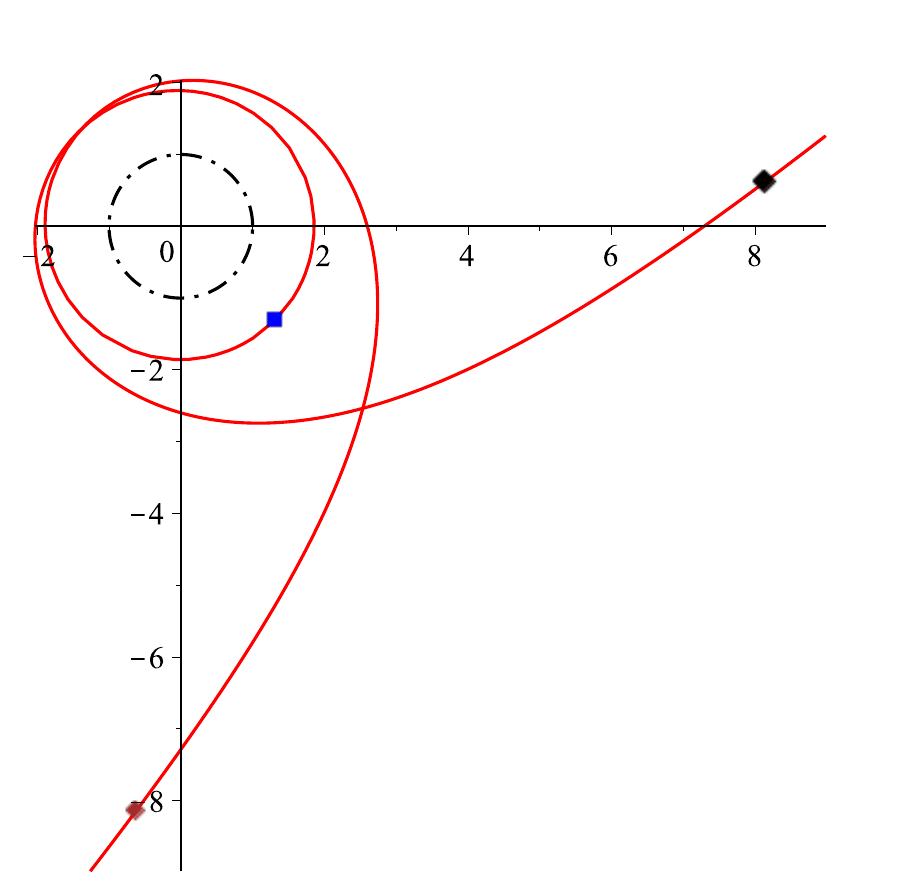}
\caption{LRL-CP angle $\Phi$ for hyperbolic-like orbits}\label{fig:IP-hyper}
\end{figure}

\begin{figure}[ht!]
\centering
\includegraphics[trim=2cm 12cm 4cm 1cm,clip,width=0.45\textwidth]{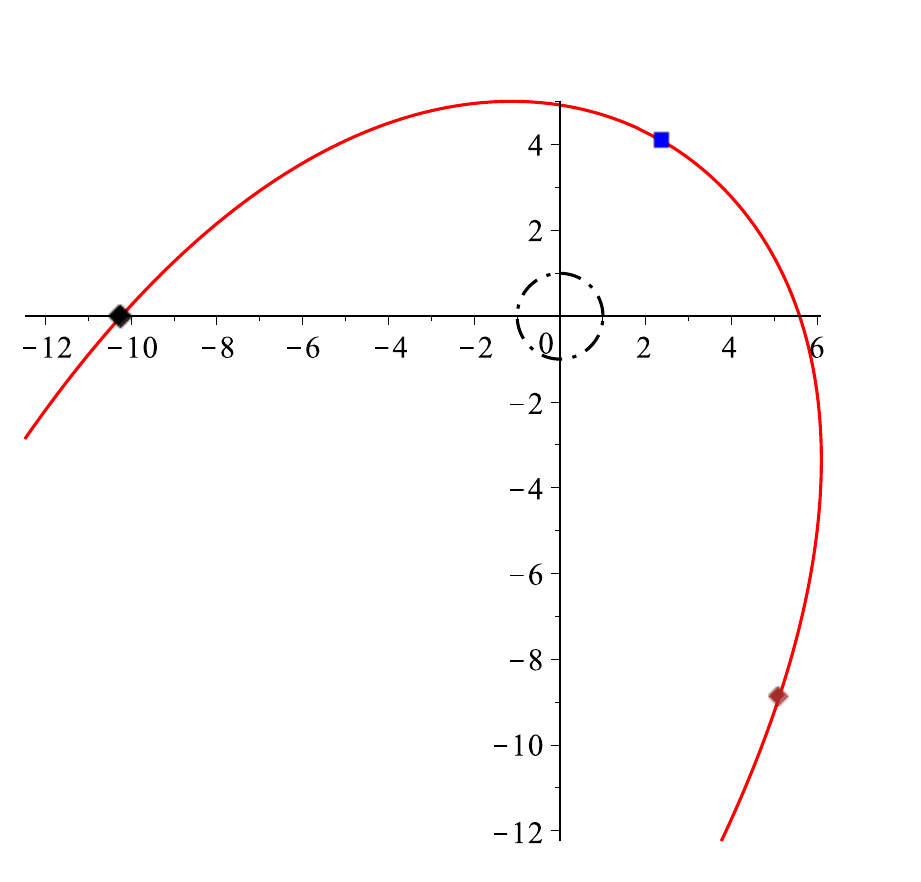}
\includegraphics[trim=2cm 12cm 4cm 1cm,clip,width=0.45\textwidth]{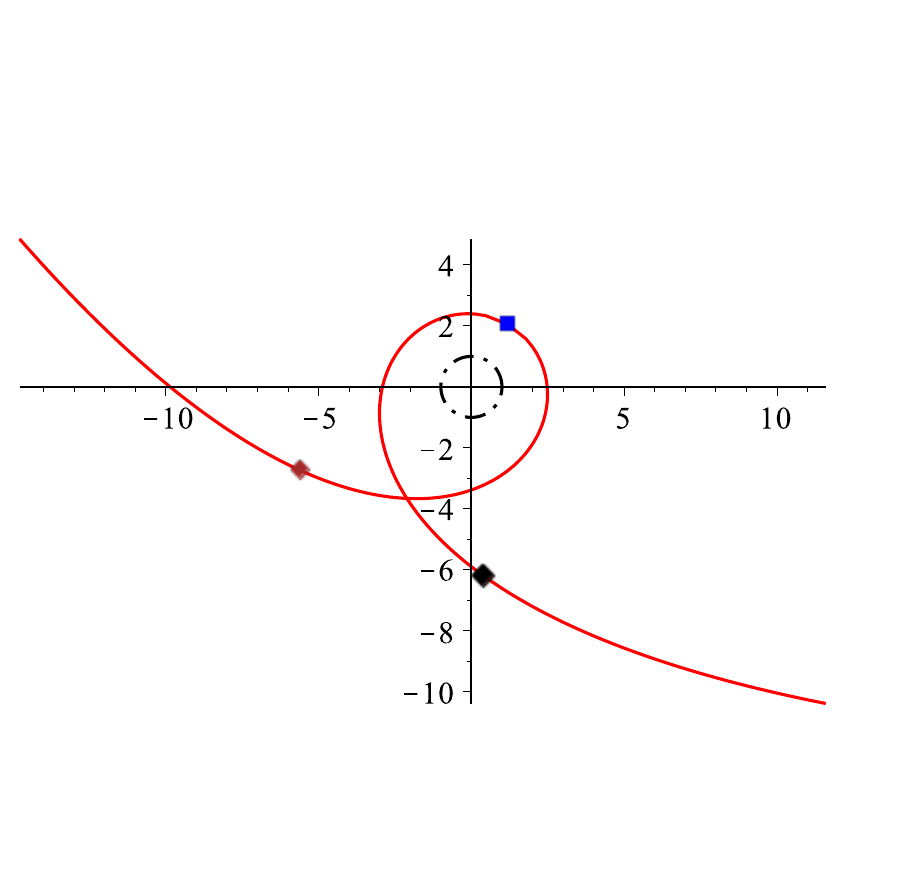}
\caption{Parabolic-like orbits: LRL-CP angle $\Phi$}\label{fig:IP-para}
\end{figure}

\begin{figure}[ht!]
\centering
\includegraphics[trim=2cm 12cm 4cm 1cm,clip,width=0.45\textwidth]{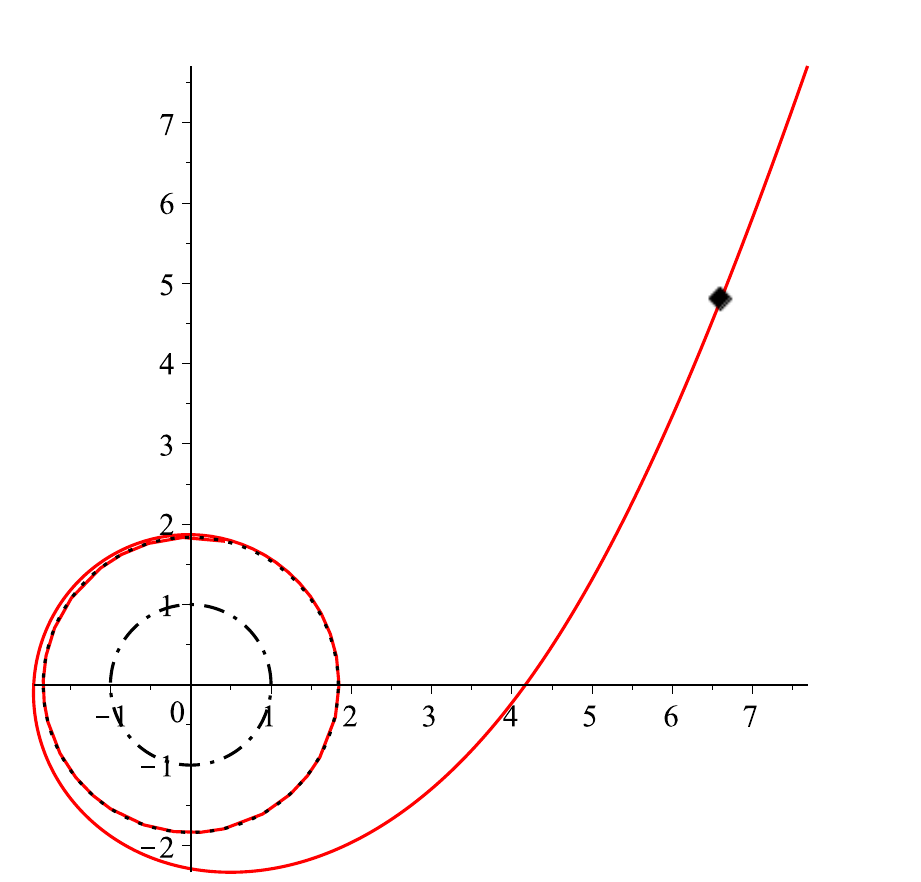}
\includegraphics[trim=2cm 12cm 4cm 1cm,clip,width=0.45\textwidth]{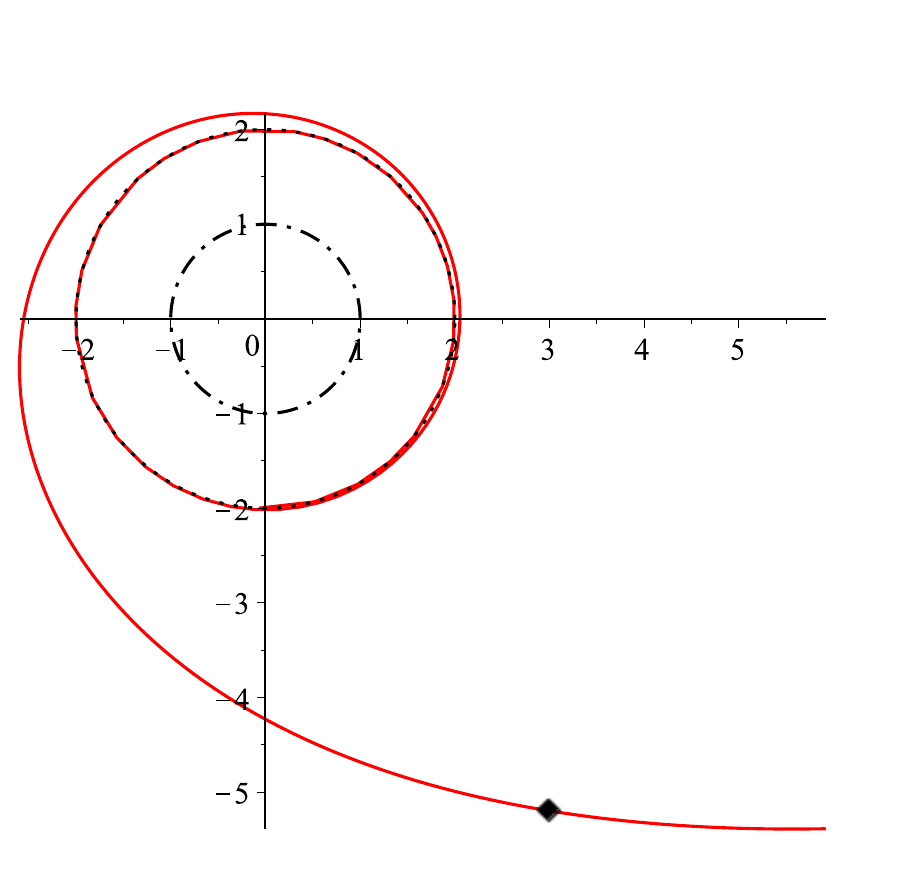}
\caption{Asymptotic circular hyperbolic-like and parabolic-like orbits: LRL-CP angle $\Phi$}\label{fig:IP-achyper-acpara}
\end{figure}

First, the universal choice of $r_0$ given by a centripetal point is illustrated
for all orbits that lie outside of the horizon. 
The conserved quantity $\Phi$ for these orbits is the analogue of 
the angle of Hamilton's vector in (post-) Newtonian gravity
which is a variant of the LRL vector. 
As seen from Tables~\ref{orbits-bounded} and~\ref{orbits-unbounded}, 
the bounded orbit types consist of 
elliptic-like and asymptotic circular, 
and the unbounded types consist of 
hyperbolic-like, parabolic-like, asymptotic circular hyperbolic-like, and asymptotic circular parabolic-like. 
These are shown in Figures~\ref{fig:IP-ellip} to~\ref{fig:IP-achyper-acpara}. 
The elliptic-like orbit illustrates precession of the centripetal point. 
In this case $\Phi$ is multi-valued and thus describes a piecewise constant of motion. 
In the other five cases, $\Phi$ is single-valued and therefore describes a global constant of motion. 

\begin{figure}[ht!]
\centering
\includegraphics[trim=2cm 12cm 4cm 1cm,clip,width=0.45\textwidth]{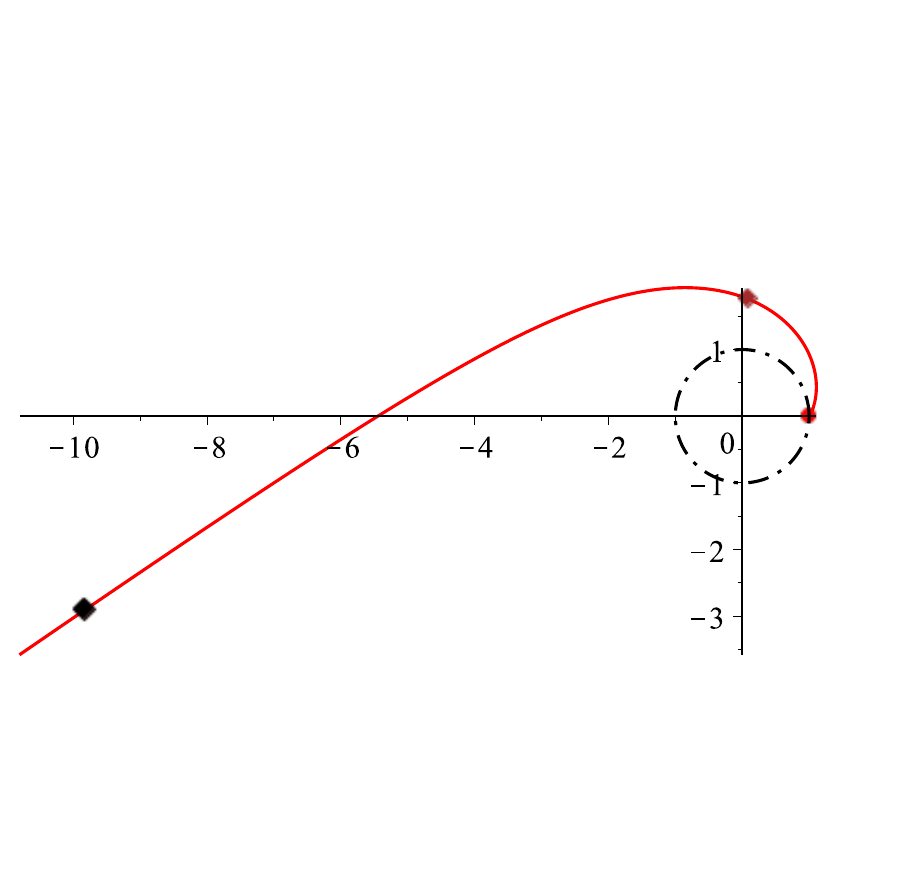}
\includegraphics[trim=2cm 12cm 4cm 1cm,clip,width=0.45\textwidth]{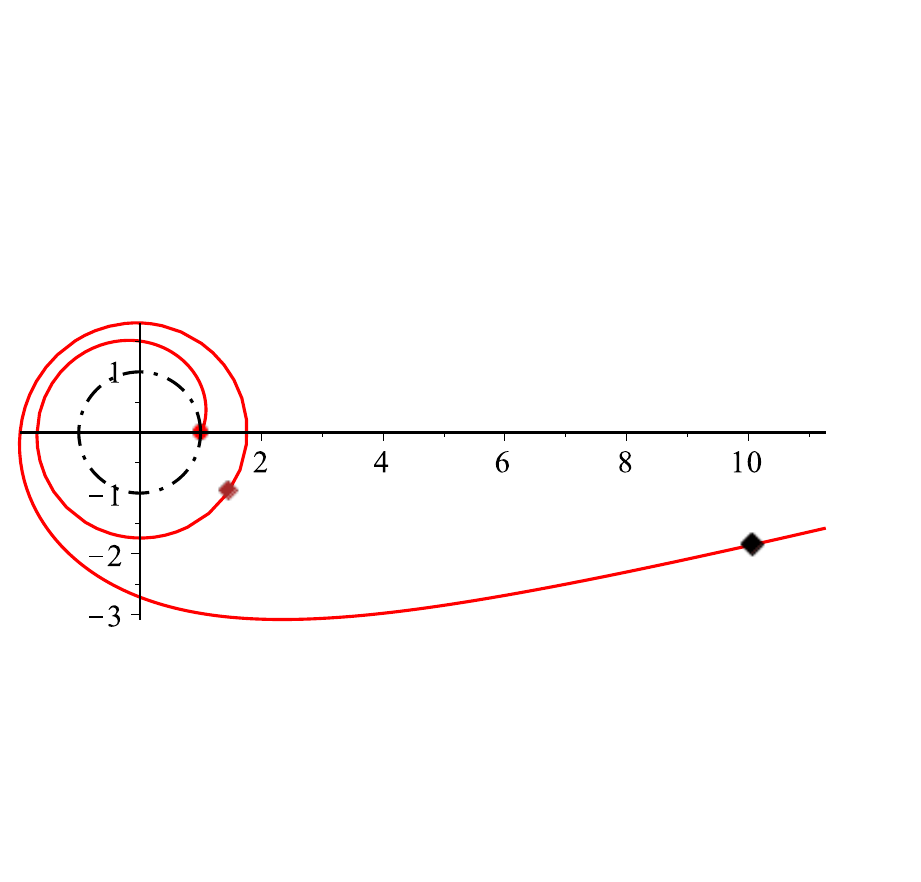}
\caption{Horizon-crossing orbits: LRL-CP angle $\Phi$}\label{fig:IP-hc}
\end{figure}

\begin{figure}[ht!]
\centering
\includegraphics[trim=2cm 12cm 4cm 1cm,clip,width=0.45\textwidth]{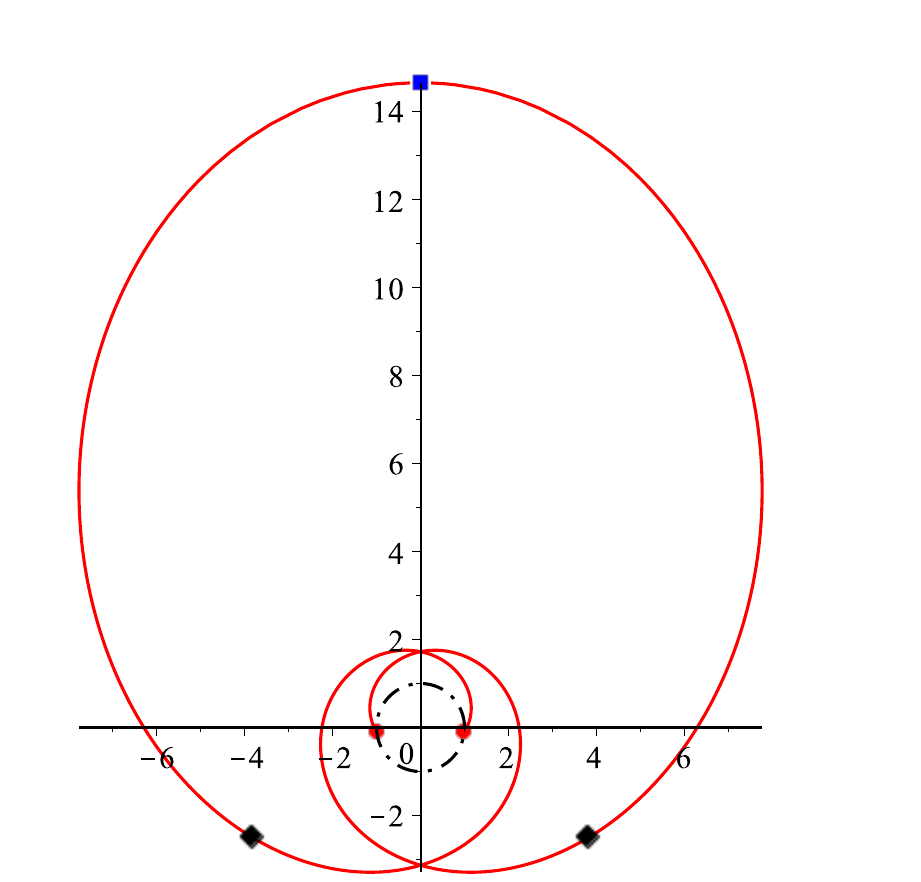}
\caption{Bounded horizon-crossing orbit: LRL-CP angle $\Phi$}\label{fig:IP-hcbdd}
\end{figure}

The choice of a centripetal point for $r_0$ can also be made in the case of 
horizon-crossing orbits that have $\bar L^2\geq 3$, 
as seen from Tables~\ref{orbits-bounded} and~\ref{orbits-unbounded}. 
For these orbits, the conserved quantity $\Phi$ is the analogue of  
the angle of Hamilton's (variant LRL) vector in (post-) Newtonian gravity 
for unbounded orbits.  
In particular, 
$\Phi$ is single-valued and therefore describes a global constant of motion. 
Figure~\ref{fig:IP-hc} shows the case of unbounded orbits,
and Figure~\ref{fig:IP-hcbdd} shows the case of bounded orbits. 

\begin{figure}[ht!]
\centering
\includegraphics[trim=2cm 12cm 4cm 1cm,clip,width=0.43\textwidth]{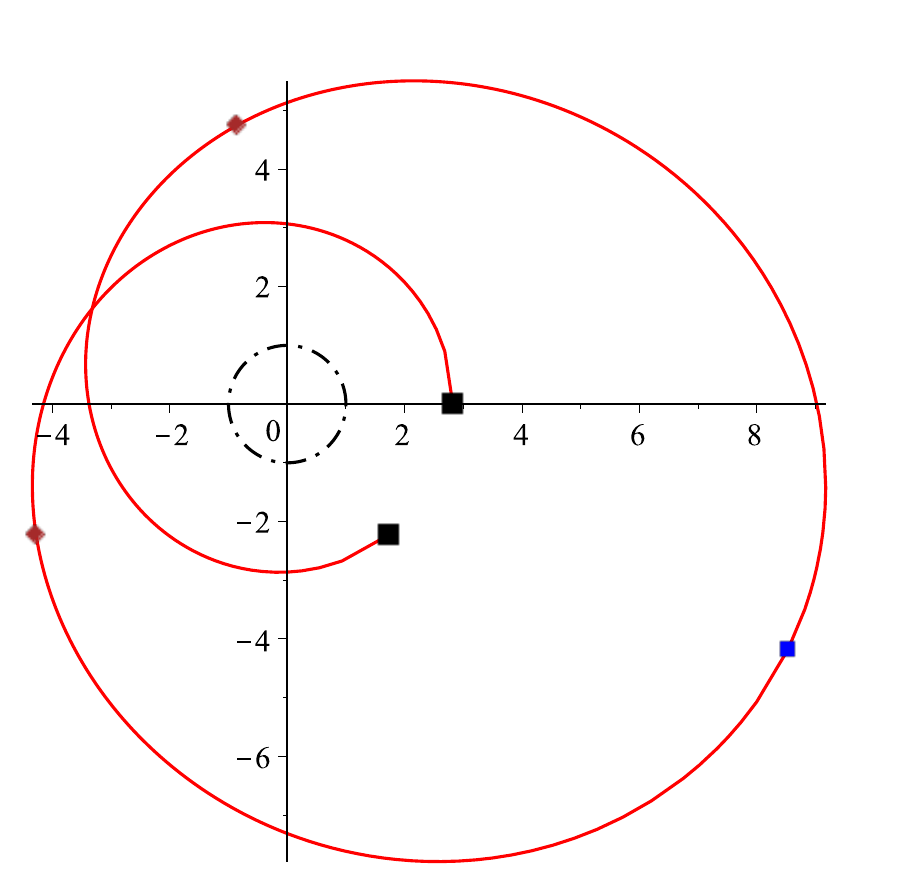}
\includegraphics[trim=2cm 12cm 4cm 1cm,clip,width=0.51\textwidth]{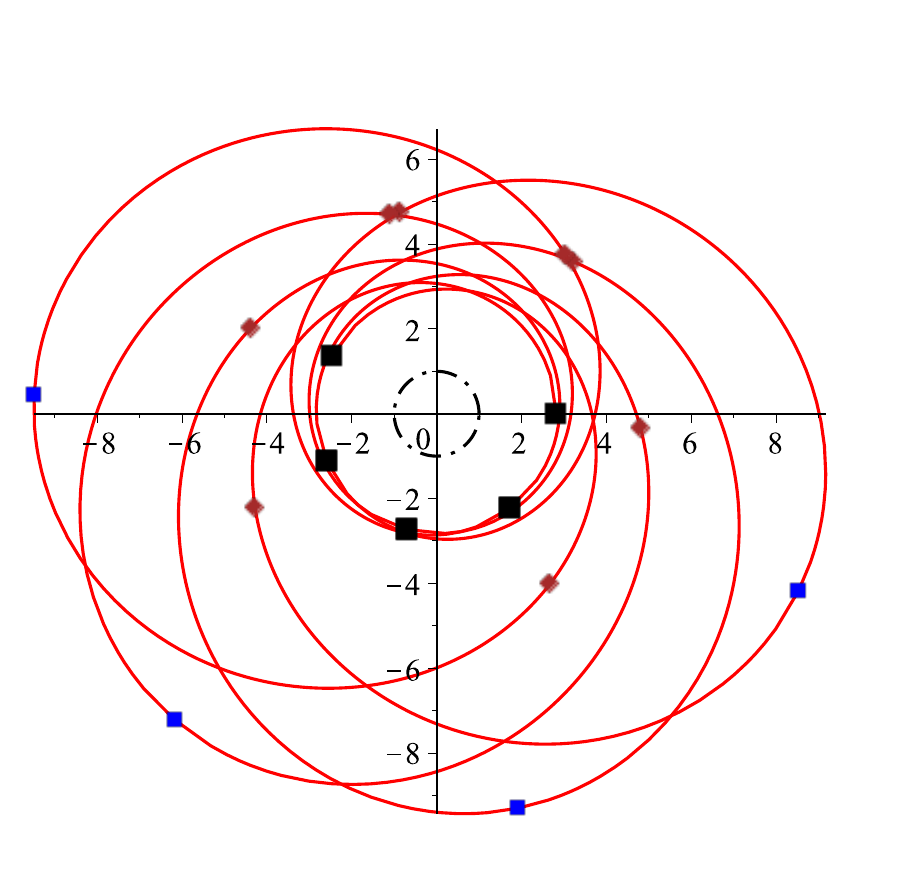}
\caption{Elliptic-like orbit: LRL-TP angle $\Phi$}\label{fig:TP-ellip}
\end{figure}

\begin{figure}[ht!]
\centering
\includegraphics[trim=2cm 12cm 4cm 1cm,clip,width=0.45\textwidth]{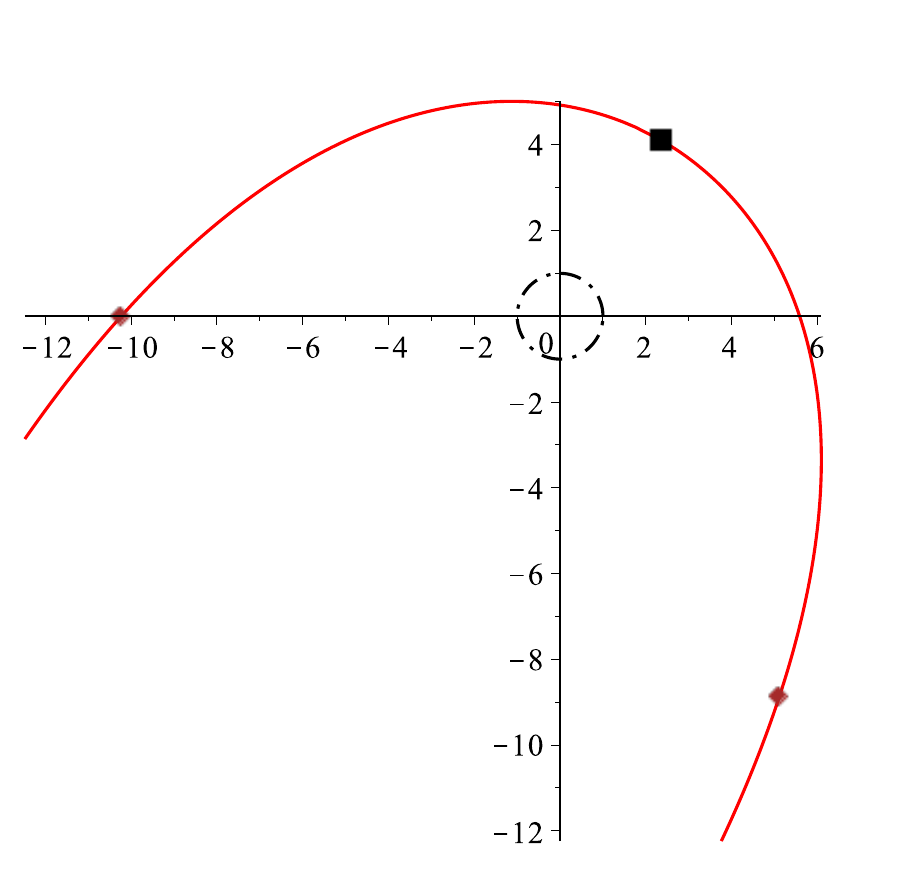}
\includegraphics[trim=2cm 12cm 4cm 1cm,clip,width=0.45\textwidth]{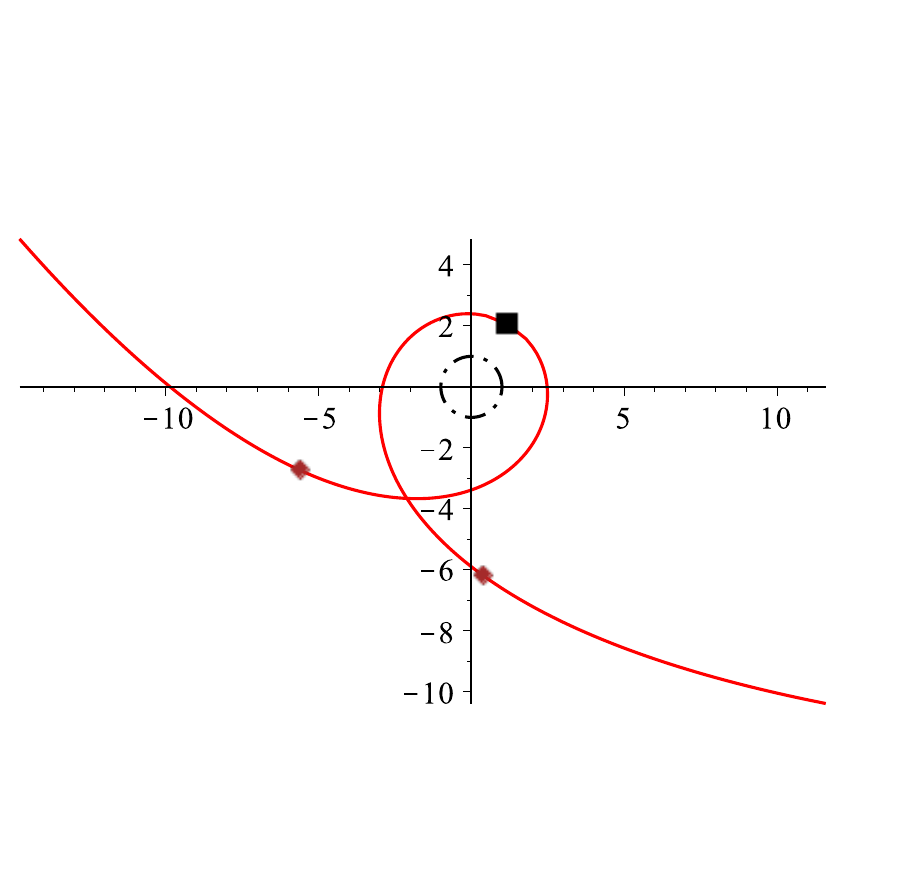}
\caption{Parabolic-like orbits: LRL-TP angle $\Phi$}\label{fig:TP-para}
\end{figure}

\begin{figure}[ht!]
\centering
\includegraphics[trim=2cm 12cm 4cm 1cm,clip,width=0.45\textwidth]{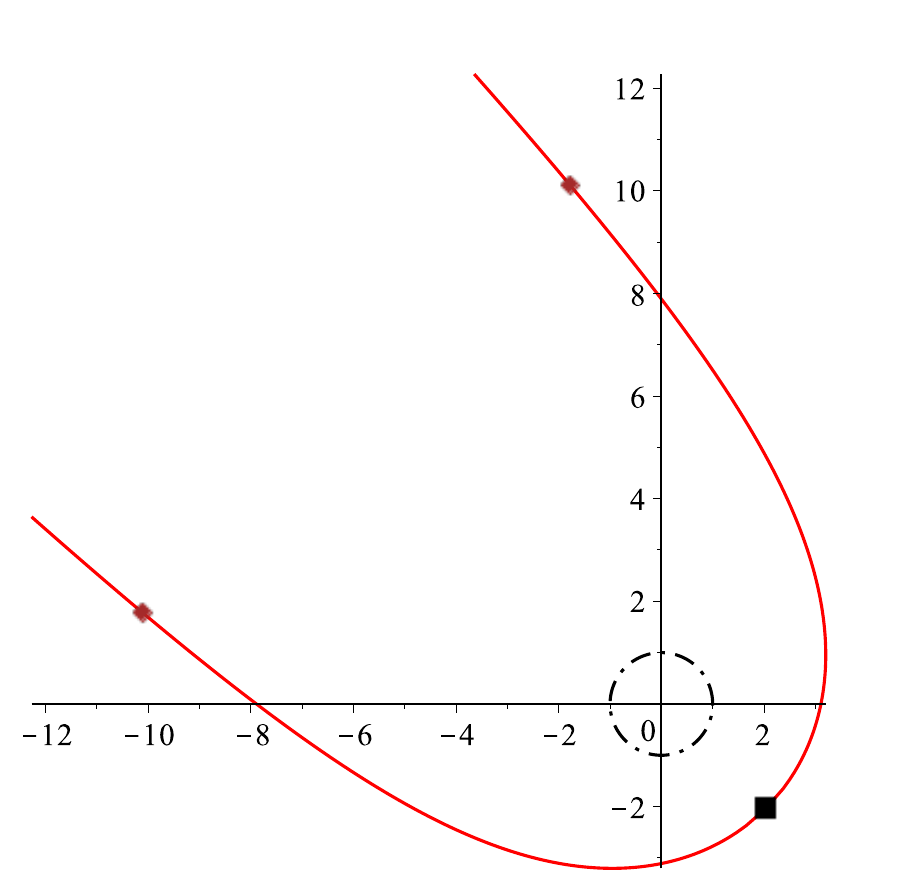}
\includegraphics[trim=2cm 12cm 4cm 1cm,clip,width=0.45\textwidth]{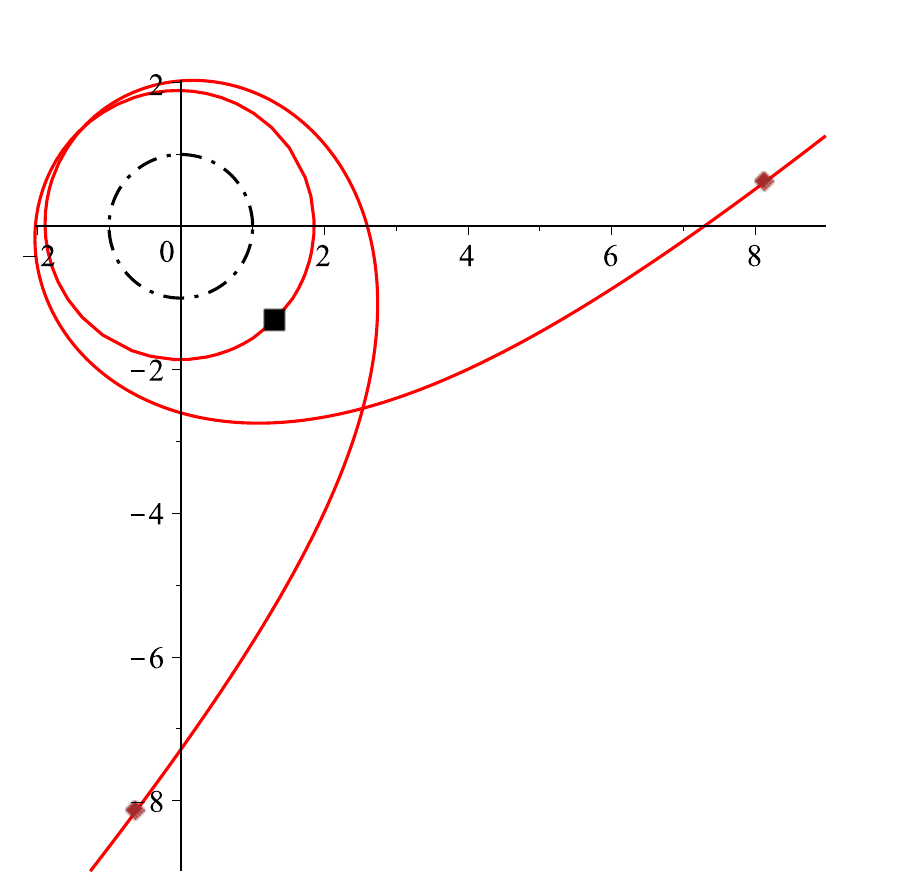}
\caption{Hyperbolic-like orbits: LRL-TP angle $\Phi$}\label{fig:TP-hyper}
\end{figure}

Next, for elliptic-like, parabolic-like, hyperbolic-like orbits 
---  which are the counterparts of Newtonian orbits --- 
the choice of $r_0$ given by a turning point is illustrated. 
The resulting conserved quantity $\Phi$ is the analogue of 
the angle of the LRL vector in (post-) Newtonian gravity. 
Figures~\ref{fig:TP-ellip} to~\ref{fig:TP-hyper} show the orbits and the turning points. 
The elliptic-like orbit illustrates precession of the turning point. 
In this case $\Phi$ is multi-valued and thus describes a piecewise constant of motion. 
In the parabolic and hyperbolic cases, $\Phi$ is single-valued and therefore describes a global constant of motion. 

\begin{figure}[ht!]
\centering
\includegraphics[trim=2cm 12cm 4cm 1cm,clip,width=0.45\textwidth]{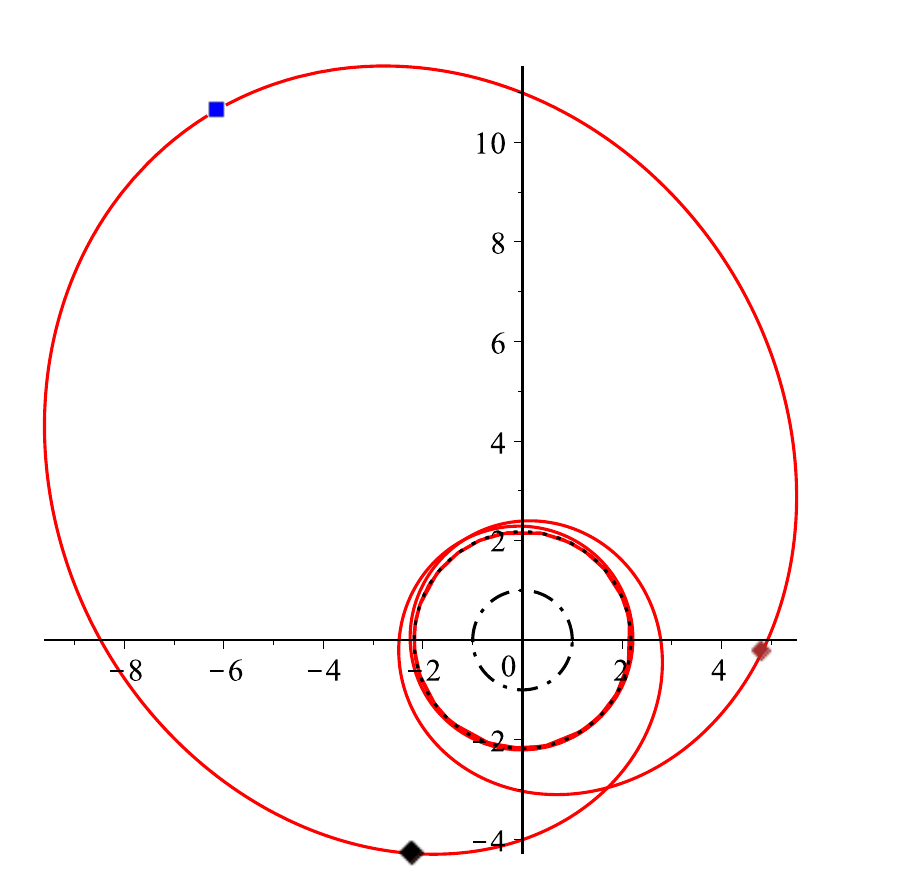}
\includegraphics[trim=2cm 12cm 4cm 1cm,clip,width=0.49\textwidth]{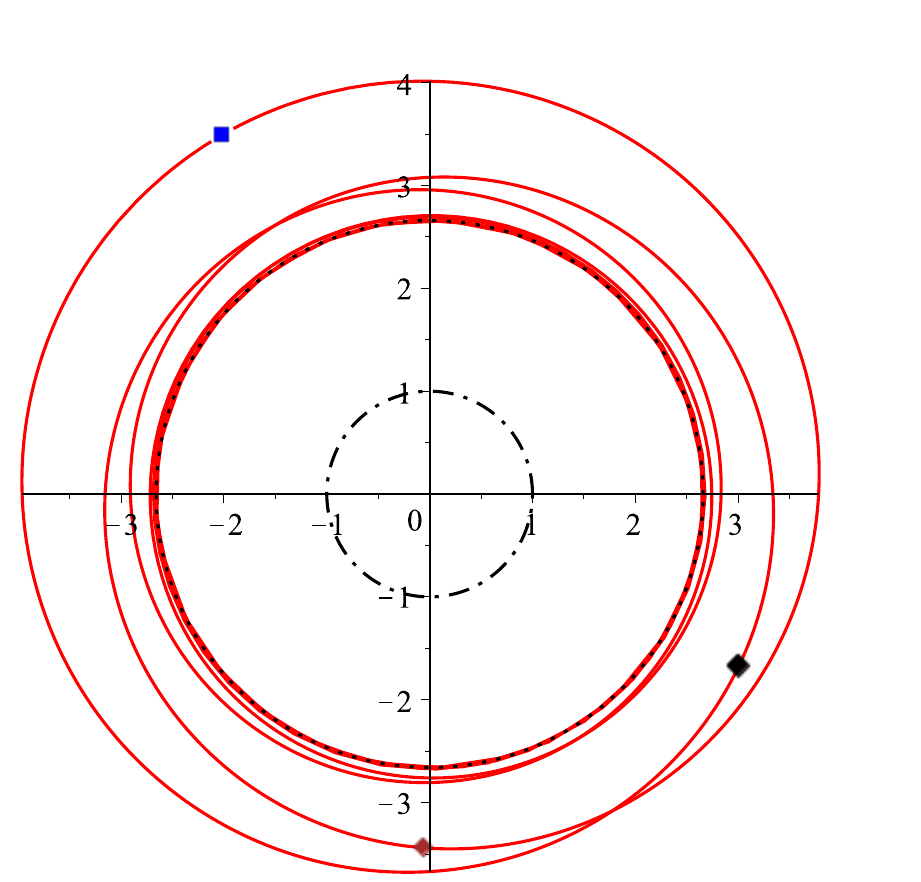}
\caption{Bounded asymptotic circular orbits: LRL-TP angle $\Phi$}\label{fig:TP-acbdd}
\end{figure}

\begin{figure}[ht!]
\centering
\includegraphics[trim=2cm 12cm 4cm 1cm,clip,width=0.4\textwidth]{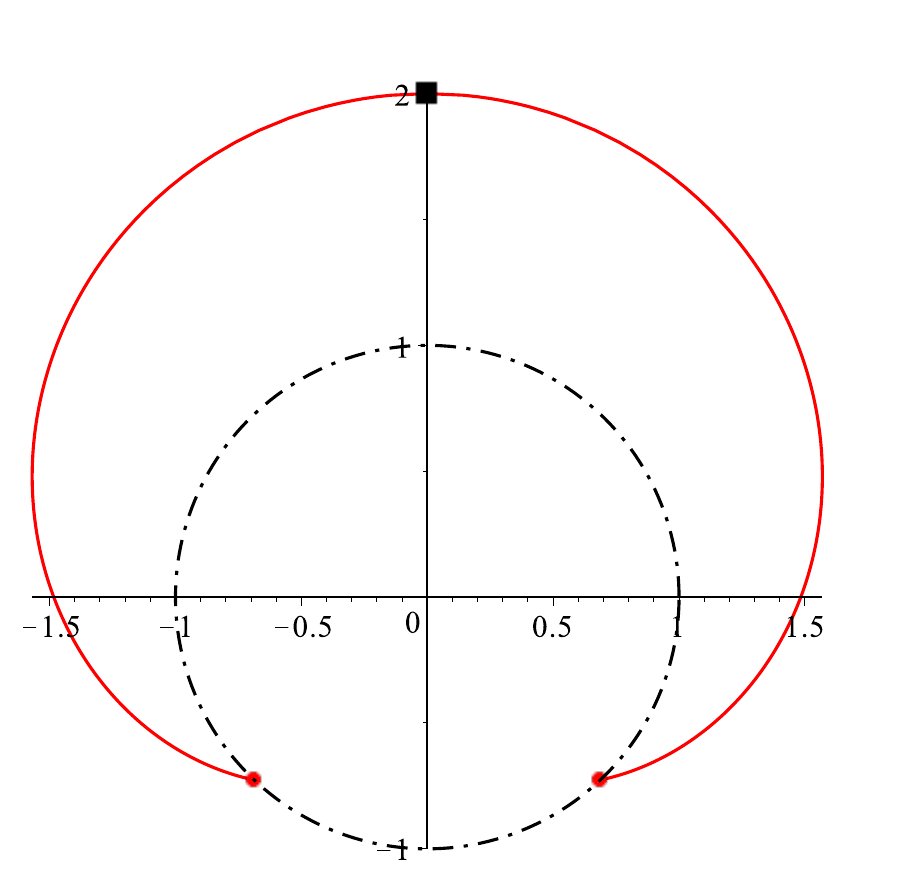}
\includegraphics[trim=2cm 12cm 4cm 1cm,clip,width=0.55\textwidth]{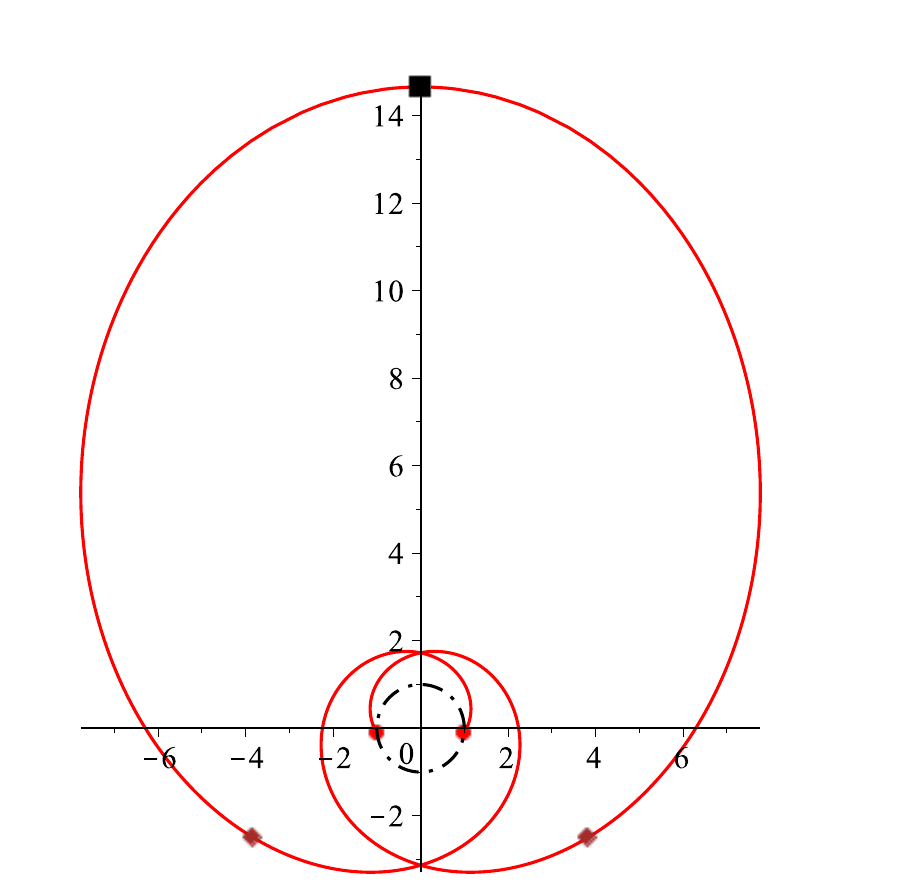}
\caption{Bounded horizon-crossing orbits: LRL-TP angle $\Phi$}\label{fig:TP-hcbdd}
\end{figure}

A turning point can also be chosen for $r_0$ in the case of 
bounded orbits that are either asymptotic circular or horizon-crossing.
The conserved quantity $\Phi$ is the analogue of  
the angle of the reflected LRL vector in Newtonian gravity,
which points in the direction of the apoapsis instead of the periapsis. 
Figures~\ref{fig:TP-acbdd} and~\ref{fig:TP-hcbdd} show the orbits and the turning points.
For all of these orbits, 
$\Phi$ is single-valued and therefore describes a global constant of motion. 

\begin{figure}[ht!]
\centering
\includegraphics[trim=2cm 12cm 4cm 1cm,clip,width=0.45\textwidth]{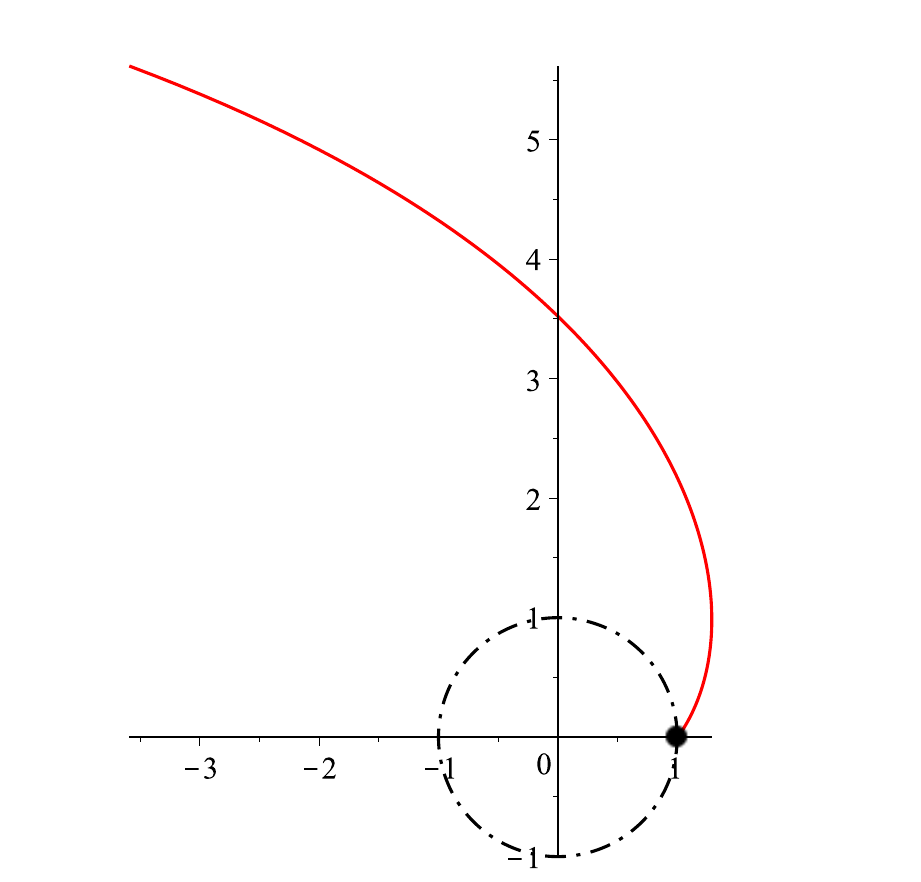}
\includegraphics[trim=2cm 12cm 4cm 1cm,clip,width=0.45\textwidth]{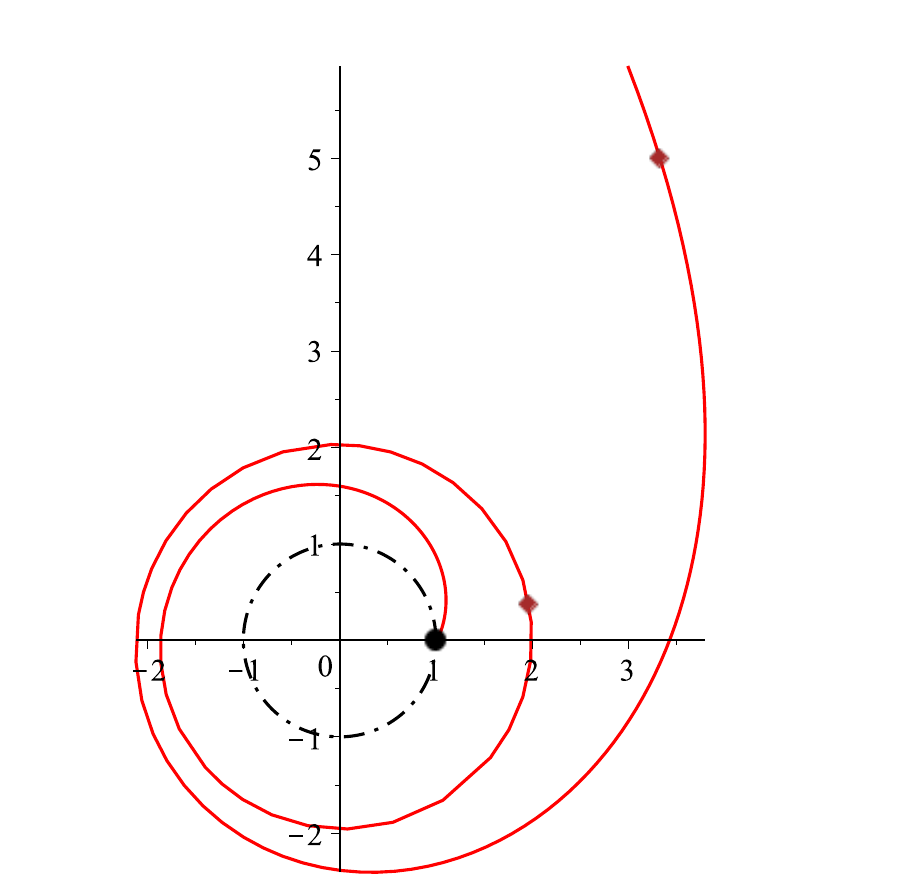}
\caption{Horizon-crossing orbits: HP angle $\Phi$}\label{fig:HP-hc}
\end{figure}

\begin{figure}[ht!]
\centering
\includegraphics[trim=2cm 12cm 4cm 1cm,clip,width=0.4\textwidth]{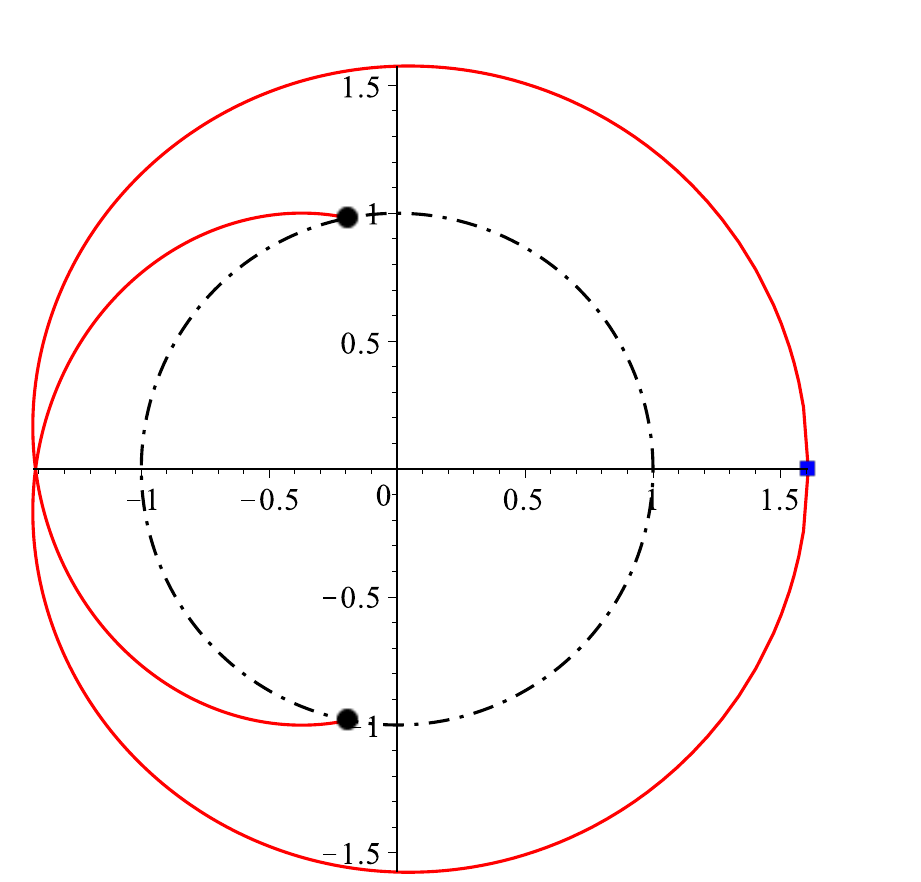}
\includegraphics[trim=2cm 12cm 4cm 1cm,clip,width=0.58\textwidth]{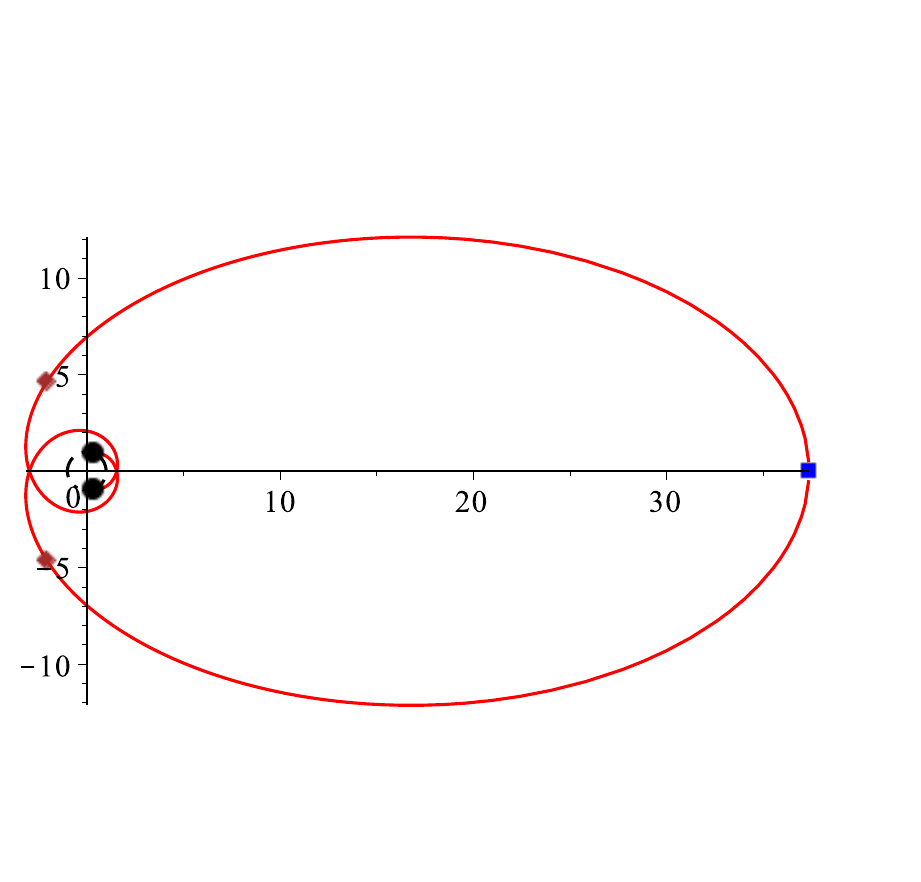}
\caption{Bounded horizon-crossing orbits: HP angle $\Phi$}\label{fig:HP-hcbdd}
\end{figure}

\begin{figure}[ht!]
\centering
\includegraphics[trim=2cm 12cm 4cm 1cm,clip,width=0.45\textwidth]{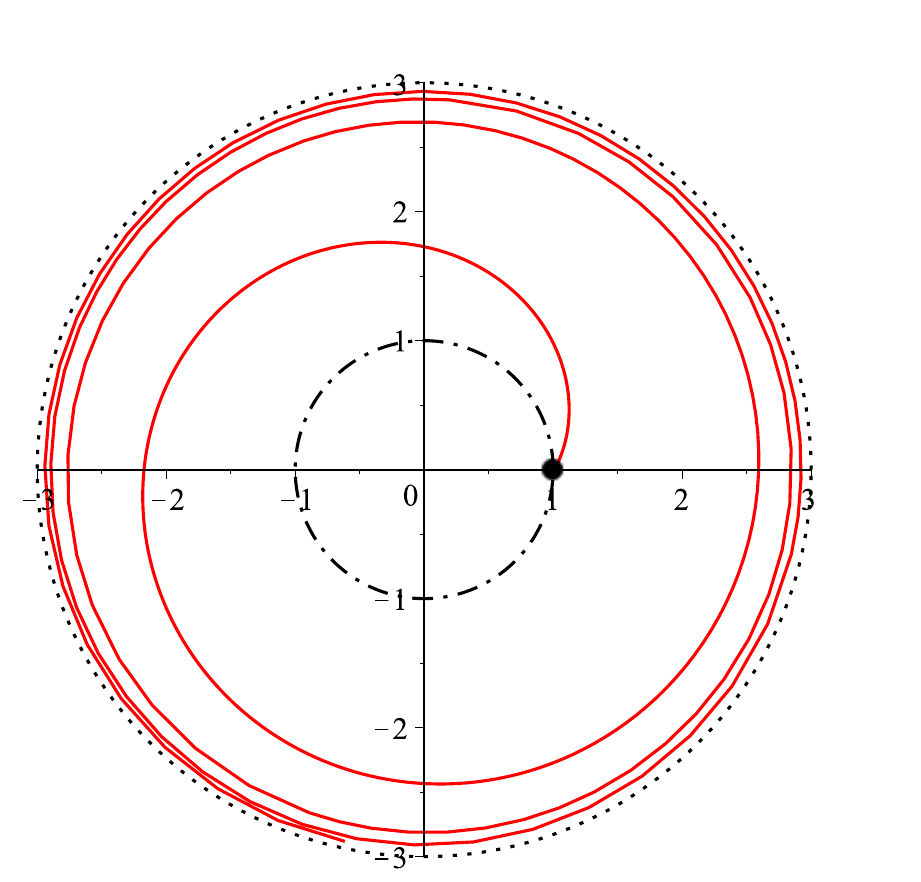}
\caption{Asymptotic circular horizon-crossing orbit: HP angle $\Phi$}\label{fig:HP-achc}
\end{figure}

Finally, for all orbits that cross the horizon, 
a universal choice of $r_0$ given by a horizon point is shown in 
Figures~\ref{fig:HP-hc} to~\ref{fig:HP-achc}. 
The conserved quantity $\Phi$ for these orbits has no analogue 
in Newtonian gravity. 
It is single-valued in the case of unbounded orbits
and double-valued in the case of bounded orbits.

\section{A generalized LRL vector at spatial infinity} \label{sec:LRLvector}

Since Schwarzschild spacetime $(M,g)$ is asymptotically flat,
the region $r\to \infty$ at finite coordinate time $t$
in spherical coordinates $(t,r,\phi,\theta)$ represents spatial infinity. 
As is well known (see e.g.\ \cite{Wal}), 
this asymptotic spacetime region can be viewed as a single point, $\i0$, 
belonging to a conformally completed spacetime manifold $(\tilde M,\tilde g)$ 
that includes future and past null infinity $\I^\pm$ of Schwarzschild spacetime. 
The completion is defined by means of a conformal transformation of 
the Schwarzschild metric $\tilde g = \Omega^2 g$, 
where the conformal factor $\Omega(r,t)\to 0$ as $r\to \infty$ for finite $t$,
such that null infinity $\I^\pm$ becomes the boundary of 
the conformal spacetime $(\tilde M,\tilde g)$
given by null curves which intersect at the point $\i0$. 
In this completion, the $SO(3)$ rotational symmetry of Schwarzschild spacetime
is manifestly preserved, namely
$\Omega(r,t)$ is a function only of $t$ and $r$. 
(Explicit expressions for $\Omega$ and $\tilde g$ can be found in \Ref{AshHan,HalLed}.)

The conformal Schwarzschild spacetime thereby 
has a tangent space at spatial infinity $T(\tilde M)_{\i0}\simeq \Rnum^{3,1}$
which is isomorphic to Minkowski space. 
This is the vector space in which the ADM energy-momentum \cite{MisThoWhe,Wal} of 
Schwarzschild spacetime sits as a 4-vector,
$\mathbf{P}= -M \hat{\mathbf{t}}$
where $\hat{\mathbf{t}}$ denotes the unit timelike vector representing 
the extension of the time-translation Killing vector $\partial_t$ to spatial infinity. 

In the equatorial plane of Schwarzschild spacetime,
consider a particle orbit $(t(\tau),r(\tau),\phi(\tau),\pi/2)$ that is non-circular 
and lies outside of the horizon. 
The locally conserved angular quantity $\Phi$ 
given by expression \eqref{LRLangle} in Theorem~\ref{thm:firstintegrals} 
describes a generalized LRL angle for the orbit,
as summarized in Theorem~\ref{thm:LRLanalogs}.
A corresponding unit radial vector, $\hat\A$, 
representing a generalized LRL vector, 
can be introduced at spatial infinity $\i0$ as follows. 

\begin{prop}\label{prop:LRLvector.spatialinfinity}
For any non-circular orbit in the equatorial plane, 
let $\hat\A$ be a spacelike unit vector given by 
the celestial angles $(\phi=\Phi,\theta=\pi/2)$ 
in the spatial 3-plane orthogonal to the asymptotic time-translation Killing vector
$\hat{\mathbf{t}}$ at $\i0$,
where $\Phi$ is the locally conserved angular quantity \eqref{LRLangle}. 
Then $\hat\A$ inherits the properties of $\Phi$ stated in 
Theorems~\ref{thm:LRLanalogs}, \ref{thm:globalproperties.TP}, \ref{thm:globalproperties.CP}, \ref{thm:globalproperties.HP}. 
\end{prop}

In particular, $\hat\A$ as a function of proper time $\tau$ for the orbit 
is (at least) piecewise constant and thus it is a locally conserved vector. 
It has the same global properties with respect to $\tau$ as $\Phi$ does, 
which depend on the choice of a dynamically distinguished radial value, $r_0$, 
consisting of either a turning point, a centripetal point, or a horizon-crossing point 
(in the case when the orbit enters the horizon)
as explained in Theorem~\ref{thm:firstintegrals}. 

When $r_0$ is a turning point on a orbit, either an apoapsis or periapsis, 
$\hat\A$ is an analogue of the LRL vector in Newtonian gravity with cubic corrections.
If the orbit has only a single turning point, it is single-valued 
(globally constant with respect to $\tau$); 
otherwise if the orbit has multiple turning points, it is multi-valued
(piecewise constant with respect to $\tau$) and undergoes a jump at proper times
given by successive apoapsis or periapsis points on the orbit. 

Alternatively, when $r_0$ is a centripetal point on a orbit with an apsis, 
$\hat\A$ is an analogue of Hamilton's vector in Newtonian gravity with cubic corrections
and undergoes a jump at the proper time given by the apsis point on the orbit.
(See Theorem~\ref{thm:globalproperties.CP} for more details.)

Finally, $\hat\A$ has no Newtonian analogue when $r_0$ is a horizon-crossing point
for an orbit that enters the horizon. 
(See Theorem~\ref{thm:globalproperties.HP} for the properties in this case.)

Regarding the geometrical status of $\hat\A$, 
some further remarks relating this vector to $\Phi$ at spatial infinity 
may be helpful.

\subsection{$\hat\A$ and the celestial sphere for Schwarzschild spacetime}

The $SO(3)$ rotational symmetry of Schwarzschild spacetime 
yields a natural correspondence between 
radial lines in the spacelike hypersurface $t=\const$ 
and points on a celestial sphere $S^2$ at spatial infinity.  

Specifically, 
first consider a congruence of radial lines in the spacelike hypersurface $t=\const$,
which extend to spatial infinity as $r\to \infty$.  
Each radial line determines a point on the celestial sphere $S^2$ 
coordinatized by the polar and azimuthal angles $(\phi,\theta)$. 
A radial line in the equatorial hypersurface, $\theta=\pi/2$, thereby determines
a point $(\phi,\pi/2)$ on the equator of the celestial sphere. 
At any point in this hypersurface, 
the tangent vector of the radial line can then be identified with the given polar angle. 
Since this identification does not depend on the norm of the tangent vector, 
it provides a one-to-one correspondence between 
unit radial vectors at every point on these lines 
and polar angles on the equator of the celestial sphere. 

Now consider temporal and radial coordinates $\tilde t$ and $\tilde r$ such that 
spatial infinity $\i0$ is the point $\tilde t =\tilde r =0$ 
in the conformal Schwarzschild spacetime $(\tilde M,\tilde g)$
and the conformally transformed Schwarzschild metric has the asymptotic form 
$d\tilde{s}{}^2 = \Omega^2 ds^2
\sim -d\tilde{t}{}^2 + d\tilde{r}{}^2 + \tilde{r}{}^2 (\sin^2\theta d\phi^2+d\theta^2))$
to leading order near $\i0$. 
The congruence of radial lines in the spacelike hypersurface $t=\const$
will get mapped into a corresponding congruence of lines through $\i0$
in the conformal spacetime, 
where the directional angle $(\phi=\const,\theta=\const)$ of each line is preserved
by the conformal transformation. 
Thus, equivalently, the celestial sphere is preserved. 

As a consequence, 
the polar angle given by $\phi=\Phi$ on the equator of the celestial sphere 
corresponds to a spacelike unit tangent vector at $\i0$ in $(\tilde M,g)$,
where this vector is orthogonal to the time-translation Killing vector 
$\hat{\mathbf{t}}$ at $\i0$,
using the Schwarzschild metric $g$ extended to $\tilde M$.

\section{Concluding remarks}\label{sec:remarks}

The main results in the present work can be viewed as being motivated by 
the similarity between the timelike equatorial geodesic equations in Schwarzschild spacetime 
formulated as a dynamical system for geodesic motion projected into an equatorial plane, 
and the equations of motion in general central force dynamics 
formulated in the plane of motion. 
Both of these systems of dynamical equations 
share invariance under the $SO(2)$ group of rotations and time-translations,
which is manifest when the respective equations are formulated 
in polar coordinates $(r,\phi)$. 
This shared symmetry structure is inherited geometrically from the fact that 
both Schwarzschild spacetime and Newtonian spacetime 
possess the same rotational Killing vectors 
as well as the same time-translation Killing vector. 

In particular, 
the dynamical equations for the spatial (projected) equatorial geodesic motion 
$(r(\tau),\phi(\tau))$
have the mathematical form of generalized central force equations of motion. 
This is what essentially gives rise to 
the conserved $\phi$-quantity and the conserved $t$-quantity 
presented in Theorem~\ref{thm:firstintegrals} for timelike equatorial geodesics. 
The additional conserved $\tau$-quantity in this Theorem, 
which has no counterpart in central force dynamics, 
arises because of the relativistic (3+1 dimensional) nature of geodesic motion,
compared to the Newtonian (3 dimensional) nature of central force motion. 

The physical meaning and mathematical properties of 
the conserved $\phi$-quantity and the conserved $t$, $\tau$-quantities  
are likewise similar to the angular and temporal conserved quantities 
that appear in central force dynamics. 
Specifically, the $\phi$-quantity is a direct analogue of 
the angle defined by the LRL and Hamilton vectors 
which are well recognized 
as locally conserved vectors in central force dynamics; 
the $t,\tau$-quantities are direct analogues of 
the time at which the angle $\phi$ coincides with the LRL/Hamilton vector's angle,
which is a less widely known conserved scalar in central force dynamics
that only recently has been studied in detail \cite{AncMeaPas,AncBalGan}. 
It is important to emphasize that each of these three conserved quantities 
do not involve initial conditions for timelike equatorial geodesics, 
and hence they have the same local dynamical status 
as the angular momentum and energy constants of motion. 
Analogs of the LRL vector and Hamilton's vector, in their normalized forms, 
are provided by the radial spacelike unit vector $\hat\A$ at spatial infinity. 

Noether's theorem \cite{Olv,BA-book,Anc-review} gives a connection between 
the conserved $\phi$, $t$, $\tau$-quantities 
and dynamical symmetries of the Lagrangian for timelike equatorial geodesics,
similarly to the symmetries associated to the LRL vector in central force dynamics 
\cite{Fra,Muk,Lev,Rog,AncMeaPas}. 
In outline, for the dynamical equations governing geodesic motion 
$(t(\tau),r(\tau),\phi(\tau))$ in the equatorial hypersurface, 
there is one-to-one correspondence between integrals of motion of these equations 
and infinitesimal symmetries of the geodesic Lagrangian. 
Hence, by Noether's theorem in reverse, 
each of the conserved quantities can be used to obtain 
a corresponding infinitesimal symmetry. 
Angular momentum and energy respectively yield 
a rotation symmetry and a time-translation symmetry,
which can be identified with the geometrical symmetries generated by 
the Killing vectors of the equatorial Schwarzschild metric.
In contrast, dynamical symmetries turn out to arise from each of 
the conserved $\phi$, $t$, $\tau$-quantities. 
These three dynamical symmetries are a hidden feature of 
the timelike equatorial geodesic equations
and neither represent nor correspond to spacetime symmetries 
or other geometrical structures.

There are several open questions for future work. 

First, the dynamical symmetries can be derived in an explicit form, 
including the associated symmetry transformation group, 
so that their features and properties can be studied
and compared with their Newtonian counterparts derived in \Ref{AncMeaPas}. 
Additionally, the integrals of motion can be studied in a Hamiltonian context, 
which will yield an isomorphism between their Poisson bracket algebra 
and the algebra of the associated dynamical symmetries. 

Second, the derivation of integrals of motion can be extended to the 
full (non-equatorial) geodesic equations. 
This would allow seeking a locally conserved geometrical vector or 1-form 
along non-circular orbits,
which would be full analog of the LRL vector in central force dynamics. 

Last, it would be interesting to explore generalizing the methods and the results 
from Schwarzschild spacetime to more general spacetimes with symmetry, 
such as non-static spherically symmetric spacetimes 
like the Friedman-Lemaitre-Robinson-Walker cosmology,
and stationary axisymmetric spacetimes like the Kerr black hole.

\appendix

\section{Evaluation of the angular and temporal integrals of motion}\label{sec:evaluate}

The angular integral of motion \eqref{LRLangle-u}
and the two temporal integrals of motion \eqref{LRLtime-u} and \eqref{LRLtau-u}
will now be evaluated with an arbitrary choice of $r_0$,
similarly to the Newtonian limit shown in the next appendix.
The technical aspect will employ the approach used in \Ref{Cha}
for integration of the timelike equatorial geodesic equations
\eqref{timelikegeodesiceqn-2ndord}--\eqref{propertimeeqn-2ndord}. 
One simplification is that the integrals will be parameterized explicitly in terms of 
the roots of the turning point cubic equation \eqref{tp-eqn} in all cases. 
This reduces the algebraic complexity of the expressions for the integrals of motion, 
especially the two temporal integrals. 
(It is straightforward to express the results in terms of 
the eccentricity and latus rectum parameters in \Ref{Cha} 
which are sometimes used for comparisons with the Newtonian case 
and for post-Newtonian approximations.)

Recall that $u$ denotes the reciprocal radial variable \eqref{u},
and likewise $u_0 = 2M/r_0$. 
Also recall that $Q(u)$ is the cubic polynomial \eqref{Q}. 

For evaluating the quadratures in expressions \eqref{LRLangle-u}, \eqref{LRLtime-u} and \eqref{LRLtau-u}, 
the root structure of $Q(u)=0$ needs to be known. 
It is determined by 
the discriminant $D$ of $Q(u)$ and the discriminant $D_0$ of $Q'(u)$: 
\begin{subequations}
\begin{gather}
\bar{L}^6 D = \tfrac{1}{27}\big( 4(\bar{L}^2-3)^3 - \bar{L}^2(27E^2-18 -2\bar{L}^2)^2 \big) 
= 27\bar{L}^2(E_+^2-E^2)(E^2-E_-^2),
\\
\bar{L}^4 D_0  =4(\bar{L}^2-3) ,
\end{gather}
\end{subequations}
where $E_\pm^2$ is given by expression \eqref{E+-}. 
As shown in \Ref{Cha}, 
the three roots $u_1$, $u_2$, $u_3$ of the cubic equation $Q(u)=0$ 
can be classified into the list of five cases shown in Table~\ref{roots}. 
In the cases where all of the roots are real, 
they will be ordered $u_3\geq u_2\geq u_1$;
in the case where only one of the roots is real, it will be designated as $u_1$.  
The orbits that occur in each case are summarized 
in Tables~\ref{orbits-bounded} and~\ref{orbits-unbounded} 
in section~\ref{sec:properties}. 

Note that the three roots are functions of the two parameters $\bar{L}^2$ and $E^2$. 
From the cubic equation $Q(u)=0$, the roots obey the relations
\begin{equation}\label{root-rel}
u_1 +u_2+u_3 =1 , 
\quad
\sgn(u_1) = \sgn(1-E^2). 
\end{equation}
In the case when the roots are real, another useful relation is that 
they lie between the roots of the quadratic equation $Q'(u)=0$:
\begin{equation}\label{u+-}
u_\pm = \tfrac{1}{3} \big( 1 \pm \smallsqrt{1-3/\bar{L}^2} \big) . 
\end{equation}

\begin{table}[h!!]
\centering
\caption{
Root structure of $Q(u)=0$ and range of $u$ for $Q(u)\geq0$.
}
\label{roots} 
\begin{tabular}{c||c|c|c|c|c}
\hline
Case & Discriminants & Root type & Range of $u$ & \parbox{0.25in}{$E^2$} & \parbox{0.25in}{$\bar{L}^2$}
\\
\hline\hline
(1) 
& 
$D = 0$, $D_0 = 0$
& 
$u_1=u_2=u_3>0$
& 
$u\geq u_1$
& 
\parbox{0.75in}{$\tfrac{8}{9}$ ($=E^2_\pm$)\strut}
& 
$3$
\\
\hline
(2a) 
& 
$D = 0$, $D_0 > 0$
& 
$u_3>u_2=u_1>0$
&
$u\geq u_3$
& 
$E^2_-$\strut
&
$>3$
\\
\hline
(2b) 
& 
$D = 0$, $D_0 > 0$
&
\parbox{1.25in}{\strut $u_3=u_2>0\geq u_1$ \\\strut $u_3=u_2>u_1>0$ }
&
\parbox{1.6in}{\strut $u\geq u_2$ or $u_2\geq u>0$ \\\strut $u\geq u_2$ or $u_2\geq u\geq u_1$}
& 
$E^2_+$\strut
&
\parbox{1.0in}{\centering $\geq 4$ \\ $>3$ and $<4$}
\\
\hline
(3) 
&
$D > 0$
&
\parbox{1.25in}{\strut $u_3>u_2>0\geq u_1$ \\ $u_3>u_2>u_1>0$ \\\quad}
&
\parbox{1.6in}{$u\geq u_3$ or $u_2\geq u\geq 0$\\ $u\geq u_3$ or $u_2 \geq u\geq u_1$\\\quad}
& 
\parbox{1.2in}{\centering $\geq 1$ and $<E_+^2$ \strut\\ $>E_-^2$ and $<1$\strut\\ $>E_-^2$ and $<E_+^2$ \strut}
&
\parbox{1.0in}{\centering $>4$ \\ $\geq 4$ \\ $>3$ and $<4$\strut}
\\
\hline
(4)
&
$D < 0$
&
\parbox{1.25in}{$u_2 = \bar{u}_3$, $0\geq u_1$ \\\quad\\ $u_2 = \bar{u}_3$, $u_1>0$\\\quad\\\quad}
&
\parbox{1.5in}{\centering $u >0$ \\\quad\\ $u \geq u_1$ \\\quad \\\quad}
&
\parbox{1.1in}{\centering $>E_+^2$ \strut\\ $>1$ \\ $>E_+^2$ and $<1$\strut\\ $<E_-^2$ \strut\\ $<1$ \strut}
& 
\parbox{1.0in}{\centering $\geq 4$ \\ $>3$ and $<4$ \\ $>3$ and $<4$ \\ $>3$ \\\strut $<3$}
\\
\hline
\end{tabular}
\end{table}


For each of the five separate cases in Table~\ref{roots}, 
the roots as well as the quadratures will be presented next.

\textbf{Case (1)} 
The discriminant conditions $D=D_0=0$ yield 
\begin{equation}\label{tripleroot-conds}
\bar{L}^2 =3, 
\quad
E^2 = \tfrac{8}{9} . 
\end{equation}
This gives a triple root:
\begin{equation}\label{triple-roots}
u_1=u_2=u_3=\tfrac{1}{3}.
\end{equation}

The quadratures \eqref{LRLangle-u}--\eqref{LRLtau-u} are given 
in terms of elementary functions: 
\begin{subequations}\label{tripleroot-integrals}
\begin{equation}\label{tripleroot-Phi-integral}
\begin{aligned}
& I^\Phi(u;u_0) = 
\Bigg( \frac{-2}{\smallsqrt{u -u_1}} \Bigg)\Bigg|_{u_0}^{u} 
;
\end{aligned} 
\end{equation}
\begin{equation}\label{tripleroot-Tau-integral}
\begin{aligned}
I^\T(u;u_0) = \Bigg( 
\frac{-3}{\sqrt{u_1}^5}\arctan\Big( \sqrt{\frac{u-u_1}{u_1}} \Big) 
+\frac{1}{u_1\sqrt{u-u_1}} \bigg( \frac{1}{u} - \frac{3}{u_1} \bigg) 
\Bigg)\Bigg|_{u_0}^{u} 
;
\end{aligned}
\end{equation}
\begin{equation}\label{tripleroot-T-integral}
\begin{aligned}
I^T(u;u_0) = \Bigg( & 
\frac{2}{\sqrt{1-u_1}^3}\arctanh\Big( \sqrt{\frac{u-u_1}{1-u_1}} \Big) 
-\frac{2u_1+3}{\sqrt{u_1}^5}\arctan\Big( \sqrt{\frac{u-u_1}{u_1}} \Big) 
\\&\quad
+\frac{1}{u_1\sqrt{u-u_1}}\Big( \frac{1}{u} -\frac{u_1-3}{u_1(u_1-1)} \Big)
\Bigg)\Bigg|_{u_0}^{u}  
.
\end{aligned}
\end{equation}
\end{subequations}

\textbf{Case (2)}
The discriminant conditions $D_0\neq D=0$ give two cases
which are distinguished by the value of $E^2$. 

\textbf{Subcase (2a)}
\begin{equation}\label{doublesingleroot-conds}
\bar{L}^2>3,
\quad
E^2 = E_-^2 . 
\end{equation}
This yields a root structure consisting of a double root and a single root:
\begin{equation}\label{doublesingle-roots}
u_1 = u_2 = \tfrac{1}{3}\big(1 -\smallsqrt{1-3/\bar{L}^2}\big) 
= \tfrac{1}{2}(1-u_3), 
\quad
u_3 = \tfrac{1}{3}\big(1 +2\smallsqrt{1-3/\bar{L}^2}\big). 
\end{equation}
They have the ranges 
\begin{equation}\label{doublesingle-roots-range}
1>u_3 >\tfrac{1}{3}, 
\quad
\tfrac{1}{3}> u_1=u_2 >0 . 
\end{equation}

The quadratures \eqref{LRLangle-u}--\eqref{LRLtau-u} are given by elementary functions: 
\begin{subequations}\label{doublesingleroot-integrals}
\begin{equation}\label{doublesingleroot-Phi-integral}
\begin{aligned}
I^\Phi(u;u_0) = \Bigg( 
\frac{2}{\sqrt{u_3 - u_1}} \arctan\Big( \sqrt{\frac{u - u_3}{u_3 -u_1}} \Big)
\Bigg)\Bigg|_{u_0}^{u} 
;
\end{aligned}
\end{equation}
\begin{equation}\label{doublesingleroot-Tau-integral}
\begin{aligned}
I^\T(u;u_0) = \Bigg( & 
{-}\frac{1}{u_1\sqrt{u_3}} \Big(\frac{2}{u_1} + \frac{1}{u_3}\Big)\arctan\Big(\sqrt{\frac{u - u_3}{u_3}}\Big)
\\&\quad
+\frac{2}{u_1{}^2\sqrt{u_3-u_1}}\arctan\Big(\sqrt{\frac{u - u_3}{u_3 -u_1}}\Big) 
-\frac{\sqrt{u-u_3}}{u_1 u_3u}
\Bigg)\Bigg|_{u_0}^{u} 
;
\end{aligned}
\end{equation}
\begin{equation}\label{doublesingleroot-T-integral}
\begin{aligned}
I^T(u;u_0) = \Bigg( &
{-}\frac{1}{u_1\sqrt{u_3}} \Big( 2+\frac{2}{u_1} + \frac{1}{u_3} \Big)\arctan\Big(\sqrt{\frac{u - u_3}{u_3}}\Big)
\\&\quad
+\frac{2}{u_1{}^2(1-u_1)\sqrt{u_3-u_1}}\arctan\Big(\sqrt{\frac{u - u_3}{u_3 -u_1}}\Big)
\\&\quad
+\frac{2}{(1-u_1)\sqrt{1-u_3}}\arctanh\Big(\sqrt{\frac{u - u_3}{1 -u_3}}\Big) 
-\frac{\sqrt{u - u_3}}{u_1u_3 u}\Big) 
\Bigg)\Bigg|_{u_0}^{u} 
.
\end{aligned}
\end{equation}
\end{subequations}

\textbf{Subcase (2b)}
\begin{equation}\label{singledoubleroot-conds}
\bar{L}^2 >3, 
\quad
E^2 = E_+^2 . 
\end{equation}
This yields a root structure consisting of a single root and a double root:
\begin{equation}\label{singledouble-roots}
u_1 = \tfrac{1}{3}\big(1 -2\smallsqrt{1-3/\bar{L}^2}\big),
\quad
u_2 = u_3 = \tfrac{1}{3}\big(1 +\smallsqrt{1-3/\bar{L}^2}\big) 
= \tfrac{1}{2}(1-u_1) . 
\end{equation}
They have the ranges 
\begin{equation}\label{singledouble-roots-range}
\tfrac{1}{3}>u_1> -\tfrac{1}{3},
\quad
\tfrac{2}{3} > u_2=u_3 > \tfrac{1}{3} .
\end{equation}

The quadratures \eqref{LRLangle-u}--\eqref{LRLtau-u} are given by elementary functions: 
\begin{subequations}\label{singledoubleroot-integrals}
\begin{equation}\label{singledoubleroot-Phi-integral}
I^\Phi(u;u_0) = \sgn(u_2 -u)\Bigg( 
\frac{2}{\sqrt{u_2 -u_1}} \arctanh\Big(\sqrt{\frac{u -u_1}{u_3 -u_1}}\Big)
\Bigg)\Bigg|_{u_0}^{u} 
;
\end{equation}
\begin{equation}\label{singledoubleroot-Tau-integral}
I^\T(u;u_0) =\begin{cases}
\begin{aligned}
\sgn(u_2 -u)\Bigg( & 
\frac{1}{u_2\sqrt{u_1}}\Big(\frac{1}{u_1} +\frac{2}{u_2}\Big)\arctan\Big(\sqrt{\frac{u -u_1}{u_1}}\Big)
\\&
+\frac{2}{u_2{}^2\sqrt{u_3-u_1}} \arctanh\Big(\sqrt{\frac{u -u_1}{u_2 -u_1}}\Big)
\\&
+\frac{1}{u_1u_2}\frac{\sqrt{u -u_1}}{u} 
\Bigg)\Bigg|_{u_0}^{u} ,
\quad
u_1>0
\end{aligned}
\\
\begin{aligned}
\sgn(u_2 -u)\Bigg( & 
\frac{1}{u_2\sqrt{|u_1|}}\Big(\frac{1}{|u_1|} -\frac{2}{u_2}\Big) \arctanh\Big(\sqrt{\frac{u -u_1}{|u_1|}}\Big)
\\&
+\frac{2}{u_2{}^2\sqrt{u_3-u_1}} \arctanh\Big(\sqrt{\frac{u -u_1}{u_2 -u_1}}\Big)
\\&
+\frac{1}{u_1u_2}\frac{\sqrt{u -u_1}}{u} 
\Bigg)\Bigg|_{u_0}^{u} ,
\quad
u_1<0
\end{aligned}
\\
\begin{aligned}
& \sgn(u_2-u)\bigg(
\frac{2}{\sqrt{u_2}^5}\arctanh\Big(\sqrt{\frac{u}{u_2}}\Big) 
-\frac{2}{u_2\sqrt{u}}\Big( \frac{1}{u_2} +\frac{1}{3u} \Big) \bigg) ,  
\quad
u_1=0 
\end{aligned}
\end{cases} 
;
\end{equation}
\begin{equation}\label{singledoubleroot-T-integral}
I^T(u;u_0) = \begin{cases}
\begin{aligned}
\sgn(u_2 -u)\Bigg( & 
\frac{1}{u_2\sqrt{u_1}}\Big( 2 + \frac{2}{u_2}+\frac{1}{u_1}\Big)  \arctan\Big(\sqrt{\frac{u -u_1}{u_1}}\Big)
\\&
-\frac{2}{u_2{}^2(u_2-1)\sqrt{u_2-u_1}} \arctanh\Big(\sqrt{\frac{u -u_1}{u_2-u_1}}\Big)
\\&
+\frac{2}{(u_2-1)\sqrt{1-u_1}} \arctanh\Big(\sqrt{\frac{u -u_1}{1-u_1}}\Big)
\\&
+\frac{1}{u_1u_2}\frac{\sqrt{u -u_1}}{u} 
\Bigg)\Bigg|_{u_0}^{u} , 
\quad
u_1>0
\end{aligned}
\\
\begin{aligned}
\sgn(u_2-u)\Bigg( & 
\frac{1}{u_2\sqrt{|u_1|}}
\Big({-}2 + \frac{1}{|u_1|} -\frac{2}{u_2}\Big)  \arctanh\Big(\sqrt{\frac{u -u_1}{|u_1|}}\Big)
\\&
-\frac{2}{u_2{}^2(u_2-1)\sqrt{u_2-u_1}} \arctanh\Big(\sqrt{\frac{u -u_1}{u_2-u_1}}\Big)
\\&
+\frac{2}{(u_2-1)\sqrt{1-u_1}} \arctanh\Big(\sqrt{\frac{u -u_1}{1-u_1}}\Big)
\\&
+\frac{1}{u_1u_2}\frac{\sqrt{u -u_1}}{u} 
\Bigg)\Bigg|_{u_0}^{u} , 
\quad
u_1<0
\end{aligned}
\\
\begin{aligned}
\sgn(u_2-u)\Bigg( & 
\frac{2}{u_2-1}\bigg( 
\arctanh\Big(\sqrt{u}\Big) 
-\frac{1}{\sqrt{u_2}^5} \arctanh\Big(\sqrt{\frac{u}{u_2}}\Big)
\bigg) 
\\&\quad
-\frac{2}{u_2\sqrt{u}}\Big(2+\frac{1}{u_2} + \frac{1}{3u}\Big)
\Bigg)\Bigg|_{u_0}^{u} , 
\quad
u_1=0
\end{aligned}
\end{cases} 
.
\end{equation}
\end{subequations}

\textbf{Case (3)}
The discriminant condition $D>0$ gives
\begin{equation}\label{distinctroot-conds}
\bar{L}^2>3,
\quad
E_-^2 < E^2 < E_+^2 .
\end{equation}
This yields a root structure consisting of three distinct roots:
\begin{equation}\label{distinct-roots}
u_{3-n}=
\tfrac{1}{3}
+\tfrac{2}{3}\smallsqrt{1-3/\bar{L}^2}\cos\bigg( 
\tfrac{1}{3}\arccos\bigg(\frac{1+9(1-\tfrac{3}{2}E^2)/\bar{L}^2}{\smallsqrt{(1-3/\bar{L}^2)^3}}\bigg)
-\tfrac{2}{3}\pi n\bigg), 
\quad 
n=0,1,2 . 
\end{equation}
They have the ranges 
\begin{equation}\label{distinct-roots-range}
u_1<\tfrac{2}{3} - u_+ < u_2 < u_+ < u_3 ,
\quad
\tfrac{1}{3} < u_+< \tfrac{2}{3} . 
\end{equation}

There are two different cases for the quadratures \eqref{LRLangle-u}--\eqref{LRLtau-u}, 
which are distinguished by the range of $u$. 
Both cases involve the Jacobian elliptic functions 
\begin{equation}\label{elliptic-FEPi}
\begin{gathered}
\F(\psi,k) = \int_{0}^{\psi}\frac{1}{\sqrt{1-k^2\sin^2\vartheta}}\;d\vartheta ,
\quad
\E(\psi,k) = \int_{0}^{\psi}\sqrt{1-k^2\sin^2\vartheta}\; d\vartheta, 
\\
\Pi(m;\psi,k) = \int_{0}^{\psi}\frac{1}{(1-m\sin^2\vartheta)\sqrt{1-k^2\sin^2\vartheta}}\;d\vartheta ,
\end{gathered}
\end{equation}
where $0<k<1$. 
(See e.g. \Ref{AbrSte} for details about these functions.)

When $u\geq u_3$, the quadratures are given by 
\begin{subequations}\label{distinctroot-integrals-range1}
\begin{equation}\label{distinctroot-Phi-integral-range1}
I^\Phi(u;u_0) = 
\frac{2}{\sqrt{u_3 - u_1}} \F(\psi(u),k) \Big|_{u_0}^{u} 
; 
\end{equation}
\begin{equation}\label{distinctroot-Tau-integral-range1}
I^\T(u;u_0) = \begin{cases}
\begin{aligned}
\Bigg( & 
\frac{u_2 +u_3}{u_2{}^2 u_3 \sqrt{u_3 -u_1}} \F(\psi(u),k) 
+\frac{\sqrt{u_3 -u_1}}{u_1 u_2 u_3} \E(\psi(u),k) 
\\&
{-}\frac{(u_3-u_2)(u_1u_2 + u_1u_3 +u_2  u_3)}{u_1 u_2{}^2 u_3{}^2 \sqrt{u_3 - u_1}} \Pi(m;\psi(u),k) 
\\&
-\frac{\sqrt{u-u_1}\sqrt{u-u_3}}{u_1 u_3 u \sqrt{u-u_2}}
\Bigg)\Bigg|_{u_0}^{u} ,
\quad
u_1\neq0
\end{aligned}
\\
\begin{aligned}
\Bigg( & 
\frac{2(u_2+2u_3)}{3u_2{}^2 \sqrt{u_3}^3} \F(\psi(u),k) 
{-}\frac{4}{3u_2{}^2 \sqrt{u_3}^3} \E(\psi(u),k) 
\\&\qquad
+\frac{2(u_3-u_2)}{3u_3\sqrt{u}}\Big( \frac{2u_2+u_3}{u_2 u_3} + \frac{1}{u} \Big) \sqrt{\dfrac{u-u_3}{u-u_2}} 
\Bigg)\Bigg|_{u_0}^{u} , 
\quad
u_1=0 
\end{aligned}
\end{cases}
;
\end{equation} 
\begin{equation}\label{distinctroot-T-integral-range1}
\begin{aligned}
I^T(u;u_0) = 
\Bigg( & 
\frac{2}{u_2(1 -u_2) \sqrt{u_3 -u_1}} \F(\psi(u),k) 
{-}\frac{2(u_3-u_2)}{u_2 u_3 \sqrt{u_3 - u_1}} \Pi(m;\psi(u),k) 
\\&
+\frac{2(u_3-u_2)}{(1 -u_3)(1- u_2)\sqrt{u_3 - u_1}} \Pi(n;\psi(u),k) 
\Bigg)\Bigg|_{u_0}^{u} 
;
\end{aligned}
\end{equation}
\end{subequations}
where 
\begin{equation}
k= \sqrt{\dfrac{u_2 - u_1}{u_3 - u_1}},
\quad
m=\dfrac{u_2}{u_3} , 
\quad
n=\dfrac{1-u_2}{1-u_3} , 
\quad
\sin\psi(u) = \sqrt{\dfrac{u-u_3}{u-u_2}} .
\end{equation} 

When $u_2 \geq u \geq \max(0,u_1)$, 
the quadratures have the form 
\begin{subequations}\label{distinctroot-integrals-range2}
\begin{equation}\label{distinctroot-Phi-integral-range2}
I^\Phi(u;u_0) = 
\frac{2}{\sqrt{u_3 - u_1}} \F(\psi(u),k) \Big|_{u_0}^{u} 
;
\end{equation} 
\begin{equation}\label{distinctroot-Tau-integral-range2}
I^\T(u;u_0) = \begin{cases}
\begin{aligned}
\Bigg( & 
{-}\frac{1}{u_1 u_2 \sqrt{u_3 -u_1}} \F(\psi(u),k) 
+\frac{\sqrt{u_3 -u_1}}{u_1u_2u_3} \E(\psi(u),k) 
\\&
+\frac{u_1u_2 + u_1u_3 +u_2  u_3}{u_1{}^2 u_2 u_3 \sqrt{u_3 - u_1}} \Pi(m;\psi(u),k) 
\\&
+\frac{\sqrt{u-u_1}\sqrt{u_3 -u}\sqrt{u_2-u}}{u_1 u_2 u_3 u}
\Bigg)\Bigg|_{u_0}^{u} ,
\quad
u_1\neq0
\end{aligned}
\\
\begin{aligned}
\Bigg( & 
\frac{2(u_2+2u_3)}{3u_2{}^2 \sqrt{u_3}^3} \F(\psi(u),k) 
{-}\frac{4}{3u_2{}^2 \sqrt{u_3}^3} \E(\psi(u),k) 
\\&
{-}\frac{2\sqrt{u_3 -u}\sqrt{u_2 -u}}{3u_2 u_3\sqrt{u}}\Big( \frac{2}{u_2 u_3} +  \frac{1}{u} \Big) 
\Bigg)\Bigg|_{u_0}^{u} , 
\quad
u_1=0 
\end{aligned}
\end{cases}
;
\end{equation} 
\begin{equation}\label{distinctroot-T-integral-range2}
I^T(u;u_0) = \begin{cases}
\begin{aligned}
\Bigg( & 
\frac{2}{u_1\sqrt{u_3 - u_1}} \Pi(m;\psi(u),k) 
\\&
+\frac{2}{(1-u_1)\sqrt{u_3 - u_1}} \Pi(n;\psi(u),k) 
\Bigg)\Bigg|_{u_0}^{u} ,
\quad
u_1\neq0
\end{aligned}
\\
\begin{aligned}
\Bigg( & 
\frac{2}{\sqrt{u_3}} \F(\psi(u),k) 
{-}\frac{2}{u_2\sqrt{u_3}} \E(\psi(u),k) 
\\&
+\frac{2}{\sqrt{u_3}} \Pi(n;\psi(u),k) 
{-}\frac{2\sqrt{u_3 -u}\sqrt{u_2 -u}}{u_2 u_3\sqrt{u}}
\Bigg)\Bigg|_{u_0}^{u} , 
\quad
u_1=0 
\end{aligned}
\end{cases}
;
\end{equation} 
\end{subequations}
where
\begin{equation}
k= \sqrt{\dfrac{u_2 - u_1}{u_3 - u_1}},
\quad
m=1 -\dfrac{u_2}{u_1},
\quad
n=\dfrac{u_2 -u_1}{1-u_1},
\quad
\sin\psi(u) = \sqrt{\dfrac{u -u_1}{u_2 -u_1}} .
\end{equation}

\textbf{Case (4)}
The discriminant condition $D<0$ splits into two cases 
\begin{subequations}
\begin{equation}\label{cc-conds1}
\bar{L}^2\leq 3,
\quad
0<E^2<\infty
\end{equation}
and 
\begin{equation}\label{cc-conds2}
\begin{gathered}
\bar{L}^2>3,
\\
E^2< E_-^2 
\text{ or } 
E^2 > E_+^2 . 
\end{gathered}
\end{equation}
\end{subequations}
In both cases, the roots have the same structure 
consisting of a real root and a pair of complex conjugate roots:
\begin{equation}
u_1 = \tfrac{1}{3}(1 + q_+ + q_-),
\quad
u_2 = \bar u_3 = \tfrac{1}{6}(2 -(q_++q_-) +i \sqrt{3}(q_+ -q_-)) ,
\end{equation}
where 
\begin{equation}
q_\pm = 
\sqrt[3]{ 
1+9(1-\tfrac{3}{2}E^2)/\bar{L}^2 \pm \sqrt{ (1+9(1-\tfrac{3}{2}E^2)/\bar{L}^2)^2 - (1-3/\bar{L}^2)^3 }
} . 
\end{equation}
They have the ranges 
\begin{equation}\label{cc-roots-range}
u_1 <1,
\quad
\Re(u_2)=\Re(u_3) > 0 .
\end{equation}

The quadratures \eqref{LRLangle-u}--\eqref{LRLtau-u} are given by 
\begin{subequations}\label{ccroot-integrals}
\begin{equation}\label{ccroot-Phi-integral}
I^\Phi(u;u_0) = 
\frac{-\sqrt{2}}{\sqrt{\alpha+\beta}}\F(\psi(u),ik)  \Big|_{u_0}^{u} 
;
\end{equation} 
\begin{equation}\label{ccroot-Tau-integral}
I^\T(u;u_0) = \begin{cases}
\begin{aligned}
\Bigg( & 
\frac{\beta^2+u_1(2 -3u_1)}{2\sqrt{u_1}^3 \sqrt{\beta^2+u_1(1 -2u_1)}^3} 
\arctan\Big( \frac{\sqrt{u-u_1}\sqrt{\beta^2+u_1(1 -2u_1)}}{\sqrt{u_1}\sqrt{\beta^2+(u-2u_1-1)(u-u_1)}} \Big)
\\&
+\frac{2} {u_1(\beta -u_1)\sqrt{\beta+2\alpha}} \F(\psi(u),ik) 
-\frac{\sqrt{\beta+2\alpha}}{2u_1(\beta^2 +u_1(1-2u_1))} \E(\psi(u),ik) 
\\&
{-}\frac{2\beta(\beta^2+u_1(2 -3u_1))}{u_1(\beta^2 -u_1{}^2)(\beta^2+u_1(1 -2u_1))\sqrt{\beta+2\alpha}} \Pi(m;,\psi(u),ik) 
\\&
+ \frac{(\beta-u_1)\sqrt{u-u_1}\sqrt{\beta^2+(u-u_1)(u+2u_1-1)}}{u_1 (\beta^2 + u_1(1 -2u_1))u(u-u_1+\beta)}
\Bigg)\Bigg|_{u_0}^{u} ,
\quad
u_1>0
\end{aligned}
\\
\begin{aligned}
\Bigg( & 
\frac{\beta^2+u_1(2 -3u_1)}{\sqrt{2|u_1|}^3 \sqrt{\beta^2+u_1(1 -2u_1)}^3} 
\arctanh\Big( \frac{\sqrt{u-u_1}\sqrt{\beta^2+u_1(1 -2u_1)}}{\sqrt{|u_1|}\sqrt{\beta^2+(u-2u_1-1)(u-u_1)}} \Big)
\\&
+\frac{2} {u_1(\beta -u_1)\sqrt{\beta+2\alpha}} \F(\psi(u),ik) 
-\frac{\sqrt{\beta+2\alpha}}{2u_1(\beta^2 +u_1(1-2u_1))} \E(\psi(u),ik) 
\\&
{-}\frac{2\beta(\beta^2+u_1(2 -3u_1))}{u_1(\beta^2 -u_1{}^2)(\beta^2+u_1(1 -2u_1))\sqrt{\beta+2\alpha}} \Pi(m;,\psi(u),ik) 
\\&
+ \frac{(\beta-u_1)\sqrt{u-u_1}\sqrt{\beta^2+(u-u_1)(u+2u_1-1)}}{u_1(\beta^2 + u_1(1 -2u_1))u(u-u_1+\beta)}
\Bigg)\Bigg|_{u_0}^{u} ,
\quad
u_1<0
\end{aligned}
\\
\begin{aligned}
\Bigg( & 
\frac{2(\beta-2)}{3\beta^3\sqrt{2\beta-1}} \F(\psi(u),ik) 
+\frac{2\sqrt{2\beta-1}}{3\beta^4} \E(\psi(u),ik) 
\\&\qquad
{-}\frac{2\sqrt{\beta^2+u(u-1)}}{3\beta^2\sqrt{u}}
\Big( \frac{1}{u} + \frac{2}{\beta(u+\beta)} \Big) 
\Bigg)\Bigg|_{u_0}^{u} , 
\quad
u_1=0 
\end{aligned}
\end{cases}
;
\end{equation} 
\begin{equation}\label{ccroot-T-integral}
I^T(u;u_0) = \begin{cases}
\begin{aligned}
\Bigg( & 
\frac{1}{\sqrt{u_1} \sqrt{\beta^2+u_1(1 -2u_1)}^3} 
\arctan\Big( \frac{\sqrt{u-u_1}\sqrt{\beta^2+u_1(1 -2u_1)}}{\sqrt{u_1}\sqrt{\beta^2+(u-2u_1-1)(u-u_1)}} \Big)
\\&
{-}\frac{1}{\sqrt{1-u_1} \sqrt{\beta^2+2u_1(1 -u_1)}^3} 
\arctanh\Big( \frac{\sqrt{u-u_1}\sqrt{\beta^2+u_1(1 -2u_1)}}{\sqrt{1-u_1}\sqrt{\beta^2+(u-2u_1-1)(u-u_1)}} \Big)
\\&
+\frac{4\beta}{\sqrt{\beta+2\alpha}}\Big( 
\frac{1}{\beta^2 -(u_1 -1)^2} \Pi(n;\psi(u),ik) 
-\frac{1}{\beta^2 -u_1{}^2} \Pi(m;\psi(u),ik) \Big)
\\&
+\frac{2} {(\beta -u_1)(\beta-u_1 +1)\sqrt{\beta+2\alpha}} \F(\psi(u),ik) 
\Bigg)\Bigg|_{u_0}^{u} ,
\quad
u_1>0
\end{aligned}
\\
\begin{aligned}
\Bigg( & 
\frac{1}{\sqrt{|u_1|} \sqrt{\beta^2+u_1(1 -2u_1)}^3} 
\arctanh\Big( \frac{\sqrt{u-u_1}\sqrt{\beta^2+u_1(1 -2u_1)}}{\sqrt{|u_1|}\sqrt{\beta^2+(u-2u_1-1)(u-u_1)}} \Big)
\\&
{-}\frac{1}{\sqrt{1-u_1} \sqrt{\beta^2+2u_1(1 -u_1)}^3} 
\arctanh\Big( \frac{\sqrt{u-u_1}\sqrt{\beta^2+u_1(1 -2u_1)}}{\sqrt{1-u_1}\sqrt{\beta^2+(u-2u_1-1)(u-u_1)}} \Big)
\\&
+\frac{4\beta}{\sqrt{\beta+2\alpha}}\Big( 
\frac{1}{\beta^2 -(u_1 -1)^2} \Pi(n;\psi(u),ik) 
-\frac{1}{\beta^2 -u_1{}^2} \Pi(m;\psi(u),ik) \Big)
\\&
+\frac{2} {(\beta -u_1)(\beta-u_1 +1)\sqrt{\beta+2\alpha}} \F(\psi(u),ik) 
\Bigg)\Bigg|_{u_0}^{u} ,
\quad
u_1<0
\end{aligned}
\\
\begin{aligned}
\Bigg( & 
{-}\frac{\sqrt{2\beta+1}}{\beta \sqrt{\beta+1}} 
\arctanh\Big( \frac{\beta\sqrt{u}}{\sqrt{\beta^2+u(u -1)}} \Big)
\\&
{-}\frac{2(2\beta+1)}{\beta(\beta+1)\sqrt{2\beta-1}} \F(\psi(u),ik) 
+\frac{\sqrt{2\beta-1}}{\beta^2} \E(\psi(u),ik) 
\\&
+\frac{4\beta}{(\beta^2-1)\sqrt{2\beta-1}} \Pi(n;\psi(u),ik) 
\\&
+\frac{\sqrt{\beta^2+u(u-1)}}{\beta^2\sqrt{u}}
\Big( \frac{u-\beta}{u+\beta}  -\frac{\sqrt{2\beta+1}}{\sqrt{\beta+1}} \Big) 
\Bigg)\Bigg|_{u_0}^{u} , 
\quad
u_1=0 
\end{aligned}
\end{cases} 
;
\end{equation} 
where
\begin{equation}
k = \smallsqrt{\dfrac{\beta -\alpha}{\beta+\alpha}},
\quad
m=\frac{(\beta-u_1)^2}{(\beta+u_1)^2},
\quad
n=\frac{(\beta-u_1+1)^2}{(\beta+u_1-1)^2},
\quad
\sin(\psi(u)) = \dfrac{\beta +u_1-u}{\beta -u_1+u},
\end{equation}
with 
\begin{equation}
\alpha = \tfrac{1}{2}(3u_1 -1) = \tfrac{1}{2}(q_+ +q_-) ,
\quad
\beta^2 =|u_2|^2 +u_1(2 u_1-1) 
= \alpha^2 + \tfrac{1}{12}(q_+ -q_-)^2 .
\end{equation}
\end{subequations}

\section{Newtonian Limit}\label{sec:newtonian}

The integrals of motion \eqref{LRLangle}, \eqref{LRLtime} and \eqref{LRLtau}
will now be evaluated in the Newtonian limit to show that 
they respectively correspond to the angle and the time
at which a particle reaches either an apsis point $r=r_0$ or a centripetal point $r=r_0$
on a non-circular orbit under Newtonian gravity. 
Their explicit relationship to the Newtonian LRL vector and eccentricity vector
also will be described.

\subsection{Newtonian LRL vector}

To begin, recall the expression for the conserved LRL vector of a particle 
in a non-circular orbit
in Newtonian gravity:
\begin{equation}
\vec A_\newt = m\vec v \times \vec L - m^2M \hat r
\end{equation}
where $\hat r$ is the unit radial vector, 
$\vec v = \dot{\vec r}$ is the velocity vector, 
$\vec L = m \vec r \times \vec v$ is the angular momentum vector,
and $m$ is the particle mass. 
As is well known, 
the orbit of the particle lies in a plane orthogonal to $\vec L$. 
Let $(r,\phi)$ be polar coordinates in this plane, 
with associated unit vectors 
$\hat r = (\cos\phi,\sin\phi)$ and $\hat \phi = (-\sin\phi,\cos\phi)$. 
Then the LRL vector at any point $(r(t),\phi(t))$ on the orbit is given by 
\begin{equation}\label{newtonianLRL}
\frac{1}{m^2}\vec A_\newt = \Big(\frac{L^2}{r(t)} -M\Big)\hat r + Lv(t)\hat\phi
= \frac{|\vec A_\newt|}{m^2}(\cos\varphi,\sin\varphi) 
\end{equation}
where
\begin{equation}
\tan\varphi = \frac{\tan\phi(t) - \alpha(t)}{1+\alpha(t)\tan\phi(t)}, 
\quad
\alpha(t) = \frac{Lr(t)v(t)}{L^2-Mr(t)} . 
\end{equation}
It is straightforward to show that $\dot{\vec A}=0$ holds
as a consequence of the Newtonian equations of motion of the particle,
and hence $\varphi$ (modulo $2\pi$) is a constant of the motion. 
A variant of the LRL vector is 
\begin{equation}\label{newtonianHamilLRL}
\vec B_\newt = \vec L \times \vec A_\newt 
= \frac{L|\vec A_\newt|}{m^2}(-\sin\varphi,\cos\varphi) 
\end{equation}
called Hamilton's eccentricity vector \cite{Cor}. 
Its associated angle is given by $\varphi+\tfrac{\pi}{2}$. 

Non-circular orbits with $L\neq 0$ are elliptic, parabolic, and hyperbolic,
such that the origin $r=0$ coincides with a focal point. 
These orbits are classified by their energy, 
which is $-M^2/(2L^2) < E_\newt < 0$ for elliptic orbits, 
$E_\newt = 0$ for parabolic orbits,
and $E_\newt > 0$ for hyperbolic orbits. 
The LRL vector $\vec A$ has the property that \cite{GolPooSaf,Cor} 
it lies on the radial line from the origin to the periapsis point on the particle's orbit. 
In particular, the angle of this apis line is given by $\varphi$. 
Similarly, the variant vector $\vec B$ lies on the perpendicular bisector line of this radial apsis line through the origin.

\subsection{Evaluation of the angular and temporal integrals of motion}\label{subsec:evaluatenewtonian}

The Newtonian limit of the integrals of motion \eqref{LRLangle}, \eqref{LRLtime} and \eqref{LRLtau}
is obtained by taking $r/(2M) \gg 1$ and $E-1\simeq E_{\newt} \ll 1$
along with $\tau \simeq t$. 
In this limit, 
the proper time equation \eqref{VE-eqn} 
reduces to the Newtonian energy equation 
\begin{equation}\label{newtonianEV}
E_\newt \simeq \tfrac{1}{2}v^2 + V_\newt
\end{equation} 
where
\begin{equation}\label{newtonianV}
V_\newt = - \frac{M}{r} + \frac{L^2}{2r^2} 
\end{equation} 
is the Newtonian potential per unit mass. 
Similarly, the Newtonian limit of the turning point equation \eqref{tp-eqn} 
and the centripetal point equation \eqref{ip-eqn} 
are respectively given by
\begin{gather}
E_\newt r_*^3 + M r_*^2 -\tfrac{1}{2} L^2 r_* \simeq 0, 
\quad
r_*>0 , 
\label{newtonian-tp-eqn}
\\
M{r^*}^2 - L^2 r^* \simeq 0 , 
\quad
r^*>0 .
\label{newtonian-ip-eqn}
\end{gather}
These roots provide a distinguished dynamical point $r=r_0$
which corresponds to defining a physically meaningful zero-point value 
for the integrals of motion \eqref{LRLangle}, \eqref{LRLtime} and \eqref{LRLtau}
in the Newtonian limit. 

It is straightforward to see that the angular integral of motion \eqref{LRLangle} 
reduces to the quantity 
\begin{equation}\label{newtonianTheta}
\Phi \simeq \phi - L\int_{r_0}^{r}\frac{\sgn(v)\, dr}{r^2\sqrt{2 (E_\newt - V_\newt)}} 
\mod 2\pi , 
\end{equation}
and that the two temporal integrals of motion \eqref{LRLtime} and \eqref{LRLtau} 
both reduce to the quantity 
\begin{gather}\label{newtonianT}
\T \simeq T \simeq t - \int_{r_0}^{r} \frac{\sgn(v)\, dr}{\sqrt{2 (E_\newt - V_\newt)}} . 
\end{gather}

To proceed with evaluating these quantities, 
introduce the reciprocal radial variable
\begin{equation}\label{newtonian-u}
u = M/r
\end{equation}
which is the same as the variable used in \cite{Cha}
for classifying orbits in Schwarzschild spacetime. 
Note that $u$ lacks a factor of $2$ compared with the notation in section~\ref{sec:evaluate}, 
which will help to simplify the subsequent equations here. 
Hereafter,  an overbar will denote a quantity or a variable divided by $M$. 
Then the Newtonian integrals of motion \eqref{newtonianTheta} and \eqref{newtonianT} 
are given by 
\begin{align}
\Phi_\newt & = \phi + \bar{L}\int_{u_0}^{u}\frac{\sgn(v)\, du}{\sqrt{Q(u)}} 
\mod 2\pi , 
\label{newtonianTheta-u}
\\
T_\newt & = t + M\int_{u_0}^{u} \frac{\sgn(v)\, du}{u^2\sqrt{Q(u)}} , 
\label{newtonianT-u}
\end{align}
where 
\begin{equation}\label{newtonianQ}
Q(u) = v^2 = 2(E_\newt-V_\newt) = 2E_\newt + 2u - \bar{L}^2 u^2 . 
\end{equation} 
Moreover, the Newtonian turning point equation \eqref{newtonian-tp-eqn}
corresponds to $Q(u)=0$ for $u=u_*>0$, 
namely
\begin{equation}\label{newtonian-tp}
0=Q(u_*)= 2E_\newt + 2u_* -\bar{L}^2 u_*^2 , 
\end{equation}
while the Newtonian centripetal point equation \eqref{newtonian-ip-eqn}
corresponds to $Q'(u)=0$ for $u=u^*>0$,
namely
 \begin{equation}\label{newtonian-ip}
0=Q'(u^*)= 2(1 -\bar{L}^2 u^*) . 
\end{equation}

The quadratures appearing in the integrals of motion \eqref{newtonianTheta-u} and \eqref{newtonianT-u}
are straightforward to evaluate by using the factorization 
\begin{equation}\label{newtonianQ-u}
Q(u) = \bar{L}^2 (u_+-u)(u-u_-)
\end{equation}
in terms of the roots 
\begin{equation}\label{newtonian-Qroots-u}
u_\pm  = \frac{1}{\bar{L}^2} \Big( 1 \pm  \sqrt{ 1 + 2E_\newt\bar{L}^2 } \Big) . 
\end{equation}
The nature of these roots also determines the type of the orbit 
with a given value of energy $E_\newt$ and angular momentum $L$. 
In particular, for elliptic, parabolic, and hyperbolic orbits, 
both roots are real and distinct. 
As a consequence, 
there is no need to break up the evaluation of the quadratures into cases 
given by the separate types of orbit. 

When the roots \eqref{newtonian-Qroots-u} are real and distinct, 
the quadratures are explicitly given by 
\begin{equation}\label{newtonianTheta-integral}
\begin{aligned}
I^\Phi_\newt(u;u_0) \equiv 
|\bar{L}| \int_{u_0}^{u}\frac{du}{\sqrt{Q(u)}} 
& =\int_{u_0}^{u}\frac{du}{\sqrt{(u_+-u)(u-u_-)}}
\\
& = \bigg({-2}\arctan \sqrt{\frac{u_+ - u}{u - u_-}}\; \bigg)\bigg|_{u_0}^u , 
\end{aligned}
\end{equation}
and 
\begin{equation}\label{newtonianT-integral}
\begin{aligned}
I^T_\newt(u;u_0) \equiv 
|\bar{L}| \int_{u_0}^{u}\frac{du}{u^2\sqrt{Q(u)}} 
& = \int_{u_0}^{u}\frac{du}{u^2\sqrt{(u_+-u)(u-u_-)}}
\\
& = \bigg( 
\frac{\sqrt{(u - u_-) (u_+ - u)}}{u_+u_- u} 
-\frac{u_+ + u_-}{\sqrt{u_- u_+}^3} \arctan \sqrt{\frac{u_- (u_+ - u)}{u_+ (u - u_-)}}\;
\bigg)\bigg|_{u_0}^u . 
\end{aligned}
\end{equation}
Here the $\tan$ function has the domain $(-\tfrac{\pi}{2},\tfrac{\pi}{2})$. 
To complete the evaluation of the integrals of motion, 
it is necessary to specify $u_0$ universally for all non-circular orbits. 
by choosing a turning point or a centripetal point. 
In particular, the resulting integrals of motion turn out to be 
related to the Newtonian LRL vector \eqref{newtonianLRL} 
when a turning point is chosen, 
and alternatively to the variant vector \eqref{newtonianHamilLRL} 
when a centripetal point is chosen. 
More details can be found in \Ref{AncMeaPas}. 

A side remark is that these quadratures \eqref{newtonianTheta-integral} and \eqref{newtonianT-integral} 
are only applicable when $E_\newt > -M^2/(2L^2)$. 
Specifically, for  $E_\newt = -M^2/(2L^2)$, 
the roots \eqref{newtonian-Qroots-u} are no longer distinct, $u_+=u_-=1/\bar{L}^2$, 
and hence $v^2 =Q(u)=-\bar{L}^2(u-u_\pm)^2$ must vanish identically 
(due to the opposite signs on the two sides). 
This implies $u=u_\pm=1/\bar{L}^2$, corresponding to $r= L^2/M$ 
which is the radius of a circular orbit 
with angular momentum $L$ and energy $E_\newt = -M^2/(2L^2)$.

\subsection{LRL quantities for elliptic orbits}

For an elliptic orbit, 
both roots \eqref{newtonian-Qroots-u} of $Q(u)$ are positive,
and so the range of $u$ is $u_- \leq u \leq u_+$. 
Hence, each root is a turning point $u_*=u_+$ and $u_*=u_-$. 
There is a single centripetal point $u^*= 1/\bar{L}^2$. 

Every elliptic orbit with a given energy $-1/(2\bar{L}^2) < E_\newt < 0$ possesses 
a single apoapsis point, which has $r=r_*^+=M/u_-$, 
and a single periapsis point, which has $r=r_*^-=M/u_+$. 
These apsis points comprise the two turning points in the Newtonian effective potential. 
The two points at an angular separation $\Delta\phi =\frac{1}{2}\pi$ 
between successive apsis points on the orbit have $r=r^*=L^2/M$
which coincides with the radius of a circular orbit having the same angular momentum
and which comprises the centripetal point in the Newtonian effective potential. 

If $u_0=u_+$ is chosen in the angular and temporal integrals of motion \eqref{newtonianTheta-u} and \eqref{newtonianT-u}, 
then the quadratures \eqref{newtonianTheta-integral} and \eqref{newtonianT-integral}
become
\begin{align}
&\begin{aligned}
I^\Phi_\newt(u;u_+) & 
= -2 \arctan \sqrt{\frac{u_+ - u}{u - u_-}}  
= -2 \arctan\Big( \frac{|\bar{L}|(u_+ - u)}{\sqrt{Q(u)}} \Big),
\end{aligned}
\label{newtonianTheta-integral-elliptic}
\\
&\begin{aligned}
I^T_\newt(u;u_+) & 
= \frac{\sqrt{(u - u_-) (u_+ - u)}}{u_+u_- u} 
-\frac{u_- + u_+}{\sqrt{u_- u_+}^3} \arctan \sqrt{\frac{u_- (u_+ - u)}{u_+ (u - u_-)}} 
\\&
= \frac{\sqrt{Q(u)}}{|\bar{L}|u_+u_- u} 
-\frac{u_- + u_+}{\sqrt{u_+ u_-}^3} \arctan\Big( \frac{|\bar{L}|\sqrt{u_-}(u_+ - u)}{\sqrt{u_+ Q(u)}} \Big) . 
\end{aligned}
\label{newtonianT-integral-elliptic}
\end{align}
Both quadratures will vanish for $u=u_+$,
corresponding to the periapsis point, 
where
\begin{equation}\label{elliptic-periapsis}
r=r_*^-=M/u_+=\frac{M}{2|E_\newt|}\Big(1 - \sqrt{ 1 - 2|E_\newt|\bar{L}^2 } \Big) . 
\end{equation}
Hence, using expressions \eqref{newtonianQ-u} and \eqref{newtonian-Qroots-u}, 
this yields
\begin{align}
&\begin{aligned}
\Phi_\newt & 
= \phi +\sgn(vL) I^\Phi_\newt(M/r;M/r_*^-) 
= \phi - 2\arctan\Big(\frac{L(r-r_*^-)}{r_*^- rv}\Big) 
\mod 2\pi , 
\end{aligned}
\label{newtonianTheta-elliptic}
\\
&\begin{aligned}
T_\newt & 
= t + \frac{M^2\sgn(v) }{|L|} I^T_\newt(M/r;M/r_*^-) 
\\&
= t +\frac{Mvr}{2|E_\newt|} - \frac{M}{\sqrt{2|E_\newt|^3}}\arctan\Big( \frac{\sqrt{2|E_\newt|}(r - r_*^-)}{r v} \Big) . 
\end{aligned}
\label{newtonianT-elliptic}
\end{align}
At the periapsis point $(r(t_*^-),\phi(t_*^-))=(r_*^-,\phi_*^-)$ on the orbit, 
$\Phi_\newt = \phi_*^-$ and $T_\newt = t_*^-$ hold, thereby showing 
that $\Phi_\newt$ is the angle $\phi_*^-$ of the periapsis line associated with the LRL vector,
and that $T_\newt$ is time $t_*^-$ at which the periapsis point is reached after $t=0$,
namely $(r(T_\newt),\phi(T_\newt)) = (r_*^-,\Phi_\newt)$. 
This conclusion can also be directly verified 
from the explicit expression for the LRL vector \eqref{newtonianLRL} 
through use of the identity $2\arctan(x)=\arctan( 2x/(1-x^2) )$. 
In particular, $\Phi_\newt=\varphi$. 

The LRL integrals of motion \eqref{newtonianTheta-elliptic} and \eqref{newtonianT-elliptic}
have the following global properties. 
On any elliptic orbit, 
$\Phi_\newt$ is locally constant and changes by $\Delta\phi = 2\pi = 0 \mod 2\pi$ 
when the apoapsis point is reached, 
while $T_\newt$ is locally constant and jumps by $\Delta t = \pi M/\sqrt{2|E_\newt|^3}$ 
which is the period of the orbit. 
Thus, $\Phi_\newt$ is single-valued, and $T_\newt$ is multi-valued. 
Hence, $\Phi_\newt$ is a global constant of motion. 

Finally, instead of the turning point $u_0=u_+$, suppose that 
the centripetal point $u_0=u^*=1/\bar{L}^2$ given by the root of equation \eqref{newtonian-ip} is chosen. 
Since 
$I^\Phi_\newt(u_+;u^*) = 2 \arctan \sqrt{\dfrac{u_+ - u^*}{u^* - u_-}}  = 2\arctan(1) = \tfrac{\pi}{2}$, 
the angular integral of motion \eqref{newtonianTheta-u} becomes 
the LRL angle plus $\tfrac{\pi}{2}$, 
which is the angle of the radial line from the origin to one of the two centripetal points on the orbit
(depending on the signs of $v$ and $L$). 
Likewise, the temporal integral of motion \eqref{newtonianT-u} becomes 
the time at which this centripetal point is reached after $t=0$. 
The global properties of these quantities $\Phi_\newt$ and $T_\newt$ 
are the same as in the LRL case. 
The conserved vector associated with them is the eccentricity vector \eqref{newtonianHamilLRL}.

\subsection{LRL quantities for hyperbolic and parabolic orbits}

For a hyperbolic orbit, 
the roots \eqref{newtonian-Qroots-u} of $Q(u)$ satisfy $u_+>0 >u_-$, 
and so the range of $u$ is $0< u \leq u_+$. 
Hence, there is a single turning point $u_*=u_+$. 
There is also a single centripetal point $u^*= 1/\bar{L}^2$. 
The turning point corresponds to the periapsis point on the orbit, 
which has $r=r_*=M/u_+$. 

If again $u_0=u_+$ is chosen in the angular and temporal integrals of motion \eqref{newtonianTheta-u} and \eqref{newtonianT-u}, 
then the quadrature \eqref{newtonianTheta-integral} evaluates to 
the same expression \eqref{newtonianTheta-integral-elliptic} as in the elliptic case, 
whereas the quadrature \eqref{newtonianT-integral} now evaluates to 
\begin{equation}\label{newtonianT-integral-hyperbolic}
\begin{aligned}
I^T_\newt(u;u_+) & 
= \frac{\sqrt{(u - u_-) (u_+ - u)}}{u_+u_- u} 
+\frac{u_- + u_+}{\sqrt{u_+|u_-|}^3} \arctanh \sqrt{\frac{|u_-| (u_+ - u)}{u_+ (u - u_-)}} 
\\&
=- \frac{\sqrt{Q(u)}}{|\bar{L}|u_+|u_-| u} 
+\frac{u_- + u_+}{\sqrt{u_+|u_-|}^3} \arctanh\Big( \frac{|\bar{L}|\sqrt{|u_-|}(u_+ - u)}{\sqrt{u_+Q(u)}} \Big) 
\end{aligned}
\end{equation}
through the identity $\arctan(ix)=i\arctanh(x)$. 
Both quadratures \eqref{newtonianT-integral-hyperbolic} and \eqref{newtonianT-integral-elliptic} 
will vanish for $u=u_+$,
corresponding to the periapsis point, 
where
\begin{equation}\label{hyperbolic-periapsis}
r=r_*=M/u_+=\frac{M}{2E_\newt}\Big(\sqrt{ 1 + 2E_\newt\bar{L}^2 } -1\Big) . 
\end{equation}
Hence, using expressions \eqref{newtonianQ-u} and \eqref{newtonian-Qroots-u}, 
this yields
\begin{align}
&\begin{aligned}
\Phi_\newt & 
= \phi +\sgn(vL) I^\Phi_\newt(M/r;M/r_*) 
= \phi - 2\arctan\Big(\frac{L(r-r_*)}{r_* rv}\Big) 
\mod 2\pi , 
\end{aligned}
\label{newtonianTheta-hyperbolic}
\\
&\begin{aligned}
T_\newt & 
= t + \frac{M^2\sgn(v) }{|L|} I^T_\newt(M/r;M/r_*) 
\\&
= t +\frac{Mvr}{2E_\newt} + \frac{M}{\sqrt{2E_\newt^3}}\arctan\Big( \frac{\sqrt{E_\newt}(r - r_*)}{r v} \Big) . 
\end{aligned}
\label{newtonianT-hyperbolic}
\end{align}
Similarly to the elliptic case, 
$\Phi_\newt$ is the angle of the periapsis line associated with the LRL vector \eqref{newtonianLRL}, 
and $T_\newt$ is time at which the periapsis point is reached after $t=0$,
namely $(r(T_\newt),\phi(T_\newt)) = (r_*,\Phi_\newt)$. 
In particular, $\Phi_\newt=\varphi$. 

For a parabolic orbit, since $E_\newt=0$, 
the roots \eqref{newtonian-Qroots-u} of $Q(u)$ satisfy $u_+=2/\bar{L}^2 >u_-=0$. 
Hence, just as in the hyperbolic case, 
there is a single centripetal point $u^*= 1/\bar{L}^2$, 
and also a single turning point $u_*=u_+$
which corresponds to the periapsis point on the orbit, 
\begin{equation}\label{parabolic-periapsis}
r=r_*=2L^2/M . 
\end{equation}

Again, take $u_0=u_+$ in the angular and temporal integrals of motion \eqref{newtonianTheta-u} and \eqref{newtonianT-u}. 
Because $u_-=0$, 
the evaluation of the quadratures \eqref{newtonianTheta-integral} and \eqref{newtonianT-integral}
now simplifies and can be obtained 
either from the leading term in an asymptotic expansion as $u_-\to0$, 
or by direct integration:
\begin{equation}\label{newtonianTheta-integral-parabolic}
I^\Phi_\newt(u;u_+) =\int_{u_0}^{u}\frac{du}{\sqrt{u(u_+-u)}}
= {-2}\arctan \sqrt{\frac{u_+ - u}{u}} , 
\end{equation}
and 
\begin{equation}\label{newtonianT-integral-parabolic}
I^T_\newt(u;u_+) = \int_{u_+}^{u}\frac{du}{u^2\sqrt{u(u_+-u)}}
= -\frac{2(u_++2u)\sqrt{u_+-u}}{3u_+^2\sqrt{u}^3} . 
\end{equation}
Both of these quadratures will vanish for $u=u_+$,
corresponding to the periapsis point \eqref{parabolic-periapsis}. 
Hence, 
this yields 
\begin{align}
\Phi_\newt & 
= \phi +\sgn(vL) I^\Phi_\newt(M/r;M/r_*) 
= \phi - 2\arctan\Big(\frac{rv}{L}\Big) 
\mod 2\pi , 
\label{newtonianTheta-parabolic}
\\
T_\newt &
= t + \frac{M^2\sgn(v) }{|L|} I^T_\newt(M/r;M/r_*) 
= t -\frac{vr(Mr+L^2)}{6L^2} . 
\label{newtonianT-parabolic}
\end{align}
As before, $\Phi_\newt$ is the angle of the periapsis line associated with the LRL vector \eqref{newtonianLRL}, 
and $T_\newt$ is time at which the periapsis point is reached after $t=0$,
namely $(r(T_\newt),\phi(T_\newt)) = (r_*,\Phi_\newt)$
with $\Phi_\newt=\varphi$. 

In both the hyperbolic and parabolic cases, 
$\Phi_\newt$ and $T_\newt$ are single-valued and globally constant. 
Hence, $\Phi_\newt$ is a global constant of motion for all hyperbolic and parabolic orbits. 

Finally, suppose that instead the centripetal point $u_0=u^*=1/\bar{L}^2$ is chosen. 
In both the hyperbolic and parabolic cases, 
$I^\Phi_\newt(u_+;u^*) = 2 \arctan \sqrt{\dfrac{u_+ - u^*}{u^* - u_-}}  = 2\arctan(1) = \tfrac{\pi}{2}$, 
and so the angular integral of motion becomes 
the LRL angle plus $\tfrac{\pi}{2}$, 
which is the angle of the radial line from the origin to one of the two centripetal points on the orbit
(depending on the signs of $v$ and $L$). 
The temporal integral of motion likewise becomes 
the time at which this centripetal point is reached after $t=0$. 
Moreover, the two centripetal points coincide with the radius of a circular orbit 
having the same angular momentum. 
Just as in the elliptic case, 
the conserved vector associated with the angular and temporal quantities 
for hyperbolic and parabolic orbits is the eccentricity vector \eqref{newtonianHamilLRL}.

\section*{Acknowledgements}

S.C.A.\ is supported by an NSERC Discovery grant. 
J.F.\ was supported by the Physics Department at Brock University during the period 
when the technical part of this work was completed. 
Barak Shoshany and the reviewer are thanked for useful remarks 
incorporated into the final version. 
Georgios Papadopoulos is thanked for stimulating discussions 
at an early stage of the work.


\begin{thebibliography}{00}


\bibitem{MisThoWhe}
C.W. Misner. K.S. Thorne, J.A. Wheeler,
{\it Gravitation}
(W.H. Freeman and Co.) 1973. 

\bibitem{Wal}
R.M. Wald,
{\it General Relativity} (The University of Chicago Press) 1984. 



\bibitem{Hag}
Y. Hagihara, 
Theory of the relativistic trajectories in a gravitational field of Schwarzschild, 
Jpn. J. Astron. Geophys. 8 (1931), 67--175.

\bibitem{Dar}
C.G. Darwin, 
The gravity field of a particle, 
Proc. Roy. Soc. A 249 (1959) 180--194; 
The gravity field of a particle II, Proc. Roy. Soc. A 263 (1961), 39--50. 

\bibitem{Cha}
S. Chandrasekhar,
{\it The Mathematical Theory of Black Holes}
(Oxford University Press) 1983. 



\bibitem{GolPooSaf}
H. Goldstein, C. Poole, J. Safko, 
{\it Classical Mechanics} (3rd ed.), (Addison Wesley) 2000.

\bibitem{Cor}
B. Cordani, 
{\it The Kepler Problem} (Birkhaeuser) 2003. 



\bibitem{BacRueSou}
H. Bacry, J. Ruegg, J.-M. Souriau,
Dynamical groups and spherical potentials in classical mechanics, 
Commun. Math. Phys. 3 (1966), 323--333. 

\bibitem{Fra}
D.M. Fradkin,
Existence of the dynamic symmetries $O_4$ and $SU_3$ for all classical central potential problems, 
Prog. Theor. Phys. 37 (1967), 798--812.

\bibitem{AncMeaPas}
S.C. Anco, T. Meadows, V. Pascuzzi,  
Some new aspects of first integrals and symmetries for central force dynamics, 
J. Math. Phys. 57 (2016) 062901. 



\bibitem{Kos}
U. Kosti\'c,
Analytical time-like geodesics in Schwarzschild space-time, 
Gen. Relativ. Gravit. 44 (2012), 1057-–1072. 

\bibitem{GomHorKos}
A. Gomboc, M. Horvat, U. Kosti\'c,
Relativistic GNSS. PECS project (ESA), Final Report (2014). 



\bibitem{SerSha}
V.B. Serebrennikov, A.E. Shabad, 
Method of calculation of the spectrum of a centrally symmetric Hamiltonian on
the basis of approximate $O_4$ and $SU_3$ symmetries, 
Theor. Math. Phys. 8, (1971) 644--653.

\bibitem{BucDen}
L.H. Buch, H.H. Denman, 
Conserved and piecewise-conserved Runge vectors for the isotropic harmonic oscillator, 
Amer. J. Phys. 43 (1975) 1046--1048.

\bibitem{Per}
A. Peres,  
A classical constant of motion with discontinuities, 
J. Phys. A: Math. Gen. 12 (1979), 1711--1713.

\bibitem{LeaFle}
P.G.L. Leach, G.P. Flessas, 
Generalisations of the Laplace–Runge–Lenz vector,
J. Nonlin. Math. Phys. 10 (2003), 340--423. 



\bibitem{Cha1995}
S. Chandrasekhar, 
{\it Newton's Principia for the Common Reader}, (Oxford University Press) 1995.

\bibitem{Lyn-Bel}
D. Lynden-Bell, R.M. Lynden-Bell, 
On the shapes of Newton's revolving orbits, 
Notes and Records of the Royal Society of London. 51(2)  (1997), 195--198.



\bibitem{MilPosWin}
W. Miller, S. Post, P. Winternitz, 
Classical and quantum superintegrability with applications, 
J. Phys. A 46 (2013), 423001.

\bibitem{AncBalGan}
S.C. Anco, A. Ballesteros, M. Gandarias, 
Global versus local (super)integrability of a nonlinear oscillator, 
Phys. Lett. A. 383 (2019), 801--807. 



\bibitem{Joh}
F. John,
{\it Partial Differential Equations} 
Applied Math. Sci. Volume 1 (Springer, New York) 1982. 



\bibitem{Olv}
P.J. Olver,
{\it Applications of Lie Groups to Differential Equations},
(Springer, New York) 1986.

\bibitem{BA-book}
G. Bluman and S.C. Anco,
{\it Symmetry and Integration Methods for Differential Equations},
Applied Math. Sci. Volume 154
(Springer, New York) 2002.

\bibitem{Anc-review}
S.C. Anco,
Generalization of Noether's theorem in modern form to non-variational partial differential equations.
In: Recent progress and Modern Challenges in Applied Mathematics, Modeling and Computational Science, 119--182,
Fields Institute Communications, Volume 79, 2017.




\bibitem{Muk}
N. Mukunda, 
Dynamical symmetries and classical mechanics, 
Phys. Rev. 155 (1967) 1383--1386. 

\bibitem{Lev}
J.M. L\'evy-Leblond, 
Conservation laws for gauge-invariant Lagrangians in classical mechanics,
Amer. J. Phys. 39 (1971), 502-–506. 

\bibitem{Rog}
H. Rodgers,
Symmetry transformations of the classical Kepler problem, 
J. Math. Phys. 14 (1973), 1125--1129.



\bibitem{AbrSte}
M. Abramowitz, I.A. Stegun,
{\em Handbook of Mathematical Functions: with Formulas, Graphs, and Mathematical
 Tables}, Volume 55, National Bureau of Standards (1964).



\bibitem{AshHan}
A. Ashtekar, R.O. Hansen, 
A unified treatment of null and spatial infinity in general relativity. I -- 
Universal structure, asymptotic symmetries, and conserved quantities at spatial infinity, 
J. Math. Phy. 19 (1978) 1542--1566.

\bibitem{HalLed}
J Hal\'a\v{c}ek and T. Ledvinka, 
The analytic conformal compactification of the Schwarzschild spacetime,
Class. Quantum Grav. 31 (2014) 015007.



\end{thebibliography}
\end{document}